\documentclass[11pt]{article}

\usepackage{amsfonts}

\usepackage{amsmath}
\usepackage{amssymb}
\usepackage{latexsym}
\usepackage{graphicx}
\usepackage[english]{babel}
\usepackage[font={small}]{caption}

\usepackage{amsfonts}
\usepackage{latexsym}
\usepackage{graphicx}
\usepackage[english]{babel}
\usepackage{tikz}

\begin{document}
\input epsf

\def\p{\partial}
\def\h{{1\over 2}}
\def\be{\begin{equation}}
\def\bea{\begin{eqnarray}}
\def\ee{\end{equation}}
\def\eea{\end{eqnarray}}
\def\d{\partial}
\def\la{\lambda}
\def\eps{\epsilon}
\def\b{\bigskip}
\def\m{\medskip}

\newcommand{\newsection}[1]{\section{#1} \setcounter{equation}{0}}

\def\q{\quad}

\def\h{{1\over 2}}
\def\t{\tilde}
\def\r{\rightarrow}
\def\nn{\nonumber\\}

\let\p=\partial

\newcommand\blfootnote[1]{%
  \begingroup
  \renewcommand\thefootnote{}\footnote{#1}%
  \addtocounter{footnote}{-1}%
  \endgroup
}

\begin{flushright}
\end{flushright}
\vspace{20mm}
\begin{center}
{\LARGE The VECRO hypothesis}
\\
\vspace{18mm}
{\bf    Samir D. Mathur }\\
\vspace{10mm}
Department of Physics,\\ The Ohio State University,\\ Columbus,
OH 43210, USA \vspace{2mm}\\ mathur.16@osu.edu

\vspace{8mm}
\end{center}

\vspace{4mm}

\thispagestyle{empty}
\begin{abstract}

\vspace{3mm}

We consider three fundamental issues in quantum gravity: (a) the black hole information paradox (b) the unboundedness of entropy that can be stored inside a black hole horizon (c) the relation between the black hole horizon and the cosmological horizon.  With help from the small corrections theorem, we convert each of these issues into a sharp conflict. We then argue that all three conflicts can be resolved by the following  hypothesis: 
{\it  the vacuum wavefunctional of quantum gravity contains  a `vecro' component made of virtual fluctuations of configurations of the same type that arise in the fuzzball structure of black hole microstates}.  Further,  if we assume that causality holds to leading order in gently curved spacetime, then  we {\it must} have such a vecro component  in order to resolve the above conflicts. The term vecro stands for   `Virtual Extended Compression-Resistant Objects', and characterizes the nature of the vacuum fluctuations that resolve the puzzles. It is interesting that puzzle (c) may relate the role of quantum gravity in black holes to observations in the sky.  

\end{abstract}
\newpage

\setcounter{page}{1}

\numberwithin{equation}{section} 

\section{Introduction}

\b

Suppose we start with two assumptions:

\b

{\bf A1:} If curvatures are low everywhere  (i.e. ${\mathcal R}\ll l_p^{-2}$) and further, we are not close to making a closed trapped or anti-trapped surface anywhere (i.e., we do not have black hole or cosmological horizons) then semiclassical gravity holds to leading order. 

\b

{\bf A2:} If curvatures are low throughout a spacetime region, then causality holds to leading order. Here causality means that signals do not propagate outside the light cone, and that there are no nonlocal interactions between points that are spacelike separated.

\b

Then we will argue that the vacuum of quantum gravity must contain an important component which is normally not part of our picture of the vacuum in a quantum field theory. This component is made of quantum fluctuations which we call `vecros'; this acronym stands for `Virtual Extended Compression-Resistant Objects', and is a notion that we will explain below. 

The presence of the vecro component in the vacuum will resolve the black hole information paradox \cite{hawking} while preserving assumptions A1 and A2. Further, we will argue that if we do not have this vecro component, then we cannot keep A1,A2 and also require that the Hawking radiation from a black hole carry information the way information is carried by the radiation from a piece of burning coal.

There are other important puzzles associated to black holes. Gedanken experiments suggest that
\be
S_{bek}={A\over 4G}
\ee
gives the entropy of a black hole \cite{bek}. If we try to interpret this entropy as the log of the number of black hole microstates, then we run up against the following problem: the traditional picture of the hole allows us to construct explicitly an {\it infinite} number of states in the region inside the horizon, with mass the same as the mass of the black hole \footnote{This construction is closely related to the `bags of gold' construction of Wheeler \cite{bags} and the construction of `monster' states \cite{monsters}. See also \cite{ho}.}

A further puzzle comes from cosmology. The big bang can be regarded as a `white hole'; i.e., if we reverse the direction of time, then a ball shaped region of an expanding dust cosmology looks like a dust ball collapsing to make a black hole. If we look back to early times, then we see this dust ball shrink to sizes much smaller than its horizon radius. But cosmological observations do not manifest any large quantum gravity effects during the process of this shrinkage.  This is a difficulty for {\it any} theory that postulates a resolution of the information paradox by invoking large quantum gravity effects at the horizon scale.  (The small corrections theorem \cite{cern} already rules out any solution to the paradox where the quantum gravity corrections at the horizon are small, and where we do not allow order unity nonlocal effects on distances much larger than the horizon scale.) How do we understand this difference between the black hole horizon and the cosmological horizon? More generally, semiclassical physics has the same behavior at different horizons: Rindler, black hole and cosmological including de Sitter.  Does the full quantum gravity theory have the same physics at all these horizons?

With help from the small corrections theorem, we will note that each of these three difficulties can be made into a  sharp contradiction. In other words,  if we assume that the traditional picture of the hole holds to leading order, then we cannot get out of the contradictions posed by the above physical situations while preserving assumptions A1,A2. 

We will then see how the vecro hypothesis can resolve all three puzzles.  In fact we will argue that if we do not accept the vecro picture of the quantum gravity vacuum,  then we cannot resolve these puzzles while preserving assumptions A1,A2.

\subsection{The vecro hypothesis: a first pass}

  \begin{figure}[htbp]
\begin{center}
\includegraphics[scale=.5]{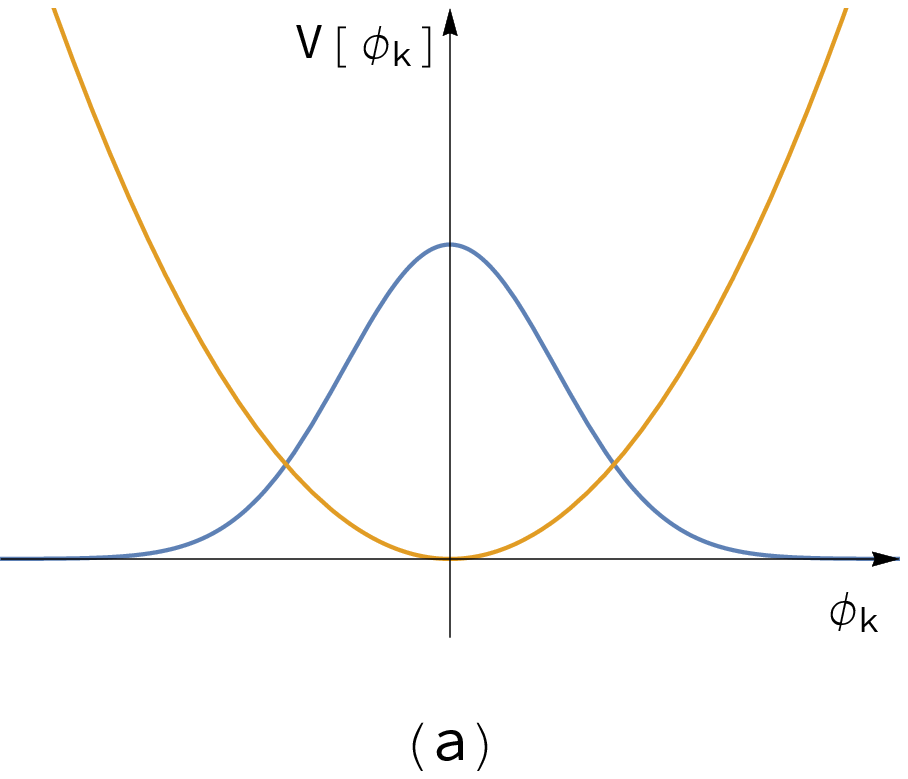}\hskip20pt
\includegraphics[scale=.5]{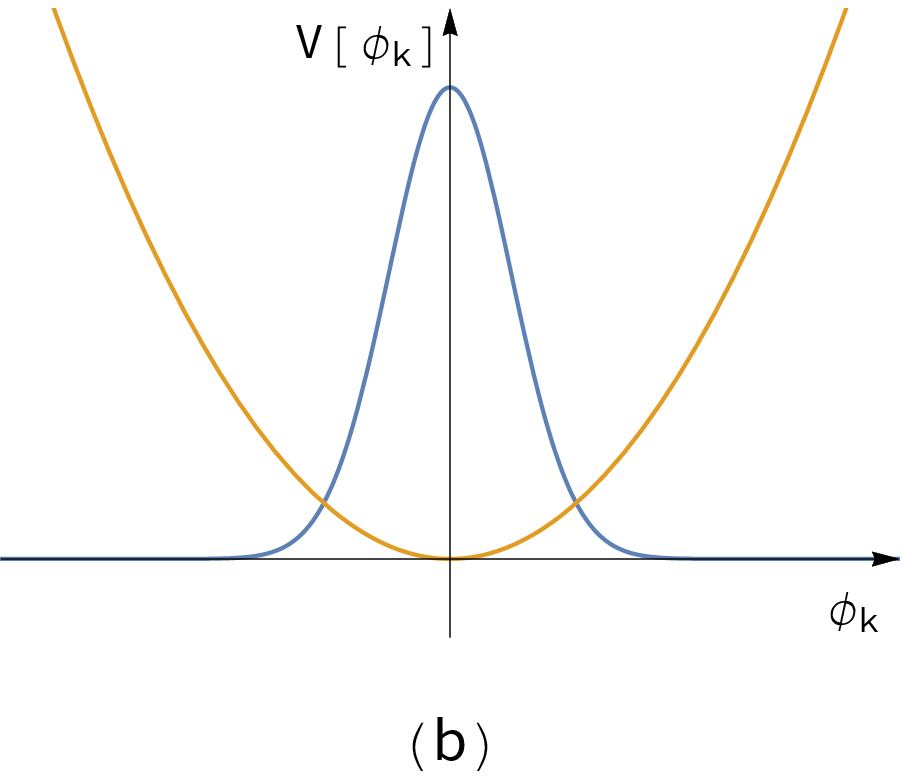}\hskip20pt
\includegraphics[scale=.5]{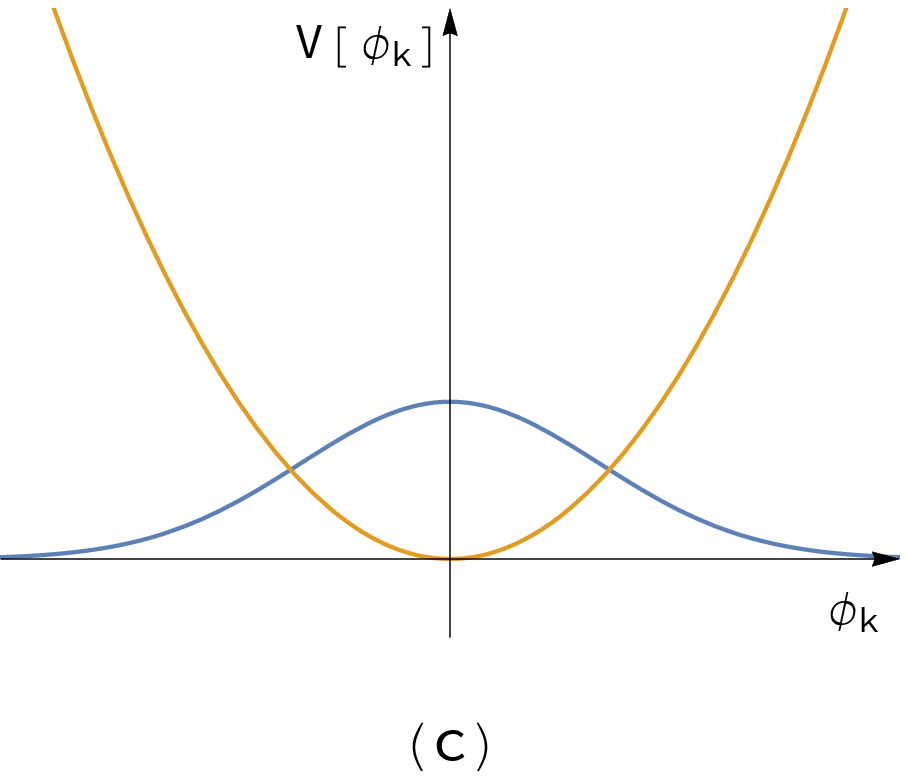}
\end{center}
\caption{(a) The potential for a Fourier mode of a scalar field, and the ground state wavefunction for this mode. (b) A narrower wavefunction has a higher energy, which can be interpreted as having pairs of quanta of the field. (c) A broader wavefunction also has a larger energy, and a similar interpretation.}
\label{fig1}
\end{figure}

Let us make a first pass at the idea of the vecro hypothesis; we will explore the idea in more detail in the next section. 

First consider the physics of quantum fields on curved spacetime. Let the quantum field be a scalar $\phi(x)$ satisfying $(\square+m^2)\phi=0$. Classically, the vacuum has $\phi=0$. Quantum mechanically, we find that the wavefunctional spreads to nonzero values of $\phi$. In flat spacetime we can break $\phi$ into its Fourier modes $\phi_k$, and find that $\phi_k$ behaves as the coordinate of a harmonic oscillator with frequency
\be
\omega_k=(\vec k^2+m^2)^\h
\ee
This potential is indicated by the quadratic potential in fig.\ref{fig1}(a).
The vacuum state of the field $\phi$  contains quantum fluctuations described by  wavefunctions for the $\phi_k$
\be
\psi(\phi_k)=\left ( {\omega_k\over \pi}\right)^{1\over 4} e^{-{\omega_k\over 2}\phi_k^2}
\label{ione}
\ee
The energy of the harmonic oscillator for $\phi_k$ is $\h\omega_k$, so the part of the wavefunction in the region
\be
|\phi_k|\gg \omega_k^{-\h}
\ee
is `under the potential barrier'. 

In fig.\ref{fig1}(b),(c) we depict squeezed states where the wavefunction has less or more spread in $\phi_k$ than the vacuum wavefunction. In either case the energy of the state is more than the energy of the ground state (\ref{ione}). We can regard the extra energy as describing pairs of particles in the mode $\phi_k$. 

Now imagine that instead of flat spacetime we have spacetime which changes on a time scale $T$.  Suppose the evolution is slow
\be
T\gg \omega_k^{-1}
\ee
then the vacuum state  evolves to the new vacuum state on the deformed manifold. But if 
\be
T\lesssim \omega_k^{-1}
\ee
then the wavefunction deforms to a squeezed state, and we say that we have particle creation.

\b

  \begin{figure}[htbp]
\begin{center}
\includegraphics[scale=.7]{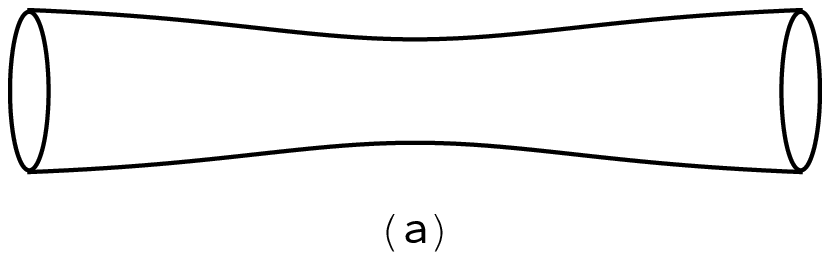}\hskip100pt
\includegraphics[scale=.7]{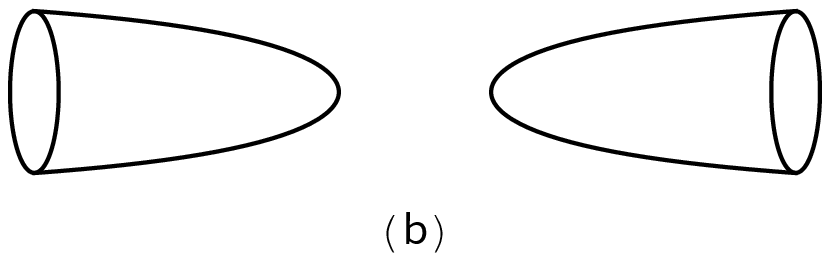}
\end{center}
\caption{(a) Fluctuations of the radius of a compact direction can be regarded as scalar fields on the noncompact spacetime; such deformations are covered by the semiclassical approximation. (b) Fluctuations where the compact directions pinch off lead to a topology where the compact directions are not a direct product with the noncompact spacetime; such configurations are not included in the usual semiclassical description of the dimensionally reduced theory.}
\label{fig2}
\end{figure}

Next, consider the situation where this scalar arises from the compactification of an extra dimension. For a compact circle parametrized by a coordinate $w$ we can define the scalar $\phi$ through
\be
g_{ww}=e^\phi
\ee
In fig.\ref{fig2}(a) we depict deformations of the circle radius which correspond to the excitations of $\phi$. The physics of such deformations is covered by our discussion above of quantum fields on curved spacetime $\phi(x)$. 

But we can also have deformations where the circle pinches off completely and we get a different topology; such a situation is depicted in fig.\ref{fig2}(b). Simple examples of manifolds with this change of topology include Euclidean Schwarzschild $\times$ time, and the bubble of nothing \cite{bubble}. As we will note below, the dynamics of the scalar field in these situations is somewhat counterintuitive. Dimensional reduction on the circle gives a scalar with the normal positive sign of energy density, but the  Euclidean Schwarzschild $\times$ time geometry does not shrink under its own gravity, and the bubble of nothing actually expands outwards. These behaviors can of course be traced to the fact that the dimensional reduction itself fails where the compact circle pinches off, so the spacetime is no longer a direct product of the noncompact directions $x$ with a compact manifold. 

Such nontrivial manifolds are of interest to us because in string theory they arise in the fuzzball solutions which give the microstates of black holes \cite{lm4,fuzzballs}. The somewhat counterintuitve gravitational behavior  noted above allows the solutions to evade the no-hair results and Buchdahl type arguments \cite{gibbonswarner,mexample}. Thus we can get horizon sized extended objects (fuzzballs) that do not have a horizon and are stable against collapse. 

For our present discussion we are not interested in the fuzzball solutions themselves, which are excitations of the theory with different masses $M$. Rather, we are interested in the structure of the {\it vacuum} wavefunctional. We have already noted that the vacuum contains fluctuations of any scalar fields $\phi(x)$ that live on the spacetime. When these scalar fields arise form the fluctuations in the radius of a compact direction, then the vacuum wavefunctional contains deformations like those in fig.\ref{fig2}(a). But by a natural extension of this argument, the vacuum wavefunctional should also have support over topologically nontrivial configurations like those in fig.\ref{fig2}(b). 

It takes more action to reach the configurations  in fig.\ref{fig2}(b) than those in fig.\ref{fig2}(a), so the amplitude of the wavefunctional will be smaller at the topologically nontrivial configurations. But against this is the fact that we have a vast space of these topologically nontrivial configurations: the black hole microstates are supported on such configurations, and the number of these states $Exp[S_{bek}(M)]$ is very large. 

The vecro hypothesis says that the part of the vacuum wavefunctional with support on this large space of nontrivial topology configurations is significant, and that this part of the vacuum wavefunctional gives rise to novel effects in situations where the classical theory would imply horizon formation. The fluctuations of interest are, by definition `virtual'  and we will see that they are  `extended' structures. We will also note that they have the property of being `compression-resistant'. Thus we call the gravitational configurations of interest `virtual extended compression-resistant objects' or `vecros' for short. 

Let us state in a little more detail when the vecro part of the vacuum wavefunctional is expected to become important.  First consider the usual quantum fluctuations of a scalar field $\phi$ on curved space, which we discussed above. There is no simple prescription for how much particle creation there will be in a given curved spacetime; one just has to evolve the vacuum wavefunctional $\Psi[\phi(x)]$ and interpret the result in terms of particles. In a similar way, the considerations of the present paper will of necessity be qualitative. For the case of the scalar field, we can, however,   note the curvature {\it scales} for which particle creation effects will be important.  If the field has a mass $m$, then $\omega_{\vec k}\ge m$ for all modes. Suppose the time scales $T$ over which the metric changes  satisfies 
\be
T\gg m^{-1}
\ee
Then there will be very little particle creation; the vacuum wavefunctional adjusts adiabatically to the changing metric of spacetime, so we do not create particles. In this situation we can absorb the small quantum effects of the  field $\phi$ into corrections to the local effective Lagrangian; i.e., the Einstein Lagrangian ${\mathcal R}$ gets corrected by higher order curvature terms. Schematically
\be
{\mathcal R}\rightarrow {\mathcal R}+c_1 {\mathcal R}^2 +c_2 {\mathcal R}^3 +\dots
\ee
where the coefficients $c_i$ depend on the mass $m$. If on the other hand $T\lesssim m^{-1}$, particles will be created with wavelengths $\sim T$, carrying an energy density $T^{-4}$. If $T\gg l_p$, the backreaction of these particles on the background is very small. Thus as long as we are in gently curved spacetime, the quantum physics of the scalar $\phi$ is a small correction to the classical  dynamics given by the Einstein action.

\begin{figure}[htbp]
\begin{center}
\includegraphics[scale=.6]{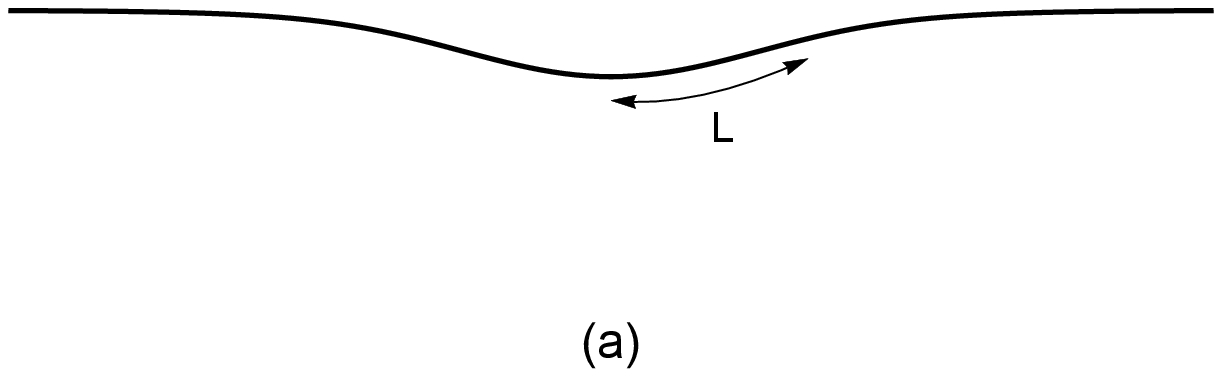}\hskip20pt
\includegraphics[scale=.6]{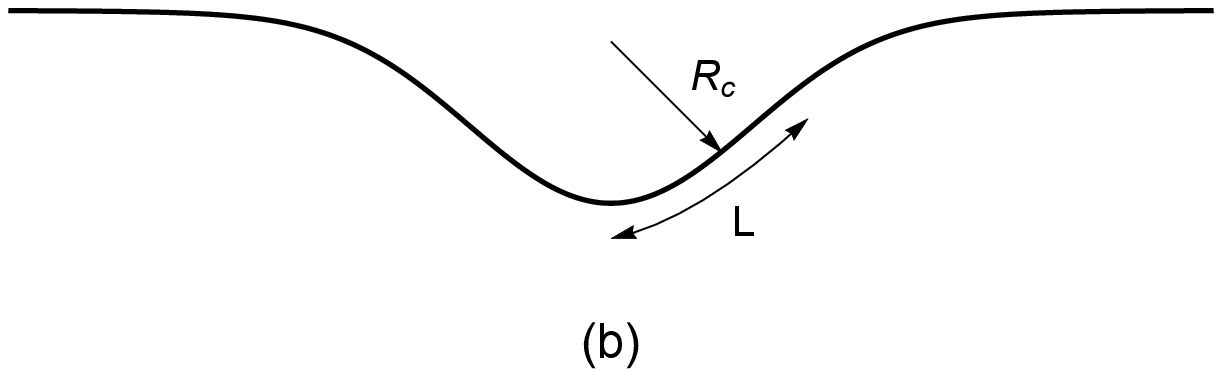}
\end{center}
\caption{(a) The length scale $L$ over which the curvature persists is much smaller than the radius of curvature $R_c$; in this situation the vecro part of the wavefunctional will not be important for the dynamics. (b) The length scale $L$ over which the curvature persists is comparable to or larger than  the radius of curvature $R_c$; in this situation the vecro part of the wavefunctional will be modified by order unity and will lead to significant corrections to the dynamics.}
\label{fig16}
\end{figure}

The situation is quite different for the vecro part of the vacuum wavefunctional. Let the curvature of space in some region be characterized by a curvature length scale ${\mathcal R}^{-\h} \sim R_c$. Now consider the length scale $L$ over which this curvature persists; outside a region of size $\sim L$, we assume that the curvature drops to much lower values (fig.\ref{fig16}(a)). If
\be
L\ll R_c
\label{none}
\ee
then there are no significant effects from the vecro part of the vacuum wavefunctional. If (\ref{none}) holds everywhere, then there is no new physics that we can expect from the vecros: the vecro part of the wavefunctional adiabatically adjusts to the changing curvature. The dynamics is given to leading order by the classical Einstein action and the small quantum corrections that do arise can be obtained from  the semiclasssical physics of quantum fields on curved spacetime. 

But now suppose that
\be
L\gtrsim R_c
\label{ntwo}
\ee
(see fig.\ref{fig16}(b)). In this situation the vecro configurations in the wavefunctional get severely affected: they are extended objects which get squeezed by  a factor of order unity, and their compression resistance leads to an increase in energy which competes with the classical energy giving rise to the curvature radius $R_c$. The vecro part of the vacuum wavefunctional -- i.e., the part supported on the vast space of configurations that are schematically like fig.\ref{fig2}(b) -- gets modified by order unity, so that instead of preserving the vacuum we create on-shell fuzzballs. This process is the analogue of particle creation, except that instead of pointlike particles we are creating the extended structures that arise in the wavefunctional describing fuzzballs. 

We will see that for gravitational fields like that of a star, we are in situation (\ref{none}), so there is no new physics expected from the vecro part of the vacuum wavefunctional.  The two principal situations where we encounter (\ref{ntwo}) are (i) the black hole  and (ii) the cosmological horizon. This is where we need novel physics to evade the paradoxes listed above, and we will see that the vecro part of the wavefunctional will help us to evade the paradoxes.

Very roughly, our situation is similar to that of superfluidity a century ago. The behavior of liquid helium was  puzzling. It was explained using a heuristic model of new component of the quantum vacuum wavefunctional, though the detailed derivation of this vacuum structure was worked out later. In our case we have sharp paradoxes posed by gravity. We will  argue that the constraint of leading order causality in gently curved spacetime forces us to a heuristic picture of a new component of the vacuum wavefunctional -- the vecro component.  We hope that a detailed derivation of this component will be worked out later through computations in string theory.

\section{What are vecros?}

In this section we introduce the idea of vecros in more detail. We will proceed in steps, examining each property of a vecro which leads to a letter in its acronym. 

\subsection{Virtual fluctuations}\label{secvv}

The vacuum of QED contains vacuum fluctuations that are composed of an electron-positron pair, or a quark-antiquark pair etc.. Could we also have  fluctuations corresponding to bound states like the  positronium or a pion?  

At some level, the answer should be yes, i.e., there should be a signal of the existence of any tightly bound state in the vacuum wavefunctional. One has to a little careful in making this precise, however. Ignoring the decay of the positronium  for the moment, a state with  one  positronium $|\psi_{positronium}\rangle$ would be orthogonal to the vacuum; i.e., $\langle 0 | \psi_{positronium}\rangle=0$, and so it is not clear what we mean by saying that there are fluctuations corresponding to positroniums in the vacuum. But on the other hand we can think of a pion as a fundamental field $\phi_\pi$ (as was done in  the early days of strong interactions), and then we would have fluctuations of pairs of pions just as we would have for the quanta of any other scalar field.

We will come to these subtleties later in section \ref{seccc}  below. For the moment let us adopt a heuristic picture that the vacuum contains fluctuations that correspond to complicated states like positroniums and pions.  By extension, we should also have vacuum fluctuations describing larger virtual objects, like atoms, and also  more extended objects, like  benzene rings.

While fluctuations describing benzene rings are {\it possible}, they are not {\it likely}, and that is why we do not normally worry about them. A fluctuation of energy $E$, existing for a time $T$, has an action $\t S\sim ET$.  We may therefore guess that the probability for such a fluctuation is suppressed as
\be
P\sim e^{-\t S}\sim e^{-ET}
\label{one}
\ee
Now consider black holes. Black holes exist for all masses M, and therefore by the above reasoning they must also exist as virtual fluctuations in the vacuum. The mass $M$ is related to the horizon radius $R_0$ in $d+1$ spacetime dimensions as 
\be
M\sim {R_0^{d-2}\over l_p^{d-1}}
\ee
Setting $T\sim R_0$ for the duration of the fluctuation, we have for the action
\be
\t S\sim M R_0 \sim \left ( { R_0\over l_p}\right )^{d-1}
\ee
Thus the probability $P$ of fluctuating to a given state of the hole is very small for $R_0\gg l_p$. But we should multiply this probability by the number of black hole microstates that we can fluctuate to. A special feature of black holes, that sets them apart from other massive objects, is that their degeneracy ${\cal N}\sim Exp[S_{bek}]$ is very large:
\be
S_{bek}\sim {A\over G}\sim \left ( {R_0\over l_p}\right )^{d-1}
\ee
We then see that it is possible to have for the total probability 
\be
P_T\sim P {\cal N} \sim 1
\ee
A similar argument was used in \cite{tunnel} to argue that a collapsing shell can violate the semiclassical approximation and tunnel into fuzzballs upon reaching the horizon. Here we are  arguing that the virtual fluctuations of black holes of mass $M$ cannot be ignored in the vacuum, for {\it any} $M$. The kinds of configurations that are involved in black hole microstates will be our vecros.   Their {\it virtual} nature gives the first letter `v' of the acronym. 

The above argument is obviously very rough, but all we wish to  take from it is that we should look seriously at the role of a certain type of virtual fluctuation in a quantum gravity theory which corresponds to large complicated structures. It is obviously vital in this argument that we have a large phase space for such structures. To see the significance of  having a large number of degrees of freedom, consider the following quantum mechanical problem \cite{masoumi1}.  Take a particle of unit mass in a 1-d square well with potential 
\be
V(x)=-V_0, ~~~|x|<a, ~~~~~~~~~~V(x)=0, ~~~|x|>a
\ee
The wavefunction of a bound state is oscillatory in a region $|x|<a$ which corresponds to the classical range of  motion inside the well. But the wavefunction also has an exponentially decaying support in the region $|x|>a$ where the energy is `under the barrier'. Suppose that 
\be
\int_{x=-a}^a  dx\,  |\psi(x)|^2  = 1-\epsilon, ~~~~~\epsilon \ll 1
\ee
so that most of the norm is in the well $|x|<a$. Now consider the particle in  $N$ dimensions with potential
\be
V(x_1, x_2, \dots x_N)= V(x_1)+V(x_2)+\dots V(x_N)
\ee
where $V(x)$ is the same potential as above, and
\be
 N\gg 1/\epsilon
 \ee
The bound state wavefunction is
\be
\Psi=\psi(x_1)\psi(x_2)\dots \psi(x_N)
\ee
The  norm in the central well $|x|< a$ is now
\be
\Big  ( \int_{x_1=-a}^a 
dx_1 |\psi(x_1)|^2\Big) \Big ( \int_{x_2=-a}^a dx_2 |\psi(x_2)|^2\Big )  
\dots \Big ( \int_{x_N=-a}^a dx_1 |\psi(x_N)|^2  \Big )=(1-\epsilon)^N 
\approx e^{-\epsilon N}\ll 1
  \label{vone}
\ee
so that most of the norm is in the region {\it under the barrier}.  This part of the wavefunction under the barrier describes virtual fluctuations: these are configurations where the particle cannot go classically. In these forbidden regions the momentum is $p^2<0$; this is what makes the wavefunction have an exponential decay behavior $e^{-\alpha x}$  instead of an oscillatory behavior $\sin(k x)$. What we see in (\ref{vone}) is that for systems with many degrees of freedom the virtual  part of wavefunction can be very significant; in fact most of the norm may live under the barrier. 

Now consider quantum gravity. The configuration space in canonically quantized 3+1 gravity is the space of 3-geometries, so the wavefunctional is $\Psi[{}^{(3)}g]$. We will actually be working with string theory, but instead of writing all the string fields we will just write these fields schematically as $\{ g \}$, so the wavefunctional will be denoted $\Psi[g]$. Let the variables conjugate to the $g$ be called $\pi$.   Let the vacuum wavefunctional describing Minkowski spacetime be $\Psi_0[g]$. Then the virtual part of the wavefunctional $\Psi_0[g]$ is the part with support over the configurations $g$ where $\pi^2<0$.  It was argued in \cite{causality1, causality2} that ignoring this virtual part is what leads to the paradoxes associated to black holes. 
We will now discuss the nature of the configurations $g$ where this virtual part is supported.

\subsection{The extended nature of black hole microstates}\label{secee}

We now discuss the nature of the configurations in string theory which are `under the barrier'  in the vacuum wavefunctional but  which will nevertheless be important for dynamical process like black hole formation. We have noted that there must be a vast number of states in quantum gravity that account for the large entropy $S_{bek}(M)$ of black holes. While the black hole states are states with energy $M$, we are interested in the gravity {\it configurations} over which these states are supported; we will then look at the vacuum wavefunctional which has energy {\it zero} but nevertheless has a tail of its wavefunctional over these same configurations.

In traditional semiclassical gravity, the black hole is `empty space' with a singularity at its center. In such a picture it is not clear what distinguishes different microstates from one another; this is the well known `no hair' feature of semiclassical black holes.  In string theory, we find however that black hole microstates are {\it fuzzballs}: horizon-sized objects which themselves possess no horizon. To understand the structure of fuzzballs, consider the main reason why the traditional picture of the hole has a vacuum region near the horizon. An object placed just outside the horizon needs an intense acceleration to stay in place; thus any sufficiently compact star tends to collapse through the horizon. In particular Buchdahl's theorem \cite{buchdahl} considers  a perfect fluid, whose density $\rho$ increases monotonically inwards. If the radius $R$ of the fluid ball satisfies
\be
R<{9GM\over 4}
\label{qthree}
\ee
then the pressure will diverge at some radius $r>0$, rendering the solution invalid. Thus any fluid ball that has been compressed to a size  (\ref{qthree}) must necessarily collapse and generate a horizon. 

To understand the structure of fuzzballs,  we recall the toy model studied in \cite{mexample}. Consider the 4+1  dimensional spacetime obtained by adding a trivial time direction to the 3+1 dimensional Euclidean  Schwarzschild solution
\be
ds^2= -dt^2 + (1-{r_0\over r})d\tau^2 + {dr^2\over 1-{r_0\over r}} + r^2 (d\theta^2+\sin^2\theta d\phi^2)
\label{metrickk}
\ee
 The `Euclidean time' direction $\tau$ is compact, with $0\le \tau < 4\pi r_0$. This  metric is a perfectly  regular solution of the 4+1 vacuum Einstein equations. The $r, \tau$  directions form a cigar, whose tip lies at $r=r_0$.  The  spacetime ends at $r=r_0$; we can say that the ball $r<r_0$ has been excised from the manifold, and the compact directions closed off to generate a geodesically complete spacetime.  The `pinch-off' of the $\tau$ circle at $r=r_0$ is an example of the structure we had schematically depicted in fig.\ref{fig2}(b) in our first pass at the problem. 

We can  dimensionally reduce on the circle $\tau$, regarding the solution (\ref{metrickk}) as a 3+1 dimensional metric in the directions $(t, r, \theta, \phi)$ coupled to a scalar field 
\be
\Phi={\sqrt{3}\over 2}\ln  (1-{r_0\over r})
\ee
describing the  radius of the compact direction $\tau$. This scalar  is a  standard minimally coupled massless field. Its stress tensor in the above solution  works out to be
\be
T^\mu{}_{\nu} =  {\rm diag }\{-\rho, p_r, p_\theta, p_\phi    \}={\rm diag }\{-f, f, -f, -f    \} 
\ee
where
\be
f= {3r_0^2\over 8r^4 (1-{r_0\over r})^{3\over 2}}
\ee
We see that the pressures do diverge at $r\r r_0>0$, and if we followed the spirit of Buchdahl's theorem, we would discard this solution. But the solution is actually  a perfectly regular solution in 4+1 dimensions; what breaks down is the dimensional reduction map when the length of the compact circle goes to zero. 

\begin{figure}[htbp]
\begin{center}
\includegraphics[scale=.5]{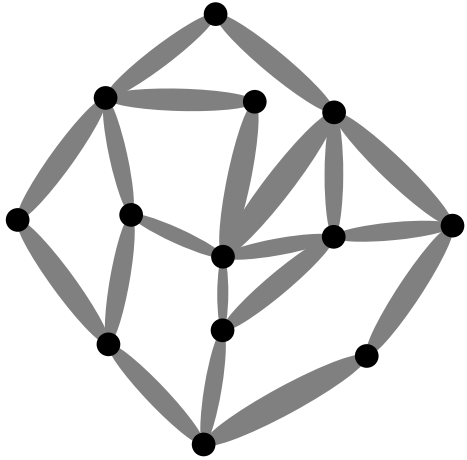}
\end{center}
\caption{A schematic depiction of a fuzzball. The dots represent KK monopoles or antimonopoles, and the grey ellipses denote $S^2$ spheres between the centers of these monopoles and antimonopoles. These spheres carry fluxes, which help maintain the rigidity of the fuzzball structure.}
\label{fig9}
\end{figure}

The above toy example helps us understand qualitatively how features of string theory like extra dimensions and extended objects allow fuzzball solutions to exist while standard 3+1 dimensional quantum gravity gave no such solutions. A large number of actual fuzzball microstates in full string theory have been constructed; many are extremal, but some nonextremal ones have been found as well. Many classes of these microstates  have the following qualitative structure. There is a set of KK monopoles and antimonopoles where the a compact $S^1$ shrinks smoothly to zero. There is topological spheres $S^2$ between any two of these topological objects, and there are gauge field fluxes on these spheres. We depict such a structure in fig.\ref{fig9}. 

We are  not saying that all fuzzball microstates will turn out have this form; for general holes there are large classes of fuzzball states that have not yet been constructed. Rather, what we are saying is that keeping in mind these explicit examples will help us to understand the vecro conjecture. Thus what we take from the above discussion is that  the field configurations on which the fuzzball wavefunctionals are supported are {\it extended}: the energy is not squeezed to an infinitesimal neighborhood of $r=0$. Note that the fuzzball cannot have a radius $R_f$ less than the horizon size $R_0$, since then the inward pointing structure of light cones would force it to collapse to a point. 
It has been argued in \cite{observe} that the radius $R_f$ of a typical fuzzball should exceed the $R_0$ by order planck length 
\be
R_f\approx R_0+\epsilon, ~~~\epsilon \sim l_p
\ee
Simple (i.e. nongeneric) fuzzball configurations can have a size much larger than $R_0$.\footnote{The existence of fuzzball structure at the horizon has led to the possibility that such structure could be observed; for discussions of this issue see \cite{observe,afshordi}.} 

The `extended' nature of the configurations we are interested in gives the second letter of our acronym.

\subsection{Compression-resistance}\label{seccc}

The last property we need is that the fuzzballs are {\it compression-resistant}. (These terms supply the letters `c,r' in our acronym; the final letter `o' just stands for `objects'.) 

Causality tells us that no object in a relativistic theory can be completely rigid. But fuzzballs are not easy to compress; their energy rises quickly if we try to squeeze them. This can be seem by looking at particular examples of fuzzball constructions. Consider the fuzzball structure depicted in fig.\ref{fig9}. There are topological spheres $S^2$ between the centers, and these spheres carry fluxes of various gauge fields in the theory. While the number of units of flux is given by an integral $\int F$ over the field strength, the energy of the field is quadratic: $E\sim \int F^2$. Thus squeezing the fuzzball makes its energy rise. It is also true that the energy will rise if we try to {\it expand} the fuzzball; the metric has a redshift at each point in the fuzzball, and expanding the fuzzball reduces this redshift, leading to a rise in energy. In other words, the fuzzball has an extended structure that has stabilized at a length scale where its energy is a local minimum. 

In this regard the fuzzball is  different from a string loop. The string loop is also an extended object, prevented from shrinking to a point by the motion of its segments. If we compress such a loop there is no resistance from the potential energy of the string; in fact the tension encourages the loop to contract. Any resistance to compression comes from the kinetic part of the energy. By contrast, the fuzzball configurations have a potential energy that rises when the fuzzball is compressed or expanded.

As noted above, our interest is not in the fuzzballs themselves but in the nature of the gravity vacuum functional which we have schematically written as $\Psi_0[g]$. How should we understand the role of vacuum fluctuations of fuzzballs in the structure of $\Psi_0[g]$? To answer this, let us first return to the case of a simple quantum field theory, and ask how the vacuum wavefunctional contains the information of bound states.

 \begin{figure}[htbp]
\begin{center}
\includegraphics[scale=.6]{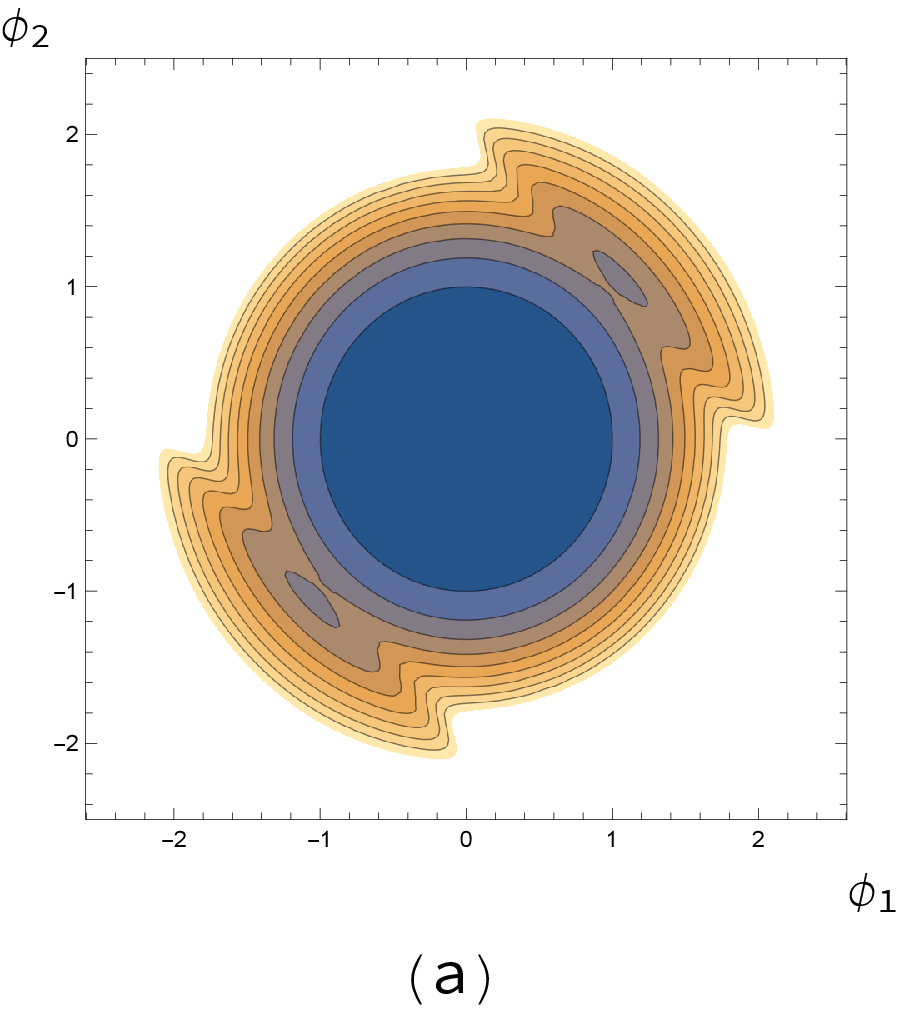}
\hskip30pt
\includegraphics[scale=.6]{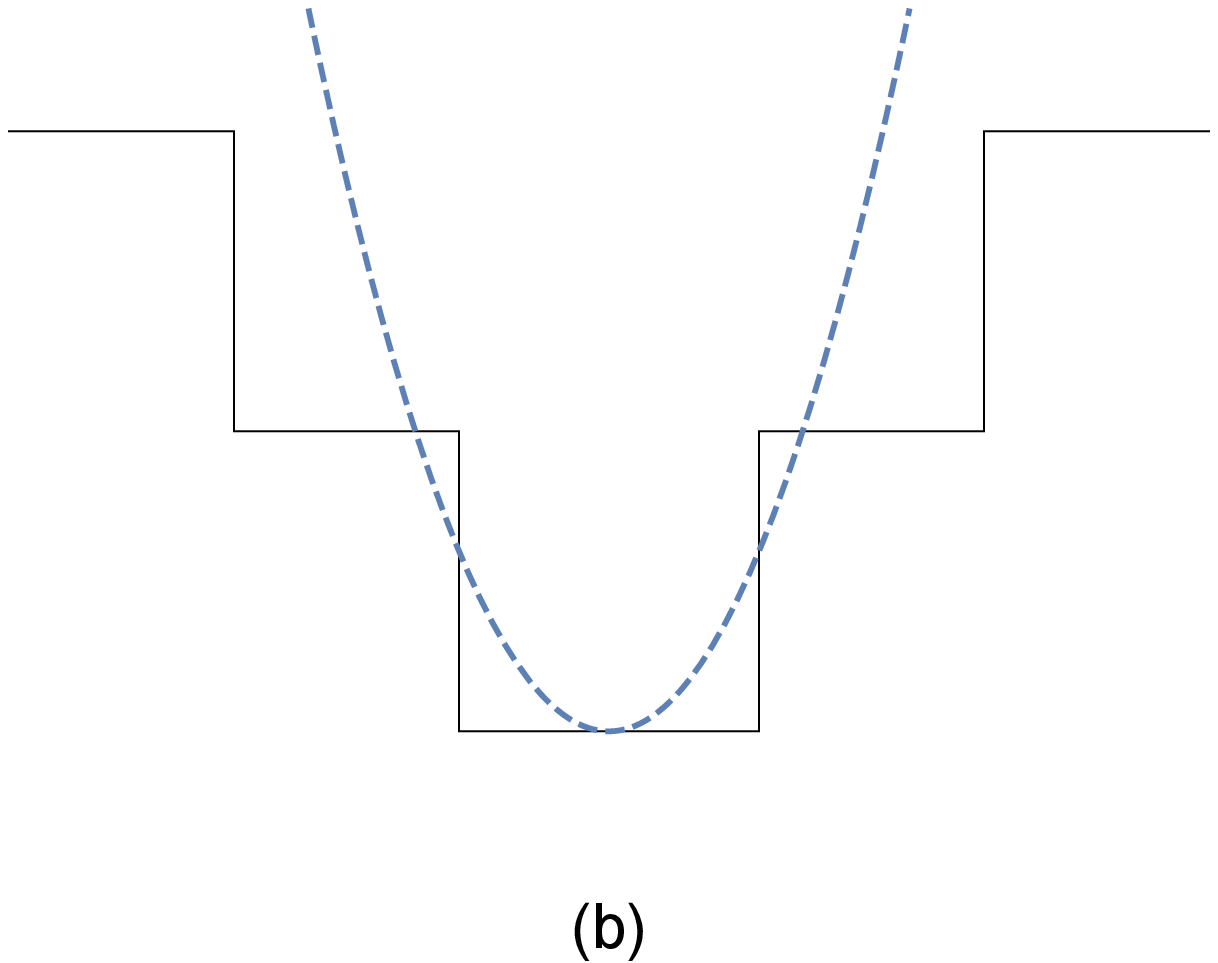}
\end{center}
\caption{(a) A contour plot of the potential energy in field space $\phi_1, \phi_2$ where the two scalar fields have an attractive interaction that leads to a bound state. The energy rises less slowly in some region along the line $\phi_1=\phi_2$ since when both these fields are present there is an  attraction giving a bound state. (b) The potential plotted along the direction $\phi_1=\phi_2$; the dips in the potential compared to the dashed line exhibit the features that leads to bound states.}
\label{fig10}
\end{figure}

Consider a field theory with two scalar fields $\phi_1, \phi_2$, each with mass $m$. Let their interaction be such  such that there is a bound state in the theory; we may roughly describe this bound state as made of one $\phi_1$ quantum and one $\phi_2$ quantum. In fig.\ref{fig10} (a) we plot schematically the potential in $\phi_1-\phi_2$ space. There is the usual quadratic part $\h m^2(\phi_1^2+\phi_2^2)$. But along the line $\phi_1=\phi_2$, there is a dip in the potential: thus if both fields are  nonzero the energy is lower, and this is what leads to the bound state. In fig.\ref{fig10}(b) we depict this potential along the line $\phi_1=\phi_2$; the various bound states of the theory manifest their presence by lowering the potential to the solid line from the dashed curve that we would get otherwise.

Now consider the vacuum wavefunctional of this theory. We are interested in the tail of the wavefunctional, which is away from the peak at $\phi_1=\phi_2=0$. This tail can be found by a WKB approximation as it is the part `under the barrier'; it is schematically given by
\be
\psi(\phi_1, \phi_2)\sim e^{-\t S}
\label{thone}
\ee
where $\t S$ is the action for a Euclidean solution to the field equations from the point $\phi_1=\phi_2=0$ to the point $(\phi_1, \phi_2)$. Since we have a  lower potential in the region of field space corresponding to a bound state, the action to reach such regions will be typically lower than the action to reach other regions at the same distance from the origin. The amplitude $\psi$ will be correspondingly less suppressed in these regions corresponding to bound states. This is the way that the bound states show up in the vacuum wavefunctional.  A detailed computation along such lines should therefore replace the relation (\ref{one}) which we had used as a rough guide on our first pass. 

Fuzzballs are bound states of the full quantum gravity theory. Their wavefunctionals are supported on a set of configurations of the gravity theory; we may take fig.\ref{fig9} as a pictorial representation of such a configuration. Our hypothesis is that the vacuum wavefunctional has an important tail that lies on these configurations; the amplitude at such configurations is suppressed because the action $\t S$ in (\ref{thone}) is large, but this suppression is counteracted by the large space of such configurations. 

Now let us come to the compression resistance of the vecro configurations. Looking ahead,  fig.\ref{fig7}(a) depicts schematically the potential in the space of vecro configurations in flat spacetime. The figure also shows the corresponding vacuum wavefunctional. Now suppose there is a star of mass $M$ centered at the origin. The gravitational field of this mass causes the vecro to compress inwards. This compression raises the energy of the vecro, so the potential becomes similar to that of fig.\ref{fig7}(b). The vacuum wavefunctional will distort under this change of potential, and this is the physical effect that we will use to resolve our paradoxes. 

\subsection{Modelling the compression-resistance of the vecros}\label{secmm}

Our considerations are obviously qualitative, but it will be useful to make a heuristic model for the compression resistance of vecros so that the ideas can be understood more easily.

 \begin{figure}[htbp]
\begin{center}
\includegraphics[scale=1]{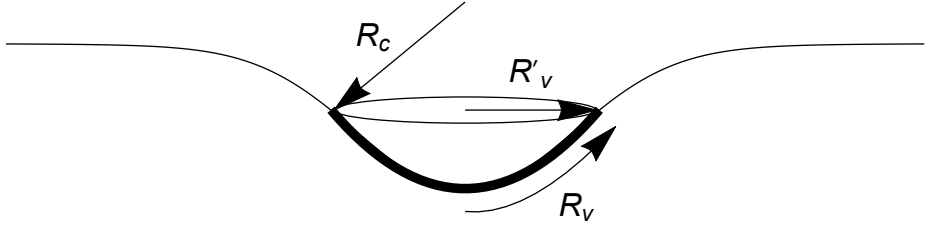}
\end{center}
\caption{A spacetime region with curvature radius $R_c$. A vecro with radius $R_v$ has its boundary points at a distance $R_v$ from its center. But the boundary is compressed to a smaller radius $R'_v$;  such a  compression raises the energy of the vecro configuration. }
\label{fig15}
\end{figure}

Consider a vecro configuration in flat space. Let it have a spherical boundary with radius $R_v$; see fig.\ref{fig15}.  Now suppose this space is deformed to be a sector of a sphere  with curvature radius $R_c$. On this spherical surface consider a ball shaped region  with radius $R_v$; this radius is defined by saying that all points on the surface of the ball are at a distance $R_v$ from the center.  The surface of the ball is now a sphere with radius  
\be
R'_v=R_c \sin{R_v\over R_c}
\ee
We thus see that this surface of the vecro is compressed by a factor
\be
 {R'_v\over R_v}\equiv 1-\delta
\ee
For $R_v\ll R_c$ the compression is small (i.e. $\delta \ll 1 $)
\be
\delta \approx {1\over 6}\left ( {R_v\over R_c}\right )^2
\label{aafive}
\ee
For such $\delta$ we model the compression resistance by a potential energy $U$ for the vecro that is quadratic in the compression 
\be
U\sim  \delta^2
\label{aaone} 
\ee
Note that we have assumed that the vecro in flat space is a configuration at the minimum of its energy: compressing or expanding the vecro raises its energy. 

The potential $U$ cannot  keep rising to arbitrarily large values. Once $\delta\sim 1$, the vecro structure breaks, and we are left with local virtual excitations that are uncorrelated across the length scale $\sim R_v$. These virtual excitations are just the normal local fluctuations of the vacuum, and do not give rise to the extra lowering of energy that comes from correlating the excitations into the vecro structure at scale $R_v$. Once the vecro is effectively broken, we assume that its energy does not rise upon further compression.  Thus we take  the potential energy function of the vecro  to level out around some critical value of the compression/expansion. Let us model this requirement by taking 
\be
U=k \sin^2 ({\pi\over 2} {\delta\over \delta_c}), ~~~~ -\delta_c<\delta<\delta_c
\label{aatwo}
\ee
where $\delta_c$ is a constant of order unity; we assume that further compression or expansion does not further raise the energy.  

To set $k$, we note that the structure of the vecro with radius $R_v$ is supported on configurations that contribute to black hole microstates with radius  $R\sim R_v$.  We assume that the energy scale $k$ in (\ref{aatwo}) corresponds to the mass of black holes with horizon radius $R_v$:
\be
M\sim {R_v^{d-2}\over G}
\label{masseq}
\ee
Thus we take
\be
k=\beta {R_v^{d-2}\over G}
\ee
where $\beta$ is a constant of order unity. This gives
\be
U(R_v, \delta)=\beta {R_v^{d-2}\over G} \sin^2 ({\pi\over 2} {\delta\over \delta_c}), ~~~~ -\delta_c<\delta<\delta_c
\label{aathree}
\ee
Thus if vecros of scale $\sim R_v$ are compressed by order unity, the energy increase of the state will be of order (\ref{masseq}).

\subsection{The scale where vecros  become relevant}\label{secrr}

Let us now see how our heuristic model above gives the scale (\ref{ntwo}) for novel physics to arise from vecros. 

Consider a  star  in $d+1$ dimensional spacetime with uniform density $\rho_0$ and radius $R_{star}$. The curvature radius of spacetime in the region of the star is
\be
R_c\sim (G\rho_0)^{-\h}
\label{aasix}
\ee
We write
\be
R_{star}=\mu R_c
\ee
We have 
\be
\mu\ll 1
\label{aafour}
\ee 
for a star, and $\mu\sim 1$ for an object that has radius of order its Schwarzschild radius. 

The energy of the star is
\be
E_{star}\sim \rho_0 R_{star}^{d}
\ee
Having the curvature radius $R_c$ will lead to a compression of vecros, and a consequent rise in their energy. The largest energy increase will arise for the largest vecros in our curvature region, which are vecros with
\be
R_v \sim R_{star}
\ee
Due to (\ref{aafour}), the compression $\delta$ is given by (\ref{aafive}).  
The compression energy for a vecro with $R_v=R_{star}$ is then
\be
U\approx \beta {R_{star}^{d-2}\over G} {\pi^2\over 144\delta_c^2 }\left ( {R_{star}\over R_c}\right )^4\sim (\rho_0 R_{star}^d)\left ( {R_{star}\over R_c}\right)^2
\ee
where we have used (\ref{aasix}). Thus
\be
{U\over E_{star}}\sim \mu^2\ll 1
\ee
so that the energy arising from the compression of vecros will not play a significant role for such a star. On the other hand for objects with $\mu\sim 1$, we will get
\be
{U\over E}\sim 1
\label{energyscale}
\ee
and this is where vecros will play a crucial role. 

\subsection{How vecros modify dynamics}

For the three puzzles we will address,  we will find in each case that  the condition (\ref{ntwo}) is met, which leads to the condition (\ref{energyscale}). Thus for each of these situations the dynamics of vecros should be  a relevant correction. Since the puzzles arose when we used the semiclassical approximation, this new source of corrections with the energy scale (\ref{energyscale}) invalidates the computations leading to the puzzles, thus resolving the sharp conflicts. We can also conjecture {\it how} vecros might affect the dynamics; let us summarize these conjectures here:

\b

(a) First consider the problem of unbounded entropy that we can get on certain kinds of hypersurfaces that can be embedded inside the semiclassical geometry of a black hole of mass $M$. We will see that these hypersurfaces have a `neck' region where the curvature satisfies (\ref{ntwo}). The vecros potential energy at this neck rises by $\gtrsim M$; thus the total energy on the hypersurface exceeds $M$, and we cannot actually embed such hypersurfaces in the black hole geometry.

\b

(b) Consider  a collapsing null shell with radius $R(\lambda)$ that passes through its horizon radius $R_0$ to make a black hole. In the region inside the horizon, light cones point inwards, so any vecro structure there is forced to keep compressing. This distorts such vecro configurations by order unity, and this in turn changes the vecro component of the vacuum wavefunctional in the region $R(\lambda)<r<R_0$ by a significant amount. Under this change, the virtual fluctuations describing vecros turn into on-shell fuzzball excitations, and we get a fuzzball instead of the traditional black hole geometry. Since fuzzballs radiate from their surface like normal bodies, we resolve the information paradox.

\b

(c) We will note that there is a significant difference in the vecro excitations between the case of a black hole in asymptotically flat spacetime and an expanding cosmology. In asymptotically flat spacetime the wavefunctional has support on vecros  with radii $R_v$ going all the way to infinity, while for the cosmology the support extends only over $0<R_v\lesssim H^{-1}$, where $H$ is the Hubble constant.  Thus quantum gravity effects differentiate between the dynamics at the black hole horizon and at the cosmological horizon, thus preventing us from mapping the black hole puzzle to a puzzle at the cosmological horizon.

\b

In each case we see that the extended nature of the vecro configurations allows the wavefunctional to `feel around' a region with any nonzero radius $R_v$ and to trigger new quantum gravity effects when this region has the structure of a trapped  surface. In the traditional picture of vacuum, quantum gravity fluctuations are confined to a fixed scale  $\sim l_p$, and there is no way for the wavefunctional to feel the existence of trapped surfaces. 

Note that with the vecro hypothesis the theory remains causal and local. The existence of extended vacuum fluctuations does not violate causality since the vacuum has been allowed to exist for times where these extended vecro correlations could develop. Signals in the theory still propagate within the light cone; thus the vecros introduce no nonlocality in the dynamics.

\section{Review of some earlier results}

The vecro picture may appear to be a radical departure from our usual picture of the quantum vacuum. But we will argue that we are {\it forced}  to this picture by the puzzles we face with quantum gravity.  To make such an argument, these puzzles must be turned into sharp contradictions. In this section we will recall some earlier results which we will need for this purpose.  

\subsection{The small corrections theorem}

The black hole information paradox \cite{hawking} says that general relativity and quantum theory are in conflict with each other. One might therefore expect that physicists working on quantum gravity -- and in particular, on string theory -- would be deeply concerned about this paradox. But for many decades the paradox remained in the background,  while work on other aspects of the theory proceeded. At least part of the reason for this was a hope that that some small, (heitherto unknown) quantum gravity effects would encode the information in a subtle way in the emitted radiation and thereby restore unitarity. After all, when a piece of coal burns away then its information does get encoded in the radiation, but it is very hard to actually unravel this information by looking at the radiation quanta. A theorem of Page \cite{page} says that one must look at more than half the photons emitted by the coal before any information of the coal can be decoded.

The small corrections theorem proves that small quantum gravity corrections to the Hawking radiation process {\it cannot} encode the information in the radiation. This result is of crucial importance to us. In our assumptions A1, A2, we have used the term `leading order'. This is important because it is always possible that quantum gravity effects induce some small  violation of what we consider `normal physics'. If such small corrections could invalidate our arguments, then we could not arrive at any firm conclusions about what is needed to resolve the information paradox. Because of the importance of the small corrections theorem to our discussions, we will present a summary of its derivation below; for more details see \cite{cern}.

\subsubsection{The nature of small corrections}

 Consider the classical metric of the Schwarzschild hole
 \be
 ds^2-(1-{2GM\over R}) dt^2+{dr^2\over 1-{2GM\over r}}+r^2d\Omega_2^2
 \ee
 Suppose that the picture of  `quantum fields on curved space' was a good approximation in all regions where the curvature was low; i.e., at all regions with ${\mathcal R}\lesssim l_p^{-2}$. Then entangled pairs will be created at the horizon. One member of the pair (called $b$) will escape to infinity as Hawking radiation while the other member (called $c$)  will fall into the hole carrying net negative energy and so lower the mass $M$. 
  
 At the first step of emission the state of the entangled pair may be schematically written as
\be
|\psi\rangle_1={1\over \sqrt{2}}\left (|0\rangle_{b_1}|0\rangle_{c_1}+|1\rangle_{b_1}|1\rangle_{c_1}\right )
\label{qone}
\ee
In the leading order evolution studied by Hawking, the evolution at the next step would be independent of the evolution at the first step, so the overall state would have the form
\bea
|\psi\rangle_2&=&{1\over 2}\left (|0\rangle_{b_1}|0\rangle_{c_1}+|1\rangle_{b_1}|1\rangle_{c_1}\right )
\left (|0\rangle_{b_2}|0\rangle_{c_2}+|1\rangle_{b_2}|1\rangle_{c_2}\right )\nn
\eea
But if we allow for small corrections, then the evolution at the next step may be slightly altered:   what happens at the second step  can depend on what happened at the first step
\bea
|\psi'\rangle_2&=&{1\over {2}} |0\rangle_{b_1}|0\rangle_{c_1}\left [
      (1+\epsilon_1)|0\rangle_{b_2}|0\rangle_{c_2}+(1-\epsilon_1)|1\rangle_{b_2}|1\rangle_{c_2}
     \right ] \nn
     &+&{1\over {2}} |0\rangle_{b_1}|0\rangle_{c_1}\left [
      (1+\epsilon'_1)|1\rangle_{b_2}|1\rangle_{c_2}+(1-\epsilon'_1)|1\rangle_{b_2}|1\rangle_{c_2}
     \right ] \nn
     \label{qonenew}
     \eea
We must require that $|\epsilon_1|<\epsilon, |\epsilon'_1|<\epsilon$ for some $\epsilon\ll 1$; otherwise the corrections are not `small'.  Let $S_N$ be the entanglement of the radiated quanta $b_1, b_2, \dots b_N\equiv \{ b \}$ with the quanta $c_1, c_2, \dots c_N$ left inside the hole. After $N$ steps of evolution, there are $\sim 2^N$ correction terms in the state. One may then ask if for large enough $N$, the largeness of $2^N$ can offset the smallness of $\epsilon$ in such a way that $S_N$ becomes close to zero. If such an offset were possible there would be no Hawking puzzle: the subleading corrections to Hawking's leading order semiclassical computation would invalidate his conclusion. 

Hawking himself, in 2004, argued that the information paradox may be resolved by small corrections to the dynamics \cite{hawkingreverse}. Using the Euclidean theory, he argued that subleading saddle points would add a contribution to the black hole path integral that was not included in the leading order semiclassical approximation. These saddle points are exponentially suppressed, by factors $\sim Exp[-S_{bek}]$, so they are not easily seen in the quantum gravity theory.

In the Lorentzian section, the effect of these subleading terms should give small corrections to the radiation process. Note that $N\sim S_{bek}$, so that corrections of order $\epsilon \sim 2^{-N}$ are exponentially small in the same way as the subleading saddle point corrections. 

As we will now see, small corrections {\it cannot} remove the problem of growing entanglement, even if we allow the  size of the corrections $\epsilon$ to be much larger than what Hawking assumed. In fact we can choose any $\epsilon\ll 1$ (i.e., we can take $\epsilon$  independent of $N$) and still prove that the  corrections cannot restore unitarity to the evaporation process.

\subsubsection{Proof of the small corrections theorem}\label{secsmall}

Consider a sphere that encloses the vicinity of the hole; for concreteness, we let it have radius $r=10GM$. We will compute the entanglement between quanta  inside and outside this sphere.

\b

(1) Let the quanta emitted at emission steps $1, 2, \dots N$ be denoted $\{ b_1, b_2, \dots b_N\} \equiv \{ b\}$. The entanglement of the radiation with the hole at step $N$ is then
\be
S_N=S(\{b\})
\ee
where $S(A)$ for any   set $A$ denotes the entanglement  of $A$ with the remainder of the system. 

\m

(2) The bits in the hole evolve to   create an `effective bit' $b_{N+1}$ and an `effective bit' $c_{N+1}$. (The bit $b_{N+1}$ has not yet   left the region $r<10GM$.) The entanglement of the earlier emitted quanta $\{ b\}$ does not change in   this evolution. (If two parts of a system are entangled, and we make a unitary rotation on one part, the entanglement between the parts does not change.) 

\m

(3) The effective   bits $b_{N+1}, c_{N+1}$ must  approximate the   properties of the Hawking pair (\ref{qone}). In (\ref{qone}) we   have $S(b_{N+1}, c_{N+1})=0$, since   the pair is not entangled with anything else. We also have $S(c_{N+1})=\ln 2$.  Thus for   our model we must have 
\be
S(b_{N+1}+c_{n+1})<\epsilon_1
\label{four}
\ee
\be
S(c_{N+1})>\ln 2 -\epsilon_2
\label{five}
\ee
for some $\epsilon_1\ll1$, $\epsilon_2\ll 1$. 

\m

(4) The bit $b_{N+1}$ now moves out to the region $r>10GM$. The value of the entanglement at timestep $N+1$ is
\be
S_{N+1}=S(\{ b\}+b_{N+1})
\label{six}
\ee
since now $b_{N+1}$ has joined the earlier quanta $\{ b\}$ in the outer region $r>10M$. 

\m

(5) We now recall the strong subadditivity relation
\be
S(A+B)+S(B+C)\ge S(A)+S(C)
\ee
We wish to set $A=\{ b\}$, $B=b_{n+1}$, $C=c_{N+1}$. We note that these sets are made of independent bits:
(i) The quanta $\{ b \} $ have already left the hole and are far away (ii) The quantum $b_{n+1}$ is composed of some bits, but as it moves out to the region $r>10GM$, it is independent of the bits remaining in the hole and also the bits $\{ b \} $ (iii) The quantum $c_{N+1}$ is made of bits which are left back in the hole. Applying the strong subadditivity relation, we get
\bea
S(\{ b_i\}+b_{N+1})+S(b_{N+1}+c_{N+1})\ge
 S(\{ b_i \})+S(c_{N+1}) 
\label{seven}
\eea
Using (\ref{four}),(\ref{five}),(\ref{six}) we get
\be
S_{N+1}> S_N+\ln 2 -(\epsilon_1+\epsilon_2)
\label{eight}
\ee
Thus for $\epsilon_1, \epsilon_2\ll 1$,  the entanglement   keeps growing monotonically; it does not behave like that for a normal body where it first rises till the halfway point of evaporation and then falls back to zero. 

\b

We can summarize this conclusion as follows:

\b

{\it If low energy physics around the horizon is `normal' to leading order, then the entanglement between the radiation and the remaining hole will keep growing.  }

\b

Note that this statement does not require us to worry about any `transplanckian physics'; we are only asking that physics at low energies appear `normal'.  Let the horizon radius be $R_0$; the typical Hawking quantum then has wavelength $\lambda\sim R_0$. Take a `good slice' through the horizon, and look at the state of an outgoing mode with wavelength, say $R_0/100$. If semiclassical physics were valid at the horizon, then this mode should be in the vacuum state to leading order. Any small corrections to this mode can be included by writing  the state for this mode as
\be
|\psi\rangle_{R_0/100}=(1-\epsilon^2-\epsilon'^2-\dots)|0\rangle+\epsilon |1\rangle+\epsilon' |2\rangle +\dots
\label{acorr}
\ee
Semiclassical evolution will make  the leading part $|0\rangle$ evolve to a state with  entangled pairs, and the corrections in (\ref{acorr}) can be absorbed into the parameters $\epsilon_1, \epsilon_2$ used above.

\subsection{The options for getting unitarity}\label{secoptions}

Hawking, in 1975, advocated that we give up on the unitarity of quantum theory due to the monotonic rise of the entanglement $S_{N}$: he argued that when the hole evaporates away then we will be left with radiation that can only be described by a density matrix \cite{hawking}. If we are not willing to give up in quantum theory in this way, then we have the following options:

\b

(i) {\bf Remnants:} The evaporation of the hole can end in a planck sized remnant. Note that we could have started with an arbitrarily large hole, and thus obtained an arbitrarily large entanglement $S_{N}$ near the endpoint of evaporation. To be able to have such an entanglement we must allow this planck sized remnant to have an infinite number of possible internal states. 
It has been suggested that the remnant could be shaped like a baby universe, thus having an adequate interior region to hold these states. 

But the  the AdS/CFT duality  \cite{adscft} that is conjectured to hold in string theory does not allow remnants. Consider global $AdS_5\times S^5$ spacetime in IIB string theory, whose dual is super Yang-Mills on the boundary of $AdS_5$. This boundary has the topology of a spatial $S^3\times R$, where $S^3$ is the compact spatial section on which the CFT lives. Let the radius of this $S^3$ be $R_{CFT}$, and the curvature radius of the $AdS_5$ be $R_{AdS}$. Then under the AdS/CFT duality, an energy $\sim 1/R_{AdS}$ in the gravity theory maps to an energy $\sim 1/R_{CFT}$ in the field theory.

Now suppose we are limited to an energy $E<E_0\sim m_p$ for our remnants in the gravity theory. This translates to
\be
E_{CFT}\lesssim E_0{R_{AdS}\over R_{CFT}}
\label{aone}
\ee
in the CFT. The super Yang Mills is an $SU(N_c)$ theory with a large but finite $N_c$. Such a theory on a compact $S^3$ will have a discrete spectrum, and thus a finite number of states at $E<E_{CFT}$. Thus a remnant at the center of $AdS$ cannot have an infinite number of states, and so cannot resolve the problem of growing entanglement.

\b

(ii) {\bf Wormholes:} In section \ref{secsmall} we have assumed that once a quantum gets sufficiently far from the hole then its state does not change further in any significant way, and it becomes a free streaming particle carrying its own bit of information. One may try to alter this assumption to evade the problem. 

One possibility is to say that the emitted bit is not independent of the degrees of freedom inside the hole \cite{pr,cool}. In the derivation of section \ref{secsmall}, this would invalidate step (5), where it is assumed that the emitted bit $b_{N+1}$ becomes independent of the bits in the hole after it gets sufficiently far from the hole. 

We recall the approach along such lines  given in \cite{cool}.\footnote{I thank Juan Maldacena for explaining the details of this idea.}   In this approach, the evolution of low energy modes around the horizon remains the one given by `normal' physics; thus entangled pairs are created at the horizon all the way up to the endpoint of evaporation. But the entanglement between the emitted quantum and the remaining hole leads to an effective `wormhole' joining the radiated quantum back to the hole. This wormhole has a planck scale throat since it describes just one bit of entanglement, but a clearer description can be obtained by condensing  the set of emitted $b$ quanta into a second hole. Now the second hole is joined to the first by a wormhole with a radius that is order the horizon size $R_0$; in fact the two holes form the right and left halves of an eternal hole joined by the Einstein-Rosen bridge. The emitted quanta $\{ b\}$ are now found to be in the causal past of the negative energy quanta $\{ c\}$ that fell into the evaporating hole. Since the sets $\{ b \}, \{ c \}$ do not live on the same spacelike slice, we cannot compute an entanglement between them; thus the problem of monotonically growing $S_N$ is bypassed. 

The nonlocalities implied in this picture are not allowed by our assumption A2, so we will not consider such approaches. It was also pointed out in \cite{dual} that such a picture of wormholes runs into difficulties with the problem of unbounded entropy inside the horizon. Finally, the weak coupling computations in string theory reproduce the greybody factors of black hole radiation, by a process which is like the burning of a piece of coal \cite{dmcompare}. Thus the quanta that are radiated in the weak coupling limit are bits that are independent of the degrees of freedom in the remaining string state. This suggests that the Hawking radiation bits that are emitted at strong coupling are also bits that are independent of the remaining hole. Thus we will not consider such nonlocalities in what follows.\footnote{There are other studies that invoke nonlocalities at the horizon scale \cite{giddings} or at the boundary of spacetime \cite{hps}. For some  other  approaches to black hole puzzle, see \cite{others}. }

\b

(iii) {\bf Fuzzballs:}  The third option is that we do not have a horizon at all; the black hole microstates are like normal bodies with no horizon, which radiate from their surface rather by by pair creation from the vacuum.  In \cite{emission} it was found that the size of bound states in string theory was always order horizon size. Detailed computations of specific states have always found such a fuzzball structure for microstates; in no case have we obtained a traditional horizon. The fuzzball conjecture says that all microstates of all black holes are fuzzballs; i.e., they are  quantum objects of size the horizon scale with no vacuum region around a horizon.

\section{The  problem of unbounded entropy in the black hole geometry}

First consider physics without gravity. Suppose we take a bounded volume $V$, and also put a bound on the total energy $E$. Then we expect to have only a finite number of allowed states; i.e., we expect the entropy $S(E,V)$ to be a finite number. Roughly speaking, the limit on energy puts a limit on the momenta of our degrees of freedom.  One cell of phase space volume $d^3 x d^3 p\sim h^3$ can hold only one quantum state, so with bounded $E, V$ we can get only a finite number of states.

The situation is less clear when we include gravity, The Newtonian potential $-Gm_1m_2/ r$ is unbounded below, so we can get  infinite momenta and thus an infinite phase space with finite $E, V$. But one might hope that quantum effects would cut this down to a finite phase space, and thus still give finite $S(E,V)$.

Bekenstein's work on black hole entropy led to the expression \cite{bek}
\be
S_{bek}={A\over 4G}
\ee
for the entropy of a black hole. Since at least classically a black hole appears to swallow all information, one might expect that it has the `maximal' possible entropy in some sense. Thus consider a system with energy  $E=M$. Let the system be confined to a spatial region with radius $R=2GM$. Then a plausible conjecture is that for any system
\be
S(E,V)\le S_{bek}(M)
\label{atwoq}
\ee
Note that we need to bound both $E$ and $V$. If we only fix $V$, then we can get much more 
entropy by taking a `black hole gas' in the volume $V$ \cite{masoumi1,gas}. If we only bound $E$, then we can spread our particles over a sufficiently large $V$ and thereby get an arbitrarily large entropy. 

The conjecture of AdS/CFT duality also supports a bound on $S(E,V)$. Consider the  states in the gravity theory with energy $E_{grav}$. By the energy relations noted in section \ref{secoptions}  these states correspond to states in the CFT with energy $E_{CFT}=E_{grav}R_{AdS}/R_{CFT}$. Since energy is bounded below in the CFT, there are only a finite number of states with $E<E_{CFT}$. Thus $S(E_{grav})$ should be bounded in the gravity theory for any $E_{grav}$. We do not need to constrain $V$ since the $AdS$ space acts like a confining box, imposing a high energy cost to any quantum that ventures too far from the center of $AdS$.

While the conjecture (\ref{atwoq}) may appear reasonable from a physical perspective, it faces an immediate difficulty. If our spacetime has a black hole horizon, then as we will now recall, we can get an {\it arbitrarily} large entropy $S$ with a finite $M$ and a finite $V$. This is the puzzle of unbounded entropy in the black hole geometry.

\subsection{The hypersurfaces admitting large entropy}

Consider the Schwarzschild metric in 3+1 dimensions
\be
ds^2=-(1-{2GM\over r}) dt^2 + {dr^2\over 1-{2GM\over r}}+r^2 d\Omega_2^2
\label{athree}
\ee
We wish to make a spacelike slice in this geometry. Outside the horizon $r=2GM$ a spacelike slice can be taken as $t=\bar t$ for some constant $\bar t$. Inside the horizon, space and time interchange roles, and a spacelike slice can be taken as $r=\bar r$ for some constant $\bar r$. As a concrete example we may take $\bar r=GM$, so that this part of the slice is neither near the horizon nor near the singularity. Note that the part  of the slice given by $r=\bar r$  has an infinite proper length. This is one of the important aspects of the construction that will allow us to get an unbounded entropy on the slice.

One might worry that the part of the slice inside the horizon cannot be smoothly connected to the part outside; in that case we would not really have a good spatial slice in the full manifold. But actually it is easy to join the inside part to the outside part in a smooth way, keeping the slice spacelike at all points. To see both the outside and inside of the horizon in a common coordinate patch, we use the Eddington-Finkelstein coordinate 
\be
u=t+r^*=t+r+2GM\log ({r\over 2GM}-1)
\ee
Then the metric (\ref{athree}) becomes
\be
ds^2=-(1-{2GM\over r})du^2+2 dudr+r^2 d\Omega_2^2
\ee
We can choose our spatial slice as follows.  For $u<u_0$, we let $r=\bar r=GM$. For sufficiently large $u$, we let the slice be  $t=\bar t$. For the connector segment in between, we take
$dr/du=\h \tanh [(u-u_0)/GM]$; note that this makes $dr/du$ continuous at $u=u_0$. The connector segment is then
\be
r(u)=\h GM \log\left[\cosh {(u-u_0)\over GM}\right ]+GM
\ee
We can join this connector segment to the $t=\bar t$ segment at a place where $dt/du=0$ on the connector segment. Since $u=t+r^*$, this condition gives $dr^*/du=1$, which is $dr/du=1-{2GM\over r}$. With the above expression for $r(u)$, we see that this condition is met at $u-u_0\approx 6.69313 \, GM$, at which location we have $r\approx 4GM\equiv r_{outer}$. Thus at $r=r_{outer}$ we join the connector segment to the hypersurface $t=\bar t$ with no discontinuity in derivatives.

 \begin{figure}[htbp]
\begin{center}
\includegraphics[scale=.5]{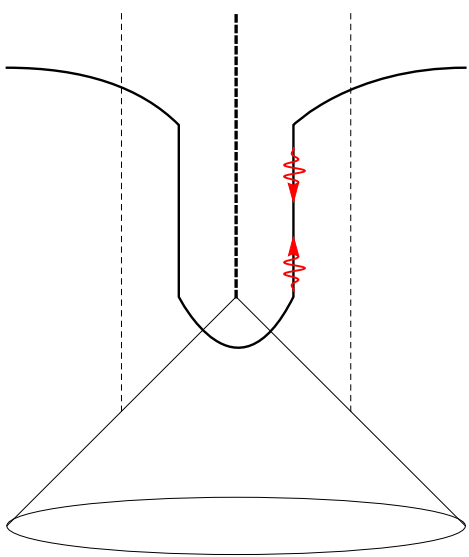}
\end{center}
\caption{The cone depicts a null shell collapsing to make a black hole. The dashed lines are the horizon, and the thick dashed vertical line is the singularity. The hypersurface has a long part $r=\bar r$ inside the horizon. On this part we can have quanta with momentum pointing upwards; these have negative energy as seen from infinity. We can also have quanta with momentum pointing downwards; these have positive energy as seen from infinity. }
\label{fig11}
\end{figure}

One may still worry that the slice does not have the usual topology $R^3$ of a spacelike slice in ordinary space, since there is a `hole' in the slice at $u\r-\infty$. To remedy this, we assume that the black hole was formed by the collapse of some matter at some early time $u_i$. Then
once our slice reaches $u=u_i$, we can just cap it off smoothly as shown in fig.\ref{fig11}. 

If we make such a slice in a  large black hole; i.e., a hole with $M\gg m_p$, then all curvatures will be small everywhere in the neighborhood of the slice. If low curvatures were a valid criterion for semiclassical gravity to hold, then we would find that physics on and around this slice is the   semiclassical physics of quantum fields on curved space. We can make such a  slice for any  black hole: all we have used is that space and time interchange roles as we move in through the outer horizon. Thus in what follows we will use the more general metric
\be
ds^2=-f(r) dt^2+{dr^2\over f(r)}+r^2 d\Omega^2
\ee
Let the outer  horizon be at $r=R_0$; thus $f(R_0)=0$.
The $r=const.$ part of the slice inside the hole will be $r=\bar r<R_0$, and we will write $f(\bar r)\equiv \bar f$ etc. We define 
\be
r^*=\int {dr\over f}, ~~~u=t+r^*
\ee
and find
\be
ds^2=-f(r) du^2+2dudr+r^2d\Omega^2
\ee

\subsection{Placing quanta on the slice}\label{secnegative}

Our next step is to add entropy to this slice. We will add quanta only to the section $r=\bar r $; thus these quanta will be inside the horizon region $r<R_0$. The spacelike segment $r=\bar r$ is parametrized by the coordinate $u$. We ignore the angular directions in what follows since the relevant vectors have no components in these angular directions. The unit tangent $\hat z$ along the slice is
\be
\hat z=( z^u, z^r) =\left ( ( -{\bar f})^\h, 0\right )
\ee
The unit normal to the slice pointing in the `forward time direction' is
\be
\hat n =(n^u, n^r)= \left ( (-{\bar f})^{-\h}, - (-{\bar f})^\h \right )
\ee
This normal points in the direction of smaller $r$, in accordance with the fact that   time evolution takes us towards the singularity. 

Now consider radially moving massless quanta on this spatial section $r=\bar r$. The momenta of such quanta are null vectors $\vec k_1, \vec k_2$ where
\be
\vec k_1=(k_1^u, k_1^r)=\left ( 0, k_1^r\right ) , ~~~~k_1^r<0
\ee
\be
\vec k_2=(k_2^u, k_2^r)=\left ( k_2^u, \h {\bar f} k_2^u\right ) , ~~~~k_2^u>0
\ee
In each case we have chosen the signs so that $\vec k_i\cdot \hat n <0$, this ensures that  the momenta  lie in the forward light cone. 

In a local frame along the slice, the momenta $\vec k_1, \vec k_2$ describe positive energy quanta. But the energy of such quanta as seen from infinity is given by
\be
E=-\vec K \cdot \vec k_i
\ee
where $\vec K$ is the Killing vector  $\p_t$ that becomes the unit  vector along time at $r\r \infty$. In the Eddington-Finkelstein coordinates, we have
\be
\vec K =  ( K^u, K^r) = \left (  1, 0  \right )
\ee
We have $K^2=f$. Thus $K$ is timelike outside the horizon and  spacelike inside the horizon. 

We now observe that
\be
E_1=-\vec K \cdot \vec k_1=-k_1^r>0
\label{aseven}
\ee
\be
E_2=-\vec K \cdot \vec k_2=\h {\bar f} k_2^u <0
\label{aeight}
\ee
Thus quanta with momenta $k_1$ contribute positively to the mass of the hole  while quanta with momenta $k_2$ contribute negatively. This is of course just a precise formulation of the statement that the gravitational potential inside the horizon is sufficiently negative to allow the existence of  states whose total energy (rest mass + kinetic + potential) is negative as seen from infinity. 

We can now make  states in the black hole geometry with arbitrarily large entropy. Consider for concreteness a solar mass Schwarzschild hole in 3+1 dimensions. The Hawking quanta have wavelength $\sim 3\,  Km$. We will let $\bar r=GM$. The  $r=\bar r$ part of the slice is parametrized by $u$. At some value of $u=u_1$, place the center of a photon wavepacket with momentum of type $\vec k_1$, with wavelength $\sim 1 \, Km$. Moving in the direction of increasing $u$,  leave a space along the slice of length $\Delta s \sim 2 \, Km$.  Then place a photon wavepacket with momentum of type $\vec k_2$, with wavelength again $\sim 1 \, Km$. Continue in this in the direction of increasing $u$, alternating between wavepackets of type $\vec k_1$ and $\vec k_2$. The energy of each of these quanta, as seen from infinity,  is of the order $(1\, Km)^{-1}\equiv\delta$, which is very small. Thus on any spacelike slice through the black hole geometry, the mass $M$ contained within the radius $R_0=2GM$ is $\approx M\pm \delta$. 

Let each photon have two spin states, so that it corresponds to one bit with entropy $\ln 2$. Let there be $N$ quanta of each momentum type $\vec k_1, \vec k_2$ on the slice. 
We can let the $r=\bar r$ segment of our spacelike slice be as long as we want, so $N$ can be arbitrarily large. We can therefore get
\be
S=2N\ln 2 > S_{bek}
\label{aten}
\ee
In fact there is no upper bound to the entropy $S$ inside the horizon $r=R_0$. This is the problem of unbounded entropy.

\subsection{Making the unbounded entropy  problem precise}

It may appear that there could be several ways around this problem, so let us now close some possible loopholes.

\subsubsection{Ambiguity in the definition of particles}

There is one issue that we have treated in an oversimplified way, so let us address that now. We are working in the neighborhood of the slice at $r=\bar r =R_0/2$. The curvature length scale here is $\sim R_0$. In a curved spacetime, the definition of particles is not unique. So which definition of particle did we use in the above discussion when we were placing quanta on the slice? 

For concreteness, we could use Gaussian normal coordinates around our slice and use it to define positive frequency modes. But with this and other similar definitions of particles we have to worry about the energy of vacuum polarization; by this we mean the stress tensor computed in the semiclassical approximation since we are not talking of the nonperturbative vecro effects for now.  Since the curvature length scale is $\sim R_0$, the stress tensor components  will be $T_{ab}\sim R_0^{-D}$ where $D$ is the spacetime dimension. Our quanta above occupied an angular volume $\sim R_0^{D-2}$ and a distance $\sim R_0$ along the slice. Thus the vacuum polarization energy over the region occupied by one of quanta is $\sim R_0^{-1}$, which is the same order as the energy we had taken for the  quantum that we placed in that region.  Can we be sure that the negative energy quanta that were vital to our argument do contribute enough negative  energy so that the overall energy of the hole remains $\approx M$, regardless of how a slice e take? 

However it is easy to dispose of this concern. One could just take wavelengths $\sim R_0/100$ 
rather than $\sim R_0$ for the quanta. The vacuum polarization energy density remains the same but the magnitude of the energy carried by the quanta go up by a factor $100$ (while still remaining very small compared to $M$). 

In fact we can go further, and ask what is the smallest wavelength $\lambda$ that we can use for the quanta on the slice and encounter the unbounded entropy puzzle the way we have outlined above. With quanta of wavelength $\sim \lambda$, spaced a distance $\sim \lambda$ apart, the energy density is $\rho_{quanta}\sim \lambda^{-(d+1)}$. We wish to embed the slice in the standard black hole geometry, so let us require that the backreaction of this energy create curvatures that are much smaller than the curvature scale of the black hole metric. This gives 
\be
{\mathcal R}_{quanta}\sim G\rho_{quanta} \lesssim R_0^{-2}
\ee
which gives
\be
\lambda \gtrsim (GR_0^2)^{1\over (d+1)}
\label{atone}
\ee
We need a number of quanta
\be
N\gtrsim  S_{bek}\sim {R_0^{d-1}\over G}
\ee
 to get (\ref{aten}). This needs a spatial volume 
\be
V\sim {N \lambda^d} \gtrsim {R_0^{d-1}\lambda^d \over G}
\ee
The angular sphere has area $A\sim R_0^{d-1}$, so we need a length along the slice
\be
L\gtrsim {1\over A} {R_0^{d-1}\lambda^d\over G}\sim {\lambda^d\over G}
\ee
From (\ref{atone}) we find
\be
L\gtrsim G^{-{1\over d+1}} R_0^{2d\over d+1}\sim \left({R_0\over l_p}\right)^{d-1\over d+1} R_0
\label{attwo}
\ee
By contrast if we had used quanta with wavelength $\lambda\sim R_0$, then we would need
\be
L\gtrsim S_{bek} R_0 \sim G^{-1} R_0^d \sim \left({R_0\over l_p}\right)^{d-1}R_0
\label{atthree}
\ee
which is parametrically  larger than (\ref{attwo}) if $R_0\gg l_p$. 

To summarize, we can put an entropy larger than $S_{bek}$ on the slice without disturbing the semiclassical black hole metric significantly if we choose the length $L$ of the $r=\bar r$ part of the slice to exceed (\ref{attwo}).

\subsubsection{Additivity of the entropy}

We have assumed an entropy  $\sim 1$ per quantum on the slice, and then assumed that this entropy adds up over the different quanta, so that overall we get the entropy (\ref{aten}). But there could be small corrections to the state on the slice due to quantum gravity effects, which would prevent the entropy from being simply additive over the quanta. Since the number of quanta involved is very large, can we be sure that we can arrange for these corrections to be controlled, so that we can indeed get an entropy satisfying the inequality (\ref{aten})?

To see that small corrections to the state cannot prevent us from realizing (\ref{aten}), we can appeal again to the small corrections theorem. In fact a simple way to get the unbounded entropy problem  is to consider the semiclassical evaporation of the hole, while feeding it at a rate that maintains its mass.  The Hawking pair consists of a particle $b$ that escapes to infinity and a negative energy particle $c$ that falls into the hole. The particles with momenta of type $\vec k_2$ are just these kind of negative energy quanta, as seen on the spacelike slice that we have taken. The quanta $b,c$ are entangled with an entanglement entropy $\log 2$. Now suppose we keep feeding the hole with quanta $a$ with energy $\sim T$ ($T$ is the temperature of the hole) at the same rate at which the hole is evaporating. Then the infalling quanta $a$ will appear on our spacelike slice as the quanta with momentum of type $\vec k_1$. Before we throw in $a$, we can entangle it with a quantum $d$ which we keep outside the hole. Then each of the $c, a$ quanta inside the hole have an entanglement $\ln 2$ with quanta outside. With $N$ quanta $c$ and $N$ quanta $a$, the entanglement entropy of the part of the slice inside the hole with the part outside is 
\be
S_{N}=2N\ln 2
\label{anine}
\ee
The mass of the hole remains $\approx M$, since we have fed the hole at the same rate at which it was evaporating. 

Eq.  (\ref{anine}) gives the leading order entanglement entropy between the quanta on the slice and quanta outside the hole. There can be small corrections to the state of each entangled pair due to hietherto unknown quantum gravity effects; let these small corrections be bounded by a parameter $\epsilon\ll 1$.   We can now  use an argument along the lines of the small corrections theorem outlines in section \ref{secsmall}, and find that
\be
{S_{N}\over 2N\ln 2}=1-O(\epsilon)
\ee
The number of states possible on the part slice inside the hole must be equal to or larger than  $Exp[S_{N}]$.  We thus see that the  problem of unbounded entropy is a robust one, not affected by some small heitherto unknown quantum gravity effects.

\subsection{Difficulties with resolving the problem}

We see have seen that once we assume that physics along the gently curved slice is usual semiclassical physics to leading order, we cannot save the conjecture (\ref{atwoq}). One might therefore try to postulate that novel quantum effects that alter semiclassical behavior even though curvatures are low everywhere along the slice. 

The principal feature of our slice is that it is very {\it long}; to get (\ref{aten}) with quanta of wavelength $\lambda\sim R_0$ we need to have the length of the $r=\bar r$ part of the slice to be 
\be
L\gtrsim S_{bek} R_0 \gg R_0
\label{ael}
\ee
Even with the smaller wavelengths (\ref{atone}), the length $L$ in (\ref{attwo}) is parametrically larger than $R_0$. Perhaps we could have a rule that disallows such long slices?

In \cite{lm4} it was noted that 2-charge extremal holes had a throat which was classically infinite, but in the full string theory all microstates of the hole had a finite depth, with the maximal depth being
\be
L_{max}={V\over 2G}
\label{atw}
\ee
where $V$ is the volume enclosed by a ball with the radius of the horizon. Since $V\sim R_0 A$, we see that the depth satisfies $L\lesssim S_{bek} R_0$. Could this be a general principle saying that hypersurfaces with deep throats have a bound on the depth of this throat?

Suppose we do conjecture that (\ref{atw}) is the maximal length to which we can stretch any slice even for nonextremal holes. Then we see from (\ref{ael}) that we would resolve the unbounded entropy problem for states made with quanta with $\lambda\sim R_0$ \cite{beyond}.

But there is a difficulty with an approach which just postulates that there is a limit to how much we can stretch a slice. Consider applying this principle to cosmology. In \cite{univexpand} it was argued that spacetime should be thought of as a rubber sheet, which is characterized not only by its shape by also by a thickness. As we stretch the slice, the rubber sheet becomes thinner, and finally when it drops below a monomolecular layer, semiclassical physics on it breaks down (even though the curvatures have not become high anywhere). Could such a picture be true? Our universe started as a marble sized ball before inflation, and stretched to a radius of $3000\, Mpc$ today. This large stretching does not appear to have  invalidated the semiclassical approximation for our cosmology. Thus it is not straightforward to simply postulate that a slice cannot be stretched too much.

A second difficulty is the following.  Look at long slice near a value $u=u_0$. The metric $g_{ab}$ at this location has no information about how long the entire $r=\bar r$ part of the slice is. Thus if new physics has to creep in when the slice becomes too long, then there must be {\it a new variable} at each point of the slice. This variable would tell us if the region we are looking at is a part of a slice has been stretched  to the point where the semiclassical approximation breaks down. But what would play the role of such a variable in string theory?

We will now see that  the vecro part of the wavefunctional creates an extra energy if we try to stretch the slice past a length $\sim R_0$; this disallows such stretched slices from being valid slices in the black hole geometry.

\subsection{How vecros resolve  the unbounded entropy problem}\label{bagresolve}

  \begin{figure}[htbp]
\begin{center}
\includegraphics[scale=.5]{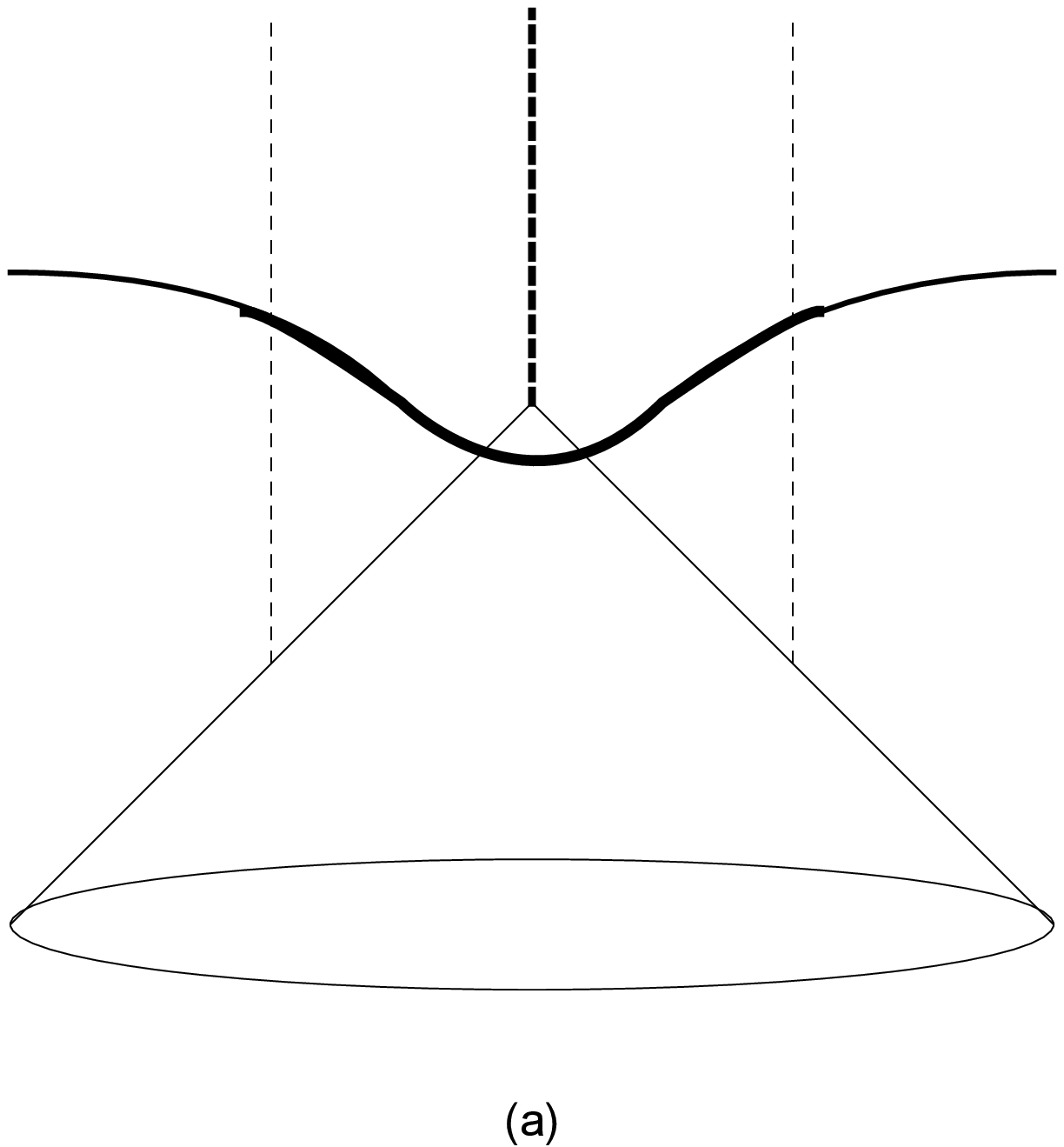}\hskip40pt
\includegraphics[scale=.5]{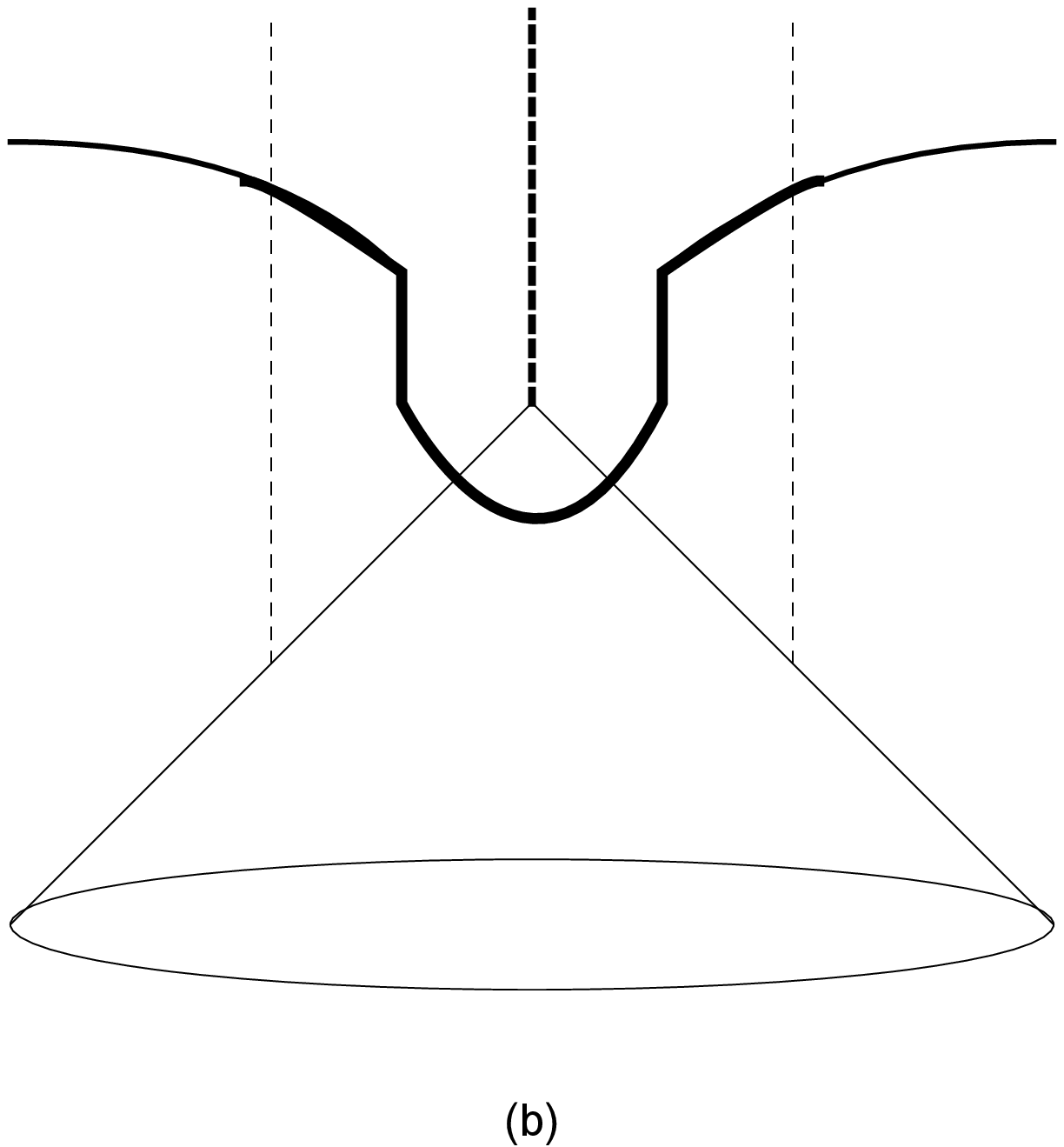}
\end{center}
\caption{(a) The same setup as in fig.\ref{fig11}, but we have not stretched the slice to have a the $r=\bar r$ part. The dark curve on the slice represents a vecro with radius $R_v$ a little larger than the horizon radius $R_0$. (b) After the slice is evolved forward in time $t$ by $\sim R_0$, the vecro with the same radius $R_v$ has had to expand to a larger internal volume.}
\label{fig12}
\end{figure}

We have already made the unbounded entropy problem precise: if we accept the traditional metric of the hole in regions where the curvature is low, then we can get an arbitrary amount of entropy inside the horizon, while keeping the mass inside the horizon region at $\approx M$. 

But now we see that the vecro hypothesis gives us a way out of this problem. We have a new ingredient: the energy of vacuum fluctuations that describe extended configurations. We do not have a detailed understanding of the behavior of vecros, but let us outline a set of steps which will show that vecro behavior gives modifications at the correct scales to resolve the unbounded entropy puzzle:

\b

(i)  In fig.\ref{fig12}(a) we depict a hypersurface where we have {\it not} stretched the slice in the way we need to for getting large entropy. On this slice consider the vecro configurations that have a radius $R_v\gtrsim R_0$ where $R_0$ is the horizon radius. The region spanned by such a  vecro configuration  is depicted  by the thick line in the figure.

\b

(ii) In fig.\ref{fig12}(b) we depict a hypersurface which has been stretched by an amount $\sim R_0$; this is not yet a slice on which we have large entropy, since by (\ref{attwo}) we need to stretch the slice by an amount that is parametrically larger than $R_0$ to get $S>S_{bek}$. We see that a vecro with outer boundary at $R_v\gtrsim R_0$ now has a larger internal volume; i.e. it has been expanded. This leads to the vecro configuration having a larger energy. Note that the shape of the region which has been stretched satisfies the condition (\ref{ntwo}) for  vecro deformations to be a significant correction to semiclassical physics. 

\b

\begin{figure}[htbp]
\begin{center}
\includegraphics[scale=.45]{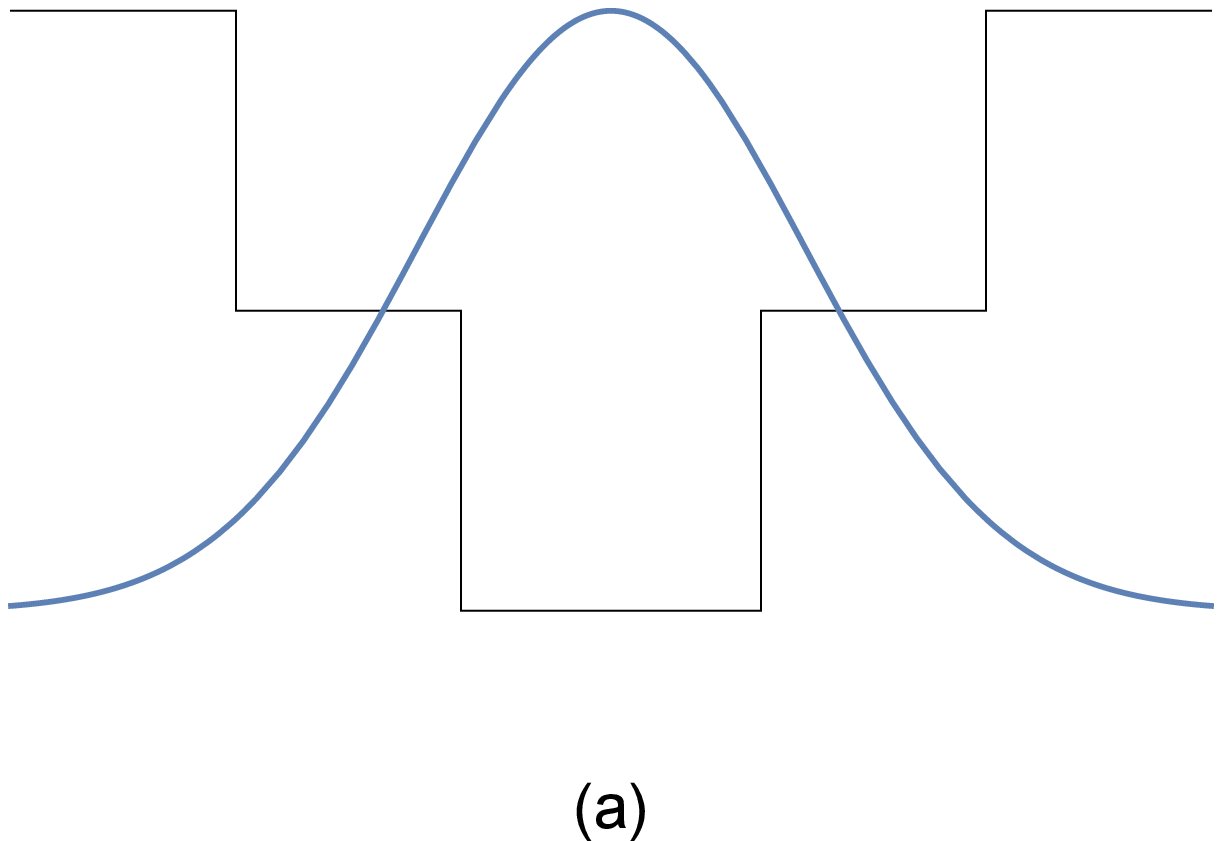}\hskip60pt
\includegraphics[scale=.45]{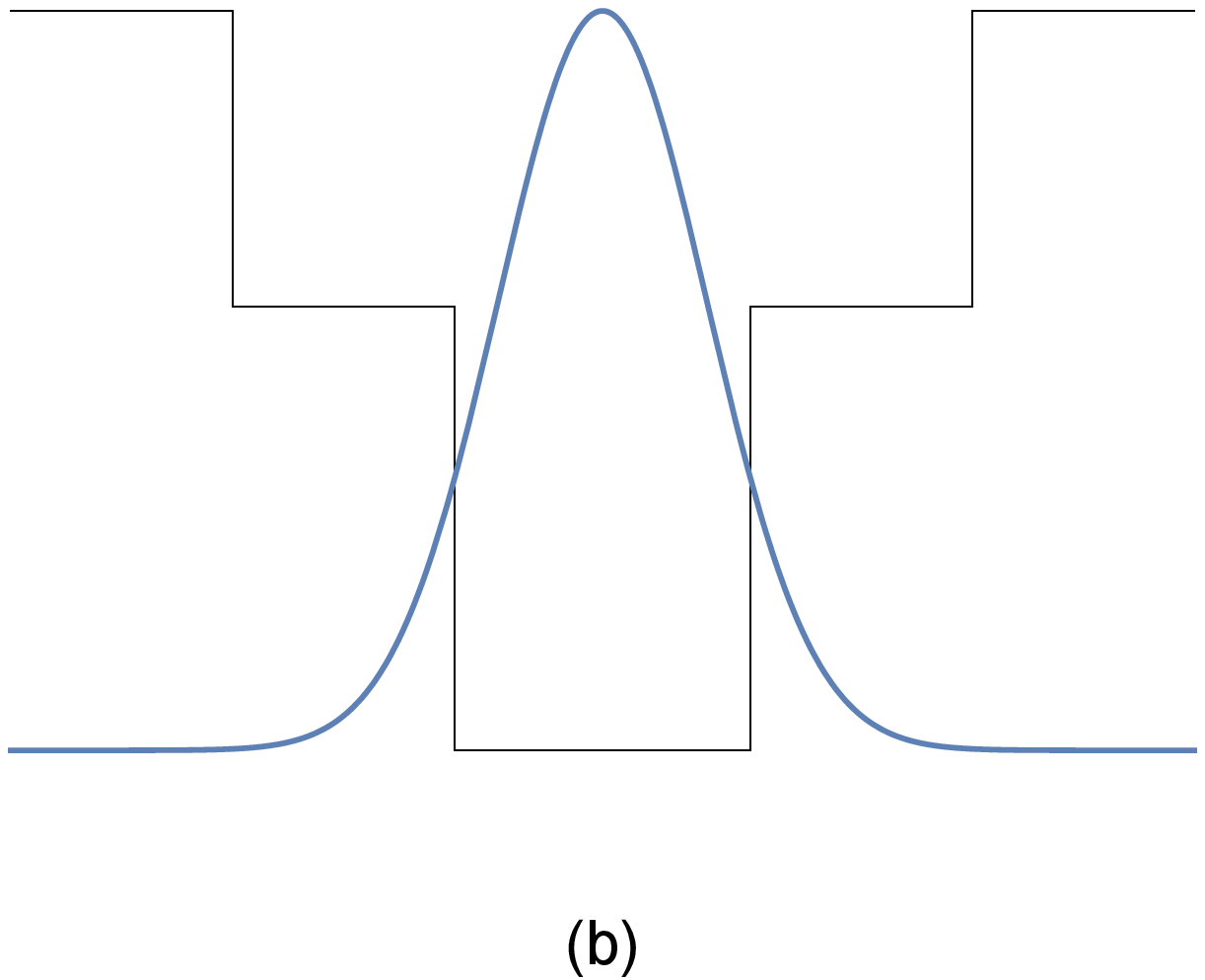}
\end{center}
\caption{(a) The vecro potential in Minkowski spacetime, and the corresponding vacuum wavefunctional $\Psi_0[g]$. (b) The vecro potential has become higher due to a deformation of spacetime; the wavefunctional now has a smaller width and a higher energy. }
\label{fig7}
\end{figure}

(iii) In fig.\ref{fig7}(a) we sketch the vecro potential and corresponding wavefunctional for the situation in fig.\ref{fig12}(a). In fig.\ref{fig7}(b) we depict the rise in energy of the vecro configurations with outer radius $R_v\gtrsim R_0$. The wavefunctional now has a smaller spread; we can see this from the expression (\ref{thone}), where the  action $\t S$ will be larger for the  vecro with radius $R_v$  as the potential energy of this vecro has increased. The wavefunctional in fig.\ref{fig7}(b) has a higher energy than the wavefunctional in fig.\ref{fig7}(a), due to the higher potential and the larger gradient terms induced by the narrower spread of the wavefunctional. 

In the potential (\ref{aathree}) if we put $\delta \sim 1$ then we get for the energy increase of such vecros
\be
 \Delta E_{vecro} \sim U \sim {R_0^{d-2}\over G} \sim M
\ee
where $M$ is the mass of the hole. We conjecture that this rise in energy satisfies
\be
\Delta E_{vecro} > M
\label{bagone}
\ee

\b

(iv) Note that the energy (\ref{bagone}) is due to the deformation of the vecro wavefunctional; it is separate from the energy $M$ of the matter on the slice which made the black hole. Thus the total energy on the slice in fig.\ref{fig12}(b) is
\be
E_{total}=M+\Delta E_{vecro} > 2M
\ee

\b

(v) We therefore see that a slice like that in fig.\ref{fig12}(b) cannot be a slice in the geometry of a black hole of mass $M$. One might try to extend the slice to be  of the kind depicted in fig.\ref{fig11}, and then place negative energy quanta on the $r=const.$ part of this slice. 
This will reduce the energy of the slice, and one might ask if it is then possible to bring $E_{total}$ back down to the value $M$. But we note that at no stage along the slice should the energy enclosed reach down to zero, as the horizon radius would then vanish and we cannot have a spacelike slice inside the horizon. Thus the total energy from the negative energy quanta on the $r=const.$ section must satisfy
\be
E_{negative}> -M
\ee
We then see that the total energy on the slice is
\be
E_{total}=M+\Delta E_{vecro} + E_{negative} >M
\ee
Thus we see that while we can have a slice like that in fig.\ref{fig12}(a), we cannot have a slice like that in fig.\ref{fig12}(b) or a slice like that in fig.\ref{fig11}. This resolves the puzzle.

\section{Resolving the information paradox}

The information paradox is intimately tied to the notion of causality. After all if we did not have to respect any notion of staying within light cones, then we could imagine novel quantum gravity  process occurring at the singularity which transport  information to, say, $r=10GM$, after which this information could freely flow out to infinity. In that case we would have no paradox, since we do not have any a priori constraints on what novel physics can happen at a singularity.

We will therefore begin in section \ref{secc1} with a discussion of causality. In section \ref{secc2} we will note  the role of this causality in the argument for the information paradox.    In section \ref{secc3}  will describe the scenario of gravitational collapse under the vecro hypothesis, and see  how we evade the  paradox while maintaining causality. In section \ref{secc4} we will study how causality constrains  modifications to this picture where we try to get fuzzball formation on timescales faster than $\sim R_0$.

\subsection{Causality}\label{secc1}

In quantum field theory on flat spacetime, causality tells us that we cannot send signals outside the light cone. Technically, one finds that the quantum field is defined by local operators $\hat \phi(x) $ which commute at spacelike separations
\be 
[ \hat \phi(x), \hat \phi(y)]=0, ~~~~(x-y)^2>0
\label{pone}
\ee
for bosonic fields and anticommute for fermionic fields. 

Quantum field theory on a curved spacetime has a similar behavior. The spacetime has a light cone at each point.  Consider two points $x,y$ such that there is no path between them that remains  timelike or null at all points along its length.  Then bosonic field operators  at $x,y$  will commute and fermionic operators will anticommute. 

The situation becomes more confusing when we get to quantum gravity, where  the metric itself fluctuates.   In the Wheeler-de Wit formalism, the state is described by a wavefunctional $\Psi[{}^{(3)}g]$. A 4-d spacetime emerges only as an approximate construct, and that too only for appropriately chosen wavefunctionals.  Since the light cones are not rigidly fixed, there cannot be a strict analogue of (\ref{pone}). 

Nevertheless, we find that some semblance of causality appears to hold even in very nonpertubative quantum gravity processes. Consider bubble nucleation in cosmology, where we tunnel  from one metric to a different metric. In such a situation, should we respect causality with respect to the light cones of the starting metric, or the final metric, or neither?  The computation shows that 
the outer surface of the expanding bubble moves at a speed less than $c$ with respect to the metric outside the bubble, and its inner surface travels moves with a speed less than $c$ with respect to the metric inside the bubble. Similarly a `bubble of nothing' created by tunneling in a spacetime $M_{3,1}\times S^1$ expands at a speed slower than $c$ in $M_{3,1}$ \cite{bubble}. So we should not be quick to discard causality in quantum gravity. 

We will adopt the following conjecture for our theory of quantum gravity:

\b

(i) Consider classical Minkowski space ${\cal M}$ labelled by points $x$. ${\cal M}$ is a maximally symmetric space with symmetry group $G=R(d,1) \rtimes SO(d,1)$. There will exist a vacuum $|0\rangle$ for the full quantum gravity theory which is invariant under the symmetry group $G$. Further, the quantum fields of the gravity theory are defined by operators $\hat\phi(x)$, where $x$ are the points of the classical space ${\cal M}$. These operators will  exactly commute/anticommute when $x,y$ are spacelike separated in the classical Minkowski metric ${\cal M}$.

\b

(ii) Suppose the quantum gravity wavefunctional  $\Psi[{}^{(3)}g]$  describes a gently curved spacetime  in a   region ${\cal W}$; i.e., we have a metric with curvature radius 
\be
R_c\gg l_p
\ee
 Now one cannot demand that relations like (\ref{pone}) be exactly true. But we assume that they are still true to leading order; i.e., if we scale all parameters so that $R_c/l_p\r\infty$ then the effect on low energy physics of any propagation of signals outside the light cones will go to zero.  
 
 \b
 
 To summarize, we assume that in Minkowski spacetime we have strict causality; i.e., we cannot send signals outside the light cones as defined by the fiducial classical metric used to define the quantum gravity theory.\footnote{A similar situation should hold for $AdS$ space since that is also maximally symmetric. de Sitter space is maximally symmetric at the classical level but it is not clear if a quantum vacuum can be found which is invariant under the full symmetry group.}   Further, we assume that if the space is gently curved, then we violate this causality only slightly; this is the content of assumption ${\bf A2}$ which says that `if curvatures are low throughout a spacetime region, then causality holds to leading order'. 
 
\subsection{The role of causality in the information paradox}\label{secc2}
 
  \begin{figure}[htbp]
\begin{center}
\includegraphics[scale=.55]{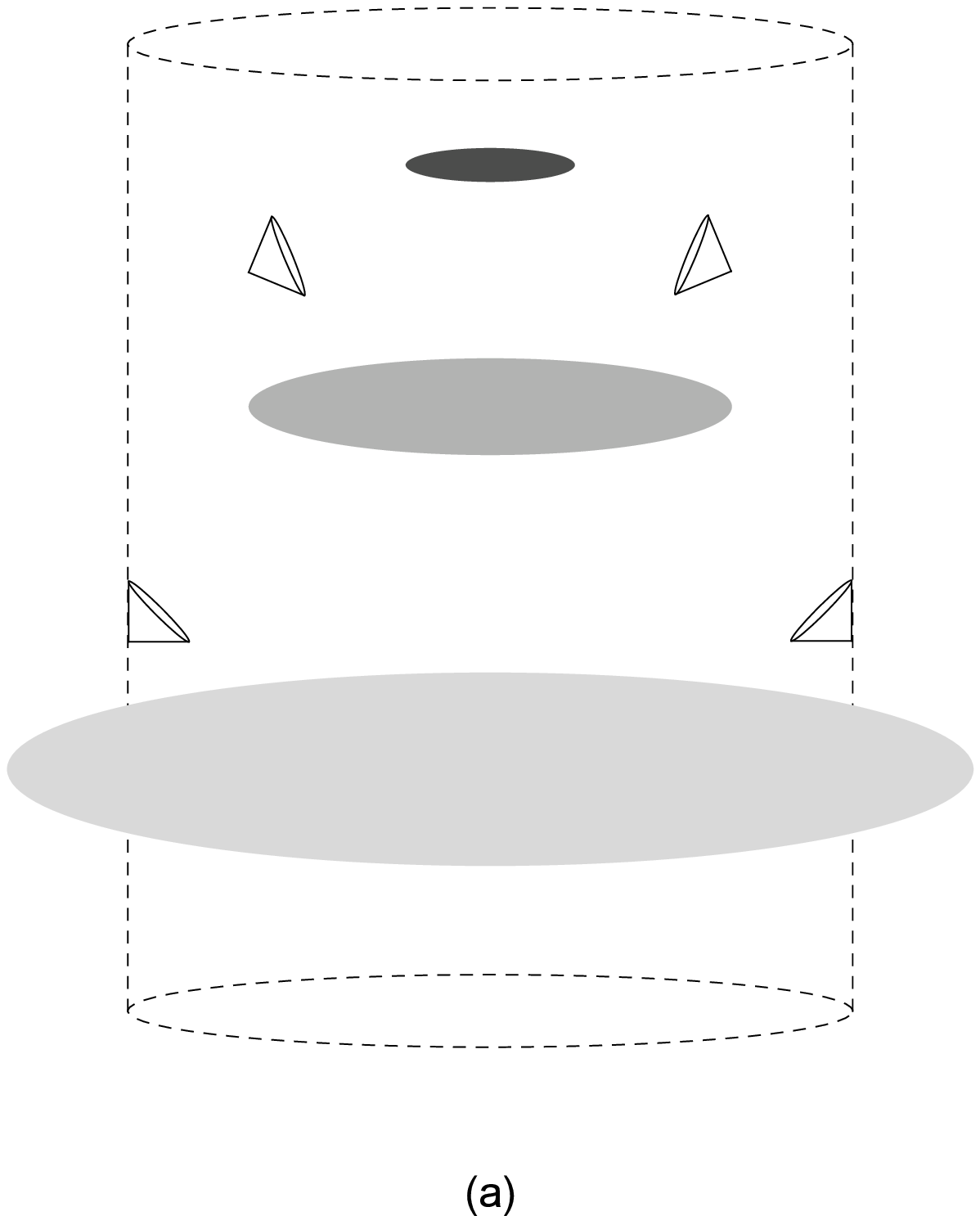}\hskip100pt
\includegraphics[scale=.55]{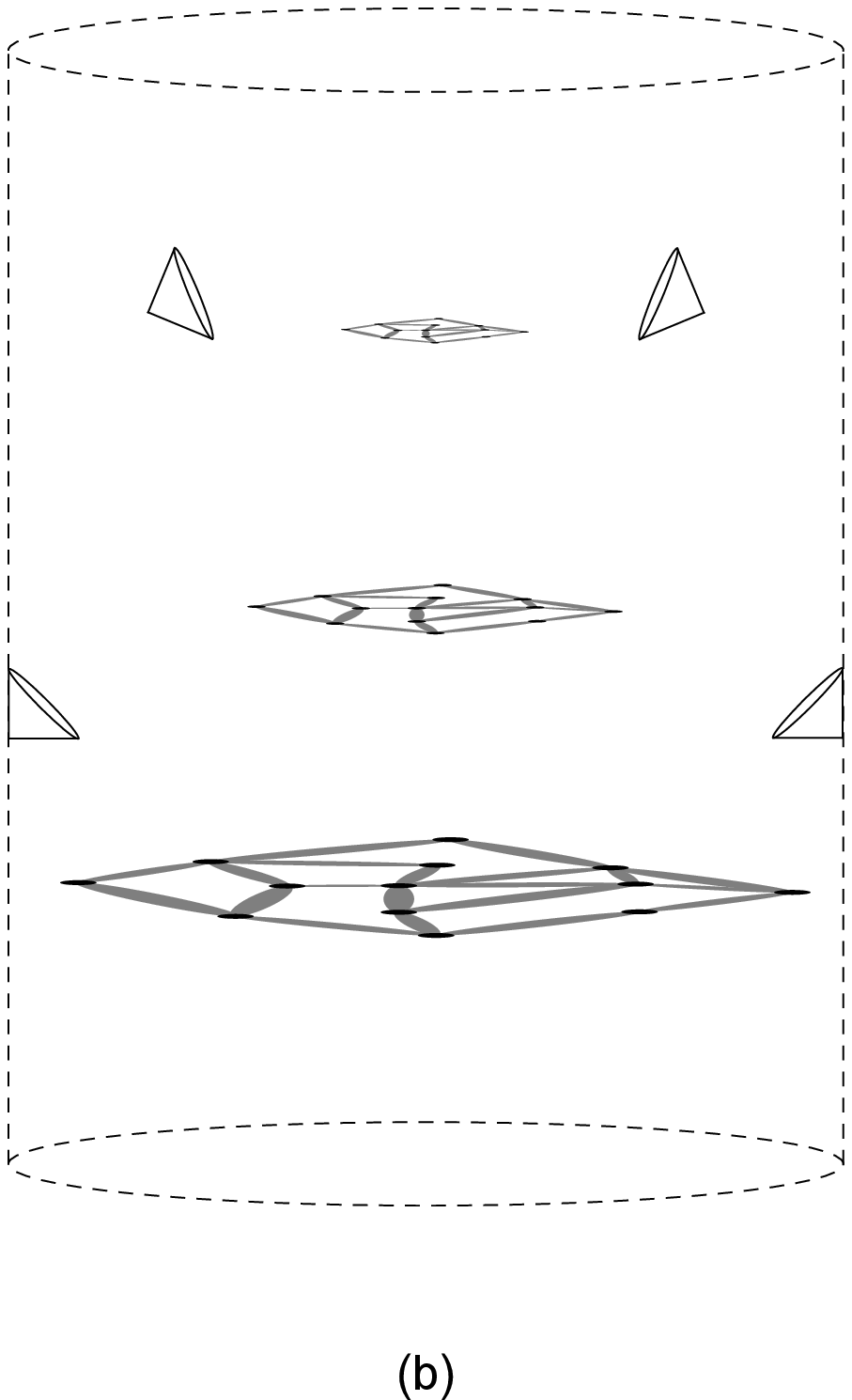}
\end{center}
\caption{(a) The gravitational collapse of a dust cloud generates a horizon; the light cones point inwards inside the horizon. (b) Vecro configurations inside the horizon are also forced to keep compressing due to the inward pointing structure of light cones.}
\label{fig20}
\end{figure}

 With this picture of causality,  we outline the steps leading to the black hole information paradox:
 
 \b
 
 (a) Consider a spherically symmetric shell of mass $M$, composed of radially moving massless particles. Let the proper radius of the shell be $R(\lambda)$ where $\lambda$ parametrizes the null trajectories.  Take one of the particles $p$. By causality, particles $p'$ on the shell which are not infinitesimally close to $p$ cannot send a signal to $p$ which would signal their existence. 
 
 \b
 
 (b) Suppose that the quantum gravity vacuum was such that all fluctuations of importance were local; for instance,  the fluctuations could be  confined to a length scale $\lesssim l_p$. 
 
 \b
 
 (c) Given (a),(b) we can say that the infall of $p$ would be, to leading order, the same as it would be in a patch of locally smooth spacetime. The reason is that at any one location on the shell, neither the classical gravity theory nor the quantum fluctuations know about the existence of the remainder of the shell.  In that case we must have the validity of the equivalence principle  which says that the motion of any particle like $p$ can be assumed to take place in a local patch of  spacetime that is approximately flat. In particular this would hold true as $p$ crosses the horizon radius $R(\lambda)=2GM$. Thus the shell must pass through its horizon without encountering any obstruction to its leading order classical infall.
 
 \b
 
 (d) Once the shell is inside its horizon, the light cones in the region 
 \be
 R(\lambda )< r<2GM
 \ee
 points `inwards'; see fig.\ref{fig20}(a).  Our assumption of causality to leading order now tells us  that   regardless of what happens to the shell when it reaches $R(\lambda)\r 0$, the shell cannot change the  leading order evolution of low energy fields around $r=2GM$. Thus the state of  Hawking pairs created  at the horizon can be modified only by $O(\epsilon)$ with $\epsilon\ll 1$. 
 
 \b
 
 (e) The small corrections theorem now tells us that these $O(\epsilon)$ corrections cannot stop the monotonic rise in entanglement between the radiation and the remaining hole. We will therefore be forced to information loss or remnants; the former is not allowed in a quantum theory of gravity like string theory,  and the latter is disallowed in string theory if we accept  AdS/CFT duality.
 
 \b

 We have given the steps of the information paradox in full detail so that we can set up the discussion of its resolution. The step we will alter is (b): vecros are extended objects that exist
 as fluctuations in the vacuum, so they are not localized to within a fixed distance like $l_p$. This extended size does not of course violate causality. The spacetime region where the hole will form has been in existence for a long time before we send in the shell of mass $M$. So there has been ample time for it to develop correlations over extended length scales, e.g. on scales of size $\sim GM$. If such correlations lower the energy in our gravity theory, then the vacuum state must have these vecros as a part of its wavefunctional. 
 
\subsection{Gravitational collapse with the vecro hypothesis}\label{secc3}

We now outline the picture of collapse in a theory where we have  the vecro component of the vacuum wavefunctional:

\b

(a') Consider again the shell of mass $M$ made of  radially infalling massless quanta, let its radius be $R(\lambda)$.  The region $0\le r<R(\lambda)$  is flat spacetime. Note that the shell is collapsing in a spacetime which at $t\r -\infty$ was Minkowski space where, by 
our assumptions in section \ref{secc1}, causality is exact. Thus there can be no effect of the shell, classical {\it or} quantum, in its interior.

\b

(b') Outside the shell, the spacetime is distorted by the gravitational field of the shell. First consider the situation where the shell is far outside its Schwarzschild radius; i.e., $R(\lambda)\gg R_0$. The gravitational attraction outside the shell is weak. The vecro excitations in the vacuum get pulled inwards by the gravitational field of the shell, but this effect is small. The vecros compress only slightly, and stabilize with a slightly smaller radius. Thus the vecro part of the vacuum wavefunctional  distorts only 
by a small amount. This distortion is part of the normal adjustment of the vacuum wavefunctional to the presence of the mass $M$, and is included in the semiclassical physics of quantum fields on curved spacetime. 

\b

(c') Now consider the situation where the shell has passed through its horizon, so $R(\lambda)<R_0$. Consider a vecro with radius $R_v$ in the range $R(\lambda)<R_v<R_0$. The light cones in this region point `inwards', so the structure in the vecro has to keep compressing (fig.\ref{fig20}(b)); it cannot resist this compression and stabilize at a slightly smaller radius.  The compression of the vecro will reach $\delta \sim 1$, and the vecro structure will be completely altered. Thus the vecro part of the vacuum wavefunctional distorts by a large amount. Just as distorting the wavefunctional of the semiclassical wavefunctional gave rise to the creation of on-shell particles, distorting the vecro part of the wavefunctional gives rise to on-shell fuzzballs.

\b

(d') Let us also consider the energy balance on a hypersurface through the spacetime. To see the creation of fuzzballs, we have to look at the region $R(\lambda)<r<R_0$, over a spacetime region of size  $\sim R_0$. Thus we have to look at hypersurfaces like that in fig.\ref{fig12}(b) which have evolved by $\Delta u\sim R_0$ to the future of the hypersurface describing the null shell. We have noted in section \ref{bagresolve} that the vecro wavefunctional acquires an extra energy $\Delta E_{vecro}$ from the region between the shell and the horizon. This energy is in addition to the energy $M$ of the shell that creates the hole. But in the present case we are limited to a total energy $M$ for the system, since we are following a process of dynamical evolution for a shell which had energy $M$ at $r\r\infty$. 

Thus we need a source of negative energy on the slice. Now we note that once we distort the vecro wavefunctional to create on-shell fuzzballs, then these fuzzballs excitations are like normal excitations of the spacetime. Such excitations carry energy that is positive in a local orthonormal frame, but since they are located in the region $R<R_0$, their energy as seen from infinity can be either positive or negative (see the discussion of section \ref{secnegative}). We see  that  the energy of these fuzzball excitations $\Delta E_{fuzzballs}$ must be negative, and we can then get
\be
E_{total}=M+\Delta E_{vecro}+\Delta E_{fuzzballs}=M
\ee
as required by energy conservation.

\b

To summarize, the gravitational field of the shell destabilizes the vecro part of the vacuum wavefunctional in the region $R(\lambda)<r<R_0$, creating fuzzballs that fill this region. This evolution to fuzzballs happens on a scale of order the crossing time, after which the fuzzballs radiate from their surface like any normal body; this resolves the information paradox. Note that the extended nature of the vecro configurations allowed them to `feel around' an extended region and thus detect the existence of the entire infalling shell. In the argument of section (\ref{secc2}) we violate the assumption in step (b) since the vecros fluctuations have all sizes upto infinity, and this in turn invalidates the equivalence principle assumed in step (c). This is what allows us to evade the paradox. 

Note that we have not violated causality or locality; signals still propagate in the spacetime with a speed less than or equal to the speed of light. In the following section we will see that this constraint of causality prevents us from having a picture where fuzzball formation takes place over timescales that are even shorter than crossing time $\sim R_0$.

\subsection{The constraint from partial shells}\label{secc4}

We have given a picture where the deformation of the vecro wavefunctional in the region $R(\lambda)<r<R_0$ changes this region from the vacuum to a region filled with fuzzball excitations. This change happens over a spacetime region of size $\sim R_0$. This scale was built into our picture by the requirement that the vecros compress by a factor of order unity; i.e., $\delta\sim 1$.  Since the vecros of interest have a radius $\sim R_0$, the spacetime region involved in this compression also has a size $\sim R_0$. 

At this point one could ask: could we instead postulate that vecros turn into on-shell fuzzballs under a parametrically {\it smaller} compression? A fundamental length scale in quantum gravity is the planck length $l_p$. Could we require that a vecro gets destroyed under the compression $R_v\r R_v-l_p$? In that case the infalling shell will turn into fuzzballs just after it enters its horizon radius $R_0$.

In this section we will argue that such a stronger requirement on vecros is not allowed by causality. In fact we cannot take any fixed length $s$ and require that the new effects of vecros arise for the compression
\be
R_v\r R_v-s
\ee
This constraint arises by requiring a physically reasonable behavior for the evolution of {\it partial shells}; i.e., shells which form a fraction of a sphere but not a whole sphere. When this fraction is small, the shell should behave like a plane wave, which should {\it not} leave a wake of fuzzballs when it passes through a region with low curvature. Causality says that one part of the shell cannot know about the existence of the remainder of the shell until light has had time to travel far enough to explore these other parts of the shell. This will force us to require a fractional compression of order unity (i.e., $\delta \sim 1$)  for the vecro configuration before it leads to novel effects like fuzzball creation. 

A given point of a vecro responds to the gravitational field present at that point. We therefore wish to understand what part of this field is produced by different regions of the shell. Since this computation is difficult for gravity, we will first perform it for the simpler case of electromagnetism, and then extrapolate the conclusions to gravity.

\subsubsection{A shell in electrodynamics}

First consider a static shell of charge $Q$ and radius $R$, in $d+1$ spacetime dimensions.  The charge per unit area is
\be
\sigma = {Q\over R^{d-1}\Omega_{d-1}}
\ee
where $\Omega_{d-1}$ is the volume of a unit $d-1$ sphere. The electric field outside the sphere is
\be
\vec E = {k Q\over r^{d-1}} \hat r= {k \sigma \Omega_{d-1}} \left ( {R\over r}\right ) ^{d-1}\hat r
\label{zonemain}
\ee
Consider this field at a point P just outside the sphere; i.e.  at a radius $r=R+a$ with $a \ll R$. This field can be broken up into contributions from two parts of the shell:

\b

(i) The part of the shell close to P. This part  looks like a flat plane. The flux from this plane is normal to the plane, and equal in magnitude on the two sides of the plane.  Thus by the Gauss law this part of the shell contributes a field 
\be
\vec E_1\approx \h{k \sigma \Omega_{d-1}}  \hat r
\ee

\b

(ii) The remainder of the shell. We get a  contribution  from all regions of this part, not just the points close to P. This contribution is 
\be
\vec E_2 \approx   \h {k \sigma \Omega_{d-1}}  \hat r
\ee
Thus  $\vec E_1+\vec E_2=\vec E$. 

\b

As we will now note, the situation is very different when we consider the shell to be collapsing at the speed of light. Suppose the shell has reached a radius $R(\lambda)=R$, and we are again looking at the field at a point $P$ with radius $r=R+a$, with $a \ll R$. The point $P$ cannot receive any information from the charges on the shell that are coming in at angular directions very different from the angular position of $P$. In fact $P$ can know only about points on the shell that are within a disc of radius $\sim a$ around the point $P'$ on the shell closest to P. 
The electric field $\vec E$ at $P$, on the other hand, is has the same value (\ref{zonemain}) as in the static case; this follows by the Gauss law. Thus this $\vec E$ must arise only from the charges in the small disc around $P'$; the other charges in the shell are irrelevant to the field at $P$.

  \begin{figure}[htbp]
\begin{center}
\includegraphics[scale=.5]{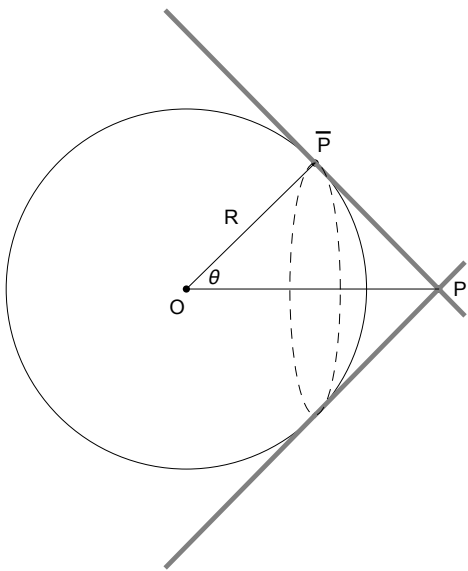}\hskip20pt
\end{center}
\caption{The circle denotes a spherical shell of charged particles falling radially inwards at the speed of light. The electric field of a charge at $\bar P$ is confined to a plane transverse to the motion of charge. Thus the field at $P$ gets contributions only from charges like $\bar P$ that lie on the dotted circle.  }
\label{fig5}
\end{figure}

In Appendix \ref{appa} we compute the field exactly for any velocity $v$ of the collapsing shell, and thereby verify the above observation.  Here we summarize the physics with the help of fig.\ref{fig5}.  If a charge is moving at a high speed, then its electric field lines get compressed to a plane that is perpendicular to the direction of motion. Since electrodynamics is linear, we can just add the fields from the different charges on the shell. From fig.\ref{fig5} we see that points that are far from $P'$ have their fields concentrated in planes that do not reach $P$. 
The field at $P$ gets contributions only from points like $\bar P$ in the figure; these points form a ring of radius $\sim a$ around $P'$.\footnote{One often thinks of a spherical shell as being made of  a large number of particles $n$, and then taking the limit of $n$ large. We see that for null shells the field $\vec E$ is not smooth in this limit. Since the field at $P$ comes only from points like $\bar P$ on a ring, the field at $P$ will be {\it zero} if there are no charges that lie precisely on this ring. This fact has its equivalent   for the gravitational field of a null shell; the nonsmoothness of this limit will be relevant for  the question of how  the information in the  shell is transferred to the fuzzballs that are created by its gravity.}

As we increase the velocity of the shell from $v=0$ to $v=1$, we find that the contribution of the part of the shell close to $P'$ rises and the contribution of the rest of the shell falls, till at $v=1$ the contribution comes just from the dotted ring in fig.\ref{fig5} which lies in  a neighborhood around $P'$ of radius $\sim a$.

This discussion is important because the constraints imposed by causality can be studied using the behavior of {\it partial shells}; i.e., shells which cover only a solid angle $\Delta \Omega$ out of the full sphere. We will examine such constraints now.

\subsubsection{Partial shells in gravity}

 First consider  a single particle moving at the speed of light in otherwise empty Minkowski space. The metric is given by the Aichelburg-Sexl shock wave 
  \be
 ds^2=-dudv+h(\rho) \delta(v) dv^2+ (d\rho^2+\rho^2 dy_idy_i)
 \label{ztwoq}
 \ee
 where $u=t+x, v=t-x$ and  $\triangle h=-C\delta (y)$ for some constant $C$.
 
 Suppose that an observer is standing at a fixed position $P=(\bar x, \bar y_i)$. Before the shock arrives, his neighborhood is a patch of flat Minkowski space, and after the shock passes, it is flat Minkowski space again.  But during the time that the shock was crossing the position $P$, the metric near $P$ was strongly time dependent. One may then ask: why did this time dependent metric not lead to particle creation? 

The answer of course is that the metric of the shock wave is fine-tuned in such a way that such particle creation will {\it not} happen. To get this metric we start with a particle of small mass $m$ at rest; this metric is weak away from the particle and also time-independent so that it does not create particles. Then we boost the particle with a boost $\gamma$. The limit $m\r 0, \gamma\r \infty$, with $m\gamma$ fixed  gives the shock wave metric. The boost results in a Lorentz contraction of the gravitational field to a plane perpendicular to the direction of motion, much like the eletromagnetic case. We therefore end up with flat Minkowski space both before and after the shock. 

 If we think of the shock as having a small width $\delta$, then the vacuum is indeed disturbed when the shock reaches $P$. But the field of the shock is arranged in such a way that these disturbances are cancelled out by the time the shock passes. This `miraculous' arrangement is, of course, a consequence of the underlying  Lorentz invariance of Minkowski space which allowed us to get the time-dependent shock as a boost from a time independent metric.
 
 Before proceeding, we note that all that was said above holds if we smear the source of the 
 Aichelburg-Sexl wave into a disc $D$ transverse to the motion. For a disc $D$ with surface energy density $\sigma$ and radius $\rho_0$ the metric would be 
 \be
 ds^2=-dudv+f(\rho) \delta(v)dv^2+ (d\rho^2+\rho^2 dy_idy_i)
 \label{ztwo}
 \ee
 with
 \bea
 \triangle f(\rho) &=& -\sigma, ~~~\rho<\rho_0\nn
 &=&0, ~~~\rho>\rho_0
 \eea
 so the space would be flat Minkowski space before and after the passage of the wave.

Now consider the case of an infalling shell in gravity, with the particles moving radially at the speed of light. As noted above  a single particle moving at the speed of light creates the Aichelburg-Sexl metric (\ref{ztwoq})  where the  gravitational field is confined to a plane transverse to the motion. This is analogous to the electromagnetic case where the field of a charge moving at the speed of light was  also confined to the transverse plane. But gravity is not linear, so we cannot superpose the shock waves from different particles to get the overall gravitational dynamics created by the shell. Thus we will proceed with qualitative arguments as follows. We will compare two cases: 

\b

{\bf Case A:}   We have a shell of mass $M$ made of radially infalling particles moving at the speed of light. The radius of the shell is $R(\lambda)$ where $\lambda$ is the affine parameter along the radial trajectories. At some point $\lambda=\lambda_0$ the shell crosses its horizon $R_0$. At a slightly larger $\lambda$, the shell is at a radius
\be
R(\lambda)=R_0-s
\label{argfive}
\ee
where we take
\be
s\ll R_0
\label{argone}
\ee
Consider a point $P'$ on the shell when the shell is at the radius (\ref{argfive}).  We are interested in the gravitational effects at  points  $P$ which is just outside the shell, after the shell has crossed its horizon. We will let $P$ have the same angular coordinates as $P'$.  We set
\be
r_P=R_0-{s\over 2}
\label{argradius}
\ee
We still need to select one more coordinate to define the point $P$. We will actually look at a set of such points $P$ defined as follows. Outside the shell we have the Schwarzschild metric, and we can place the usual Kruskal $u,v$ coordinates on this region. Take a ball in $u,v$ space of size $\sim s$ around the point $P'$ on the shell; i.e., we take the region around $P'$ outside the shell with distance from $P'$ given through $\Delta u\sim s, \Delta v\sim s$. We are interested in points $P$ in this ball with the radius (\ref{argradius}) and angular coordinate the same as $P'$. 

In semiclassical gravity there is nothing special at such  points $P$  near the horizon; we just have gently curved spacetime there with curvature radius $\sim R_0$. The question we have is the following: with the vecro hypothesis, will the spacetime at $P$ be altered from the local vacuum to a region having fuzzball excitations?

\b

{\bf Case B:} To answer the above question, we also consider a second physical system.  We again start with the shell of mass $M$ as in case A. When the shell reaches a radius
\be
 r=R_0+D
 \ee
  we alter the dynamics of the shell as follows. Select a point $P'$ on the shell as in case A. Mark out a sector of the shell whose boundary is a $d-2$ dimensional sphere centered at $P'$. Let the solid angle enclosed by this sector be   $\Delta\Omega$.  This sector is a  `partial shell' which we will denote by ${\cal S}_{\Delta\Omega}$. We assume that  ${\cal S}_{\Delta\Omega}$  continues moving inwards at the speed of light as before.The rest of the shell we will denote by ${\cal S}'$. We assume that ${\cal S}'$ stops its infall at $r_{stop}=R_0+D$ and stays at that location.\footnote{We assume that the change in the nature of the infalling particles  required to make this alteration of dynamics arises from potentials that are small enough to not affect the argument.} We assume that $D\sim R_0$. Further, let 
  \be
\Delta\Omega\ll 1
\ee
so that ${\cal S}_{\Delta\Omega}$ behaves like an almost flat disc. The radius of this disc is
\be
l\sim R_0 (\Delta\Omega)^{1\over d-1}
\ee
We assume that
\be
s\ll l\lesssim D
\label{argfour}
\ee
where $s$ was the distance inside the horizon that was selected in (\ref{argone}). 

We now ask the same question that we did in case A. Do we have fuzzball excitations at the points $P$ just outside the partial shell?

\b

To answer the above question, we begin by asking: is the gravitational dynamics the same at the location $P$ in both cases A and B? We argue that it should indeed be the same. The reason is that for the dynamics to be different, the point $P$ must have the information about whether the part ${\cal S}'$ of the shell did or did not continue its journey past $r_{stop}=R_0+D$. We argue as follows:

\b

(i) Before any part of the shell reaches the radius $r_P$, the metric at $P$ is flat; there is no signal of any of the incoming particles since these particles are coming in at the speed of light from infinity and none of them has reached $P$.

\b

(ii) After particles of the shell have crossed the radius $r_P$, there is a nontrivial gravitational effect at $P$. For case B, consider a particle    on the boundary of the segment ${\cal S}'$; such a particle is the closest to $P$ and so has the best change of sending information to $P$.  Suppose a signal is dispatched from the point $P''$ on this particle's trajectory where the particle stops at $r_{stop}$. Let us ask if such a signal can reach the point $P'$ on the shell. The answer to this is obviously {\it no}; this follows since the point $P'$ is reached by a radial null trajectory along an infalling particle, while the signal from $P'''$ has to cover the same radial distance as well as an angular distance $\sim l$ in the same time.  

We have taken $s$ to be much smaller than all other length scales involved; we think of it as a fixed length (say planck length $l_p$) and then scale up all other lengths in the problem by taking $R_0\r\infty$. Then if a signal cannot reach from $P''$ to $P'$, then it cannot reach from $P''$ to $P$ either. The field at $P$ is thus the same in cases A and B, and so can be said to come from the partial shell ${\cal S}_{\Delta\Omega}$, with no contribution from the remainder ${\cal S}'$. this is analogous to the situation with the null shell on electromagnetism, where we could carry out the computation of the field explicitly.

\b

We can now conclude our argument. In case B we do not expect fuzzball excitations to arise at $P$. We have not made a black hole, and the gravitational effects at $P$ arise from a disc which is very gently curved, and so is almost like the Aicehberg-Sexl disc (\ref{ztwo}). For such a disc we expect that the spacetime will return to the local vacuum after the disc has passed. But if there are no fuzzball excitations at $P$ in case B, then there cannot be any fuzzball excitations in case A either, since the field at $P$ cannot, by causality, know about the difference between cases A and B. 

This concludes the argument. The lesson from this argument is an important one for the dynamics of vecros. A vecro configuration is an extended object with some radius $R_v$. But the dynamics of the theory is local, and signals cannot propagate faster than the speed of light. Thus we cannot get novel physics like fuzzball formation by compressing a vecro by an amount $s\ll R_v$. To get these novel effects we must compress the vecro by a distance $s\sim R_v$; over such times the entire vecro structure can receive a signal that the compression is taking place.  Equivalently, the time scales for which  vecros of radius $R_v$  must be deformed to create a permanent deformation in their wavefunctional is $\Delta t\sim R_v$; deformation on times much shorter than this lead to semiclassical physics without any creation of fuzzballs.\footnote{One may also ask about the constraint from causality in the process of colliding two oppositely moving Aichelberg-Sexl shock waves. The spacetime in this case differs in an essential, way from the case of the collapsing shell, as noted in the discussion of fig.7 of \cite{causality1}. For the collapsing shell we have a region $R(\lambda)<r<R_0$ where the light cones have turned `inwards', and no signal can escape from this region to infinity; this is also the region where fuzzball formation will happen. By contrast, for the colliding shock waves there is no such region; signals from the shocks can escape to infinity for all times of emission upto the moment of collision. Thus the spacetime remains Minkowski space (with no fuzzball formation) in the neighborhood of each shock wave; fuzzball formation happens only after their actual collision when nontrivial effects could be expected to occur anyway.}

\section{Cosmology}

The information paradox poses a  puzzle for  black holes, and we have noted that using the small corrections theorem we can convert this puzzle to a precise contradiction.  We will now recall how we can map this puzzle to an equally precise puzzle for cosmology \cite{cosmoessay1}. 

Suppose we have (i) CPT invariance (ii)  the Birkoff theorem of classical gravity holding to leading order in  gently curved spacetime (iii) our assumption of causality to leading order in gently curved spacetime. Then as we will recall below, we can map  a ball of dust in an expanding cosmology to the gravitational collapse of a dust ball in asymptotically flat spacetime. The information paradox  requires order unity quantum gravity effects at the black hole horizon to preserve unitarity. But as we will note below, we do not see evidence for large violations of semiclassical physics in the sky. How should we account for this difference between the black hole horizon and the cosmological horizon?

 We will  see that the vecro hypothesis   resolves this puzzle by noting a basic difference in the structure of the  quantum gravity  wavefunctional between the two cases. More precisely, the  vecro distribution function will differ between asymptotically flat spacetime and an expanding cosmology.  In asymptotically flat spacetime we have vecros fluctuations of all radii $R_v$, while in a cosmology these fluctuations exist only with radii smaller than the cosmological horizon radius; i.e. $R_v\lesssim H^{-1}$. 
 
 This difference will resolve the contradiction noted above. What is interesting is that this resolution of the puzzle brings in the role of quantum gravity at the cosmological horizon scale, just as the fuzzball resolution of the information paradox brought in the role of quantum gravity for   black holes at the black hole horizon scale. Thus we open  the door to many possible effects of quantum gravity in today's universe, in particular for the cosmological constant problem. 

\subsection{Mapping the  information paradox to cosmology}\label{cosmopuzzle}

\b

\b

\begin{figure}[htbt]
\begin{center}
 \includegraphics[scale=1.1] {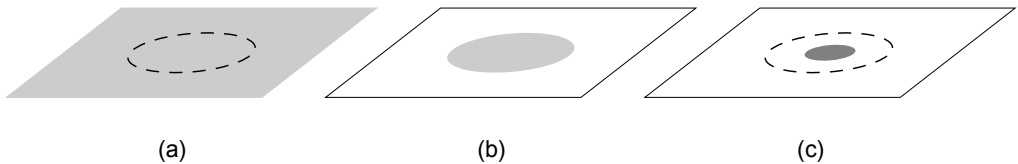}
\end{center}
\caption{ (a) A constant time slice in a homogeneous collapsing flat cosmology. (b) By the Birkoff theorem,  the dynamics of the marked ball cannot change if we remove the matter outside. (c) The collapsing ball then reproduces the semiclassical picture of the black hole, a situation which leads to the information paradox.} 
\label{fig31}
\end{figure}

We proceed in the following steps:

\b

(a) Consider a flat dust cosmology in $d+1$ spacetime dimensions
\be
ds^2=-dT^2+a^2(T)(dr^2+r^2 d\Omega_{d-1}^2)
\label{qtenpmain}
\ee
The energy density of the dust, measured in a local orthonormal frame where the dust is at rest,   has the form
 \be
 T^0{}_0=\rho_0={C\over a^d}
 \label{qtenqmain}
 \ee
 The cosmological metric has  the expansion
 \be
 a=a_0 T^{2\over d}=\left ( {4\pi dGC\over d-1}\right )^{1\over d} T^{2\over d}
 \label{qtenmain}
 \ee

\b

(b) We map the big bang (\ref{qtenpmain})  to a big crunch by the map $T\r -T$. We assume that our physical theory is invariant under CPT; the  standard model and string theory both have  this invariance. The physical theory will then allow  a big crunch solution if it allows  a big bang solution.

\b

(c) Consider the spacelike slice $T=T_0<0$. Let the density of dust on this slice  $\rho_0$ be  low, very much below the planck scale.  Mark out a  ball around $\vec r=0$ of proper radius $R_b=a(T) r_b$. The mass of dust  in this ball is
\be
M_b=\rho_0 {\Omega_{d-1}\over d}(R_b)^{d}={C} {\Omega_{d-1} \over d}r_b^{d}
\label{qfourtbmain}
\ee

\b

(d) As the universe heads towards the big crunch, the proper radius of the ball $R_b(T)$ shrinks towards zero. At least at the classical level, the evolution inside this ball is identical to the gravitational collapse of a dust ball of mass $M_b$ in asymptotically flat spacetime. The reason the two evolutions are similar is the Birkoff theorem, which holds in classical general relativity. This theorem says that the metric inside a spherically symmetric shell is {\it flat}.  Thus a spherical shell should have no gravitational influence on its interior. This implies that  if matter is distributed in a spherically symmetric way outside a shell $r>R_b$, then we can replace the matter at $r>R_b$ by  empty space.

We use this result as follows. On the slice $T=T_0$ we  discard the dust in the region outside our selected ball and replace it with empty space.  The surface of the collapsing ball is matched onto  an exterior Schwarzschild metric
 \be
 ds^2_S=-(1-\left ({R_0\over R}\right ) ^{d-2})dT^2+ {dR^2\over (1-\left ({R_0\over R}\right ) ^{d-1}} + R^2 d\Omega_{d-2}^2
 \label{qtenqqmain}
 \ee
 with
 \be
R_0=\left ({16\pi GM_b\over (d-1)\Omega_{d-1}}\right )^{1\over d-2}
\label{qthirmain}
\ee

\b

(e) We have now mapped the cosmological situation to the black hole problem. Suppose we observe the cosmology and notice no order unity violations of semiclassical physics at the radius $R_0$ in (\ref{qthirmain}). Then we would not be able to get a violations of semiclassical physics at the black hole horizon either. But if we keep semiclassical physics at the horizon to leading order, then by the small corrections theorem we cannot resolve the information paradox while maintaining causality in gently curves spacetime (fig.\ref{fig31}).

\b

(f)  One might argue that the Birkoff theorem does not need to strictly hold in quantum theory. For example the Casimir effect can give a small nonzero energy density inside a conducting shell, while the classical electromagnetic theory would have predicted a zero energy density. But here we can again use the small corrections theorem. Suppose the process of replacing the dust cosmology (\ref{qtenpmain})  by empty spacetime (\ref{qtenqqmain}) in the region $R>R_b(T)$ creates a small error to the evolution of the ball in the region $R<R_b(T)$. We can include this error as a small correction to the Hawking pairs that will be created in the Schwarzschild metric (\ref{qtenqqmain}), as indicated in (\ref{qonenew}). But then the small corrections theorem (\ref{eight}) tells us that we will continue to have the problem of growing entanglement; i.e., these small corrections will not get us out of the paradox. Thus in mapping cosmological evolution to a collapsing dust ball we just need the Birkoff theorem to hold to `leading order'; not as an exact result in the quantum gravity theory. 
 
 \b
 
 This concludes the argument. Note that we used a dust cosmology in our argument, as opposed to a cosmology with a more general fluid having a nonzero pressure $p$. If we had $p\ne 0$, then the part of the cosmology (\ref{qtenpmain}) at $R>R_b(T)$ would exert a pressure on the dust ball $R<R_b(T)$. In that case we cannot replace the outside of the ball by empty space without altering the dynamics of the ball itself. We will see later that the vecro fluctuations of the quantum gravity  wavefunctional  connect the physics across the radius $R_b$, and this is what will resolve the contradiction.

 \subsection{Observing the horizon scale}
 
 Let us now note what we can learn from observations of the sky. We will note  that we can observe matter that is squeezed past the density that would  make a horizon in asymptotically flat space, and we can also observe dynamics at  cosmological horizons. These facts will confirm that we indeed have the puzzle in section \ref{cosmopuzzle} above; i.e., we need to reconcile the need for order unity quantum gravity effects at the black hole horizon with observations that show no violations of semiclassical physics in the sky.
 
 \subsubsection{Squeezing of matter to a size less that its horizon radius}
 
 We see galaxies when the light they emit reaches us. If we look at objects that are further out, then our vision encompasses a larger amount of matter. Moreover, we are looking at matter at earlier times $T$, so this matter would have been denser. Thus if we look at a large enough sphere in the sky today, then we could be looking at a ball of matter which was heavy enough and dense enough to be within its own Schwarzschild radius. 
 
Let us make this more precise. Suppose we are   looking at light emitted from a sphere of dust  at time $T_1$. Let the proper radius of this sphere  be $R_p$ (as measured at this time $T_1$). Let the mass of this dust  ball be $M$, and the Schwarzschild radius for this mass be $R_s$. If $R_p\ll R_s$, then we are observing a sphere of dust that has been compressed to a size much smaller than its horizon radius. 
   
 The details of this computation are in Appendix \ref{appb}; here we summarize the results.   
 We find that
 \be
{\cal C}\equiv {R_p\over R_s}=\left ({(d-2)\over 2}\right )^{2\over d-2}\left ( {T_1\over T_0}\right )^{2\over d}
\ee
Suppose  we set
 \be
T_0\approx 1.4\times 10^{10}\, yrs, ~~~T_1\approx 4\times 10^5 \, yrs
\ee
This gives
\be
{\cal C}\sim 2\times 10^{-4}\ll 1
\ee
Thus we can observe balls of matter that have been squeezed much past their horizon radius.  If we do not observe violations of semiclassical physics for balls of matter that have been squeezed in this way,  then we have to reconcile this fact with the need for quantum gravity effects at the black hole horizon. 

Note that  as far as the information paradox black hole it is still possible for the a dust ball to squeeze over crossing time to a radius that is much smaller than its horizon size, and then relax to a fuzzball type object on a longer time scale; as long as the horizon gets removed in a time shorter than the Hawking evaporation time scale we will avoid the problem of monotonically growing entanglement of Hawking pairs. But in our picture of gravitational collapse we have  fuzzball formation in a time of order the crossing time. Thus we need to reconcile this behavior of dust balls squeezed in gravitational collapse with the existence of highly squeezed dust balls in cosmology.

\subsubsection{Observing the cosmological horizon}

The black hole horizon has a definite location: it is a sphere of radius $2GM$ around the center of the hole. By contrast the cosmological horizon is not a particular place; rather, it is a  is a length scale, and we can draw a sphere of this radius around any point in our homogenous universe. Since we see no large quantum gravity effects in our solar neighborhood, we could use this to conclude that there should be no large quantum gravity effects at cosmological horizons. But one might still argue that the relevant effects can only be seen only upon observing regions as large as the horizon. Can we observe regions this large?

We cannot see past the cosmological horizon today, since light from distances of this order   would barely have reached us today. But in the past the cosmological horizon was smaller, and looking at the sky we can indeed see regions larger than the cosmological horizon at those times. 

Consider an early time $T_1$ at which we can observe the universe through light emitted at $T=T_1$. In Appendix \ref{cosmohorizon} we find that the cosmological horizon radius at this time is, in comoving coordinates
\be
r_1(t_1)={d\over 2 a_0} T_1^{d-2\over d}
\ee
Now consider the comoving radius of  the region that we can observe; this is given by $r_p(T_1)$. The ratio of these two radii is
 \be
\t  {\cal C}\equiv {r_1(T_1)\over r_p(T_1)}\sim \left ( {T_1\over T_0}\right ) ^{{d-2\over d}}
 \ee
 If we set
 \be
T_0\approx 1.4\times 10^{10}\, yrs, ~~~T_1\approx 4\times 10^5 \, yrs
\ee
we find
\be
{\cal C}'\sim {1\over 30}
\ee
Thus we can observe regions at early times $T_1$ that are larger than the cosmological horizon at that time; if we do not see evidence of large quantum gravity effects at these cosmological horizon scales then we have to reconcile this fact with the need for quantum gravity effects at the black hole horizon. 

\subsection{Resolving the cosmological horizon problem}

  \begin{figure}[htbp]
\begin{center}
\includegraphics[scale=1.05]{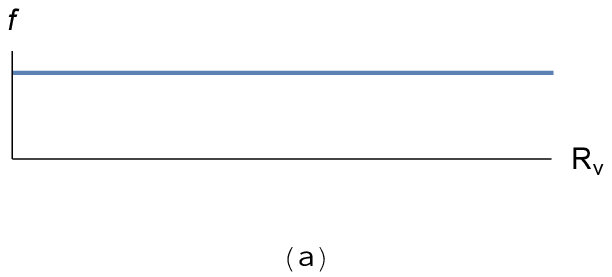}\hskip30pt
\includegraphics[scale=1.05]{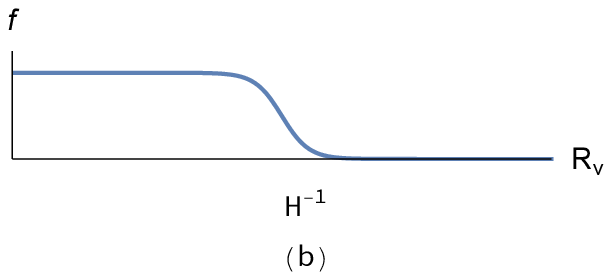}
\end{center}
\caption{(a) The vecro distribution function $f(R_v)$  for Minkowski space; we have vecros of all radii $0<R_v<\infty$ (b) The vecro distribution function for a cosmology; we have support only on vecros with radii $0<R_v\lesssim H^{-1}$.}
\label{fig24}
\end{figure}

We have seen that once we accept that the Birkoff theorem is true to leading order in gently curved spacetime, then the cosmological situation can be mapped to the black hole puzzle, and we would then be forced to require large quantum gravity effects like fuzzball creation at the cosmological horizon. Since we do not see such effects, we must find a way to argue that the Birkoff theorem does {\it not} apply to  cosmology.

We see the the vecro hypothesis gives us a natural way out of the puzzle. Vecro fluctuations correspond to extended configurations in the gravity theory. A region of the cosmology with radius $R\gtrsim H^{-1}$ satisfies the condition (\ref{ntwo}), so we expect nontrivial effects from the vacuum wavefunctional at this scale. We argue as follows:

\b

(i)  For the black hole, we had noted that the vecro wavefunctional is not stable inside the horizon, since the inward pointing structure of light cones forces the vecro configurations to keep compressing without bound. This inward pointing structure of light cones is of course the consequence of having a closed trapped surface. In our cosmology, the horizon $R=H^{-1}$  is an `anti-trapped surface' i.e., light cones point {\it outwards} at this surface; see eq. (\ref{eqhor}). Thus any vecro configuration with radius $R_v>H^{-1}$ must continue to stretch without bound, and will get destroyed. Thus in a cosmology we will not have support for the wavefunctional $\Psi[g]$ on vecro configurations of radius $R_v\gtrsim H^{-1}$. 

We illustrate this situation in fig.\ref{fig24}. We define a vecro distribution function $f(R_v)$ describing the probability of vecro configurations of radius $R_v$  in the wavefunctional. In fig.\ref{fig24}(a) we depict schematically the situation with Minkowski  spacetime. There are vecros of all radii $0<R_v<\infty$, so we depict the vecro distribution function as flat. In fig.\ref{fig24}(b) we depict the vecro distrbution function for a cosmology; now the vecros lie only in the range
$0<R_v\lesssim H^{-1}$.

\b

(ii) We now see that the Birkoff theorem is not able to map the cosmological situation to the situation of gravitational collapse. At the classical level we can indeed discard the dust in the region outside finite ball; under the map $T\r -T$ this makes  the cosmology look like a dust ball collapsing to make a black hole. But the vecro part of the quantum wavefunctional does not map between the two cases. The cosmology had no vecros with $R_v\gtrsim H^{-1}$, while if we wanted to replace the region outside the cosmological horizon by asymptotically flat space then we would need $\Psi[g]$ to have support on vecro configurations with radii $R_v$ going to infinity. This resolves the puzzle.

\subsection{Some speculations on other puzzles in cosmology}
 
 While we have resolved the puzzle we had started with, there are many other deep puzzles associated to cosmology, and we can wonder if the vecro hypothesis will be able to resolve these as well. We do not have a detailed understanding of vecro behavior  at present, but we speculate about how the existence of the vecro part of the wavefunctional can impact these issues:
 
 \b
 
 (a) String theory admits a very large number of compactifcations from 10-d to 4-d, and one question has always been: what selects a particular compactification for our universe?
 
 Fuzzball microstates are configurations where the compact directions do not factor into a 3+1 spacetime and  a fixed internal space. Rather, the compact directions mix nontrivially with the noncompact ones to produce the overall microstate. Since fuzzball wavefunctionals are supported on the space of vecro configurations, vecros have a similar structure; i.e., there is no unique shape for the compact directions that exists everywhere over a vecro configuration; the compact directions keep fluctuating from position to position and mixing with the noncompact directions.  
 
  This suggests that the full vacuum wavefunctional should not be peaked around a given compactification, but should be a wavefunctional that is spread over the entire space of compactifications. It is possible that the spreading of the wavefunctional over this space of compactifications drives the cosmological constant $\Lambda$ towards zero.\footnote{Tunneling between flux compactifications has been studied; see for example \cite{tunnelcompact}.}
 
 \b
 
 (b) Inflation requires a field -- the inflaton -- to provide the energy to drive the exponential expansion of the universe. It is not clear which fundamental field could play this role. But it is possible that the vecro part of the wavefunctional could itself provide a vacuum energy that would drive inflation.
 
We have noted that in a cosmology the vecro wavefunctional is limited to the range $0<R_v\lesssim H^{-1}$. Suppose we modify this wavefunctional to one that has a smaller spread; i.e., it is supported on the range 
\be
0<R_v\lesssim \alpha H^{-1}, ~~~\alpha<1
\label{cosmoone}
\ee
 The energy of this wavefunctional is higher than the energy of the vacuum. We have adopted the scale (\ref{masseq}) for the energy contributed by order unity deformations of the vecro distribution function. Setting $R_v\sim H^{-1}$ we find  that the extra energy inside a cosmological horizon is
\be
\Delta E \sim {H^{-(d-2)}\over G}
\ee
This corresponds to an energy density
\be
\Delta \rho \sim {1\over H^{-d}}{H^{-(d-2)}\over G}\sim {H^2\over G}
\ee
which is of order the closure density (eq. (\ref{eqcosmobh})).  In \cite{nagpur} it was conjectured that this extra energy density could be the driver of inflation.

One might ask why the distribution function (\ref{cosmoone}) does not relax to a distribution with $\alpha=1$, which corresponds to the case where we have no additional energy from the vecro part of the wavefunctional. The answer is that if the universe is expanding exponentially on a timescale $\Delta T \sim H^{-1}$, then it is difficult for vecro correlations to grow on scales $\sim H^{-1}$. Thus we can get stuck in a phase where the vecro distribution function has the form (\ref{cosmoone}), and thus provides the energy needed for the universe to inflate.\footnote{For discussions of inflation and its alternatives, see for example \cite{inflation}.}

\b

(iii) In cosmology we also have the `coincidence' problem: why is the cosmological constant of a magnitude that is relevant to the expansion dynamics at the present epoch, and not at any earlier epoch? 
 
A special aspect of the current epoch is that matter has clumped, and so is not very uniform unless we go out to supercluster scales. With the vecro hypothesis, we have a new ingredient to consider in cosmological dynamics: the evolution of the vecro distribution function $f$. In a homogeneous cosmology we took $f$ to be only a function of the vecro radius $R_v$. In an inhomogeneous cosmology $f$ could also depend on the position $x$: the width of the function in the variable $R_v$ could be smaller in regions with higher density. This could lead to an effective $f(R_v)$ that is of the type (\ref{cosmoone}), which in turn is equivalent to an effective cosmological constant.\footnote{It has been debated whether inhomogeneities in the matter distribution could furnish an effective $\Lambda$ \cite{buchert,branmajum,waldrebut}. Note that our question here is a little different: we are asking if the inhomogeneities of matter lead to inhomogeneities in the vecro distribution function $f$ which describes {\it vacuum fluctuations}, and if this alteration of $f$ can act as an effective $\Lambda$.}

\section{Discussion}
  
  We have proposed a picture of the quantum vacuum that resolves several puzzles associated with the physics of horizons. We have proposed that the vacuum wavefunctional $\Psi_0[g]$ has a part supported on configurations that are extended compression-resistant structures. Oscillatory wavefunctionals over this space of structures gives the fuzzball microstates of black holes. But the vacuum wavefunctional also has an `under the barrier' tail  -- the `vecro' part -- that has support over such configurations. The amplitude at any point of such a  tail will be small since the configurations of interest typically have a high potential energy.  But the phase space of these configurations is  vast since they have to support the $Exp[S_{bek}(M)]$ microstates of black holes.  Thus a large part of the norm of the vacuum wavefunctional can be in the tail, and this is the part that gives rise to nontrivial quantum gravity effects in situations where  the classical theory predicts a horizon. 
  
We have considered three puzzles that give rise to sharp contradictions; thus we {\it must} resolve these conflicts   in our theory of quantum gravity. Let us summarize the puzzles and the resolutions we find under the vecro hypothesis: 

\b

(i) {\bf The problem of unbounded entropy:} Many arguments suggest that the Bekenstein relation
\be
S_{bek}={A\over 4G}
\ee 
gives an upper bound on the entropy $S$ that can be put in the region $r\le 2GM$ with an energy budget $M$.  But with the traditional black hole metric, we can construct a spacelike slice that holds an {\it arbitrarily} large entropy $S$ in the region $r<2GM$ while keeping the energy at $M$. Such a slice  is smooth everywhere and the matter on it satisfies all the usual requirements for semiclassical physics to be valid. The puzzle then is: should we give up the idea that $S$ is bounded when we bound the energy and the radius of a region? Or does some feature of the quantum gravity theory disallow the slices used in the construction of unbounded entropy states?

\b

{\it Resolution:} We have to look at the vecro part of the vacuum wavefunctional on  slices with the shape needed to get unbounded entropy. We focus on a class of vecro fluctuations  with radius $R_v$ of order the horizon radius $R_0$. When we stretch the slice  to create the space needed to hold the entropy carrying quanta, we increase the energy of this class of vecros, and deform the vacuum wavefunctional to one with higher energy. This energy increase is of the same order as the mass $M$ of the hole. Since the total energy on the slice is thus more than the mass $M$ of the hole, such slices are not  valid hypersurfaces  in the black hole geometry.  

We also note that if we do {\it not} have extended size fluctuations (vecros) in the vacuum, then the energy of a smooth hypersurface will  be given by integrating over contributions from  local regions. In this case there will be no large correction to the semiclassical energy $M$ on the slice; moreover, all niceness conditions characterizing the validity of the semiclassical approximation will be local ones that hold  on the unbounded entropy slices. Since we have already checked that the problem  is stable against all sources of small corrections to semiclassical physics, we cannot resolve the puzzle if we do not have extended correlations like those invoked in the vecro hypothesis.

\b

(ii) {\bf The information paradox:} A collapsing shell appears to feel nothing special as its particles pass through the horizon. Once the shell is inside the horizon, causality prevents any novel physics at the shell from altering the evolution of fields at the horizon. Entangled pairs are created at the horizon, and the entanglement entropy $S_N$ of the emitted radiation with the remaining hole keeps rising. The small corrections theorem  tells us that no small correction to this emission process can prevent this monotonic rise of entanglement. This leads to a sharp problem when the mass of the remaining hole approaches zero.

\b

{\it Resolution:} When the shell of radius $R(\tau)$ passes through its horizon radius $R_0$, the vecro component of the vacuum wavefunctional in the region $R(\tau)<r<R_0$ gets destabilized. The vecros are virtual fluctuations that are extended objects with different radii $R_v$. Any vecro structure inside the horizon must keep squeezing to smaller radii due to the inward pointing structure of light cones in this region. The extended nature of the vecro fluctuations allows them to recognize the existence  of a trapped surface, thus bypassing the equivalence principle which focuses only on local dynamics. The compression-resistance of the vecros causes their wavefunctional to deform significantly, changing the spacetime in the region $R(\tau)<r<R_0$ to a linear superposition of fuzzballs in a timescale of order the crossing time $\sim R_0$. The fuzzballs have no horizon and radiate from their surface like normal objects, so there is no information paradox. 

Note that the underlying theory maintains causality and locality; interactions are local and signals do not propagate faster than light. The vecro structures are extended scale correlations in the vacuum wavefunctional that have had time to form since the spacetime existed for a long time before the collapse of the shell. 

We have noted in section \ref{secc2} that there is no resolution to the puzzle if we assume that all quantum fluctuations in the vacuum are confined to a bounded length scale like $\sim l_p$, and if we assume causality to leading order in gently curved spacetime. Thus if we wish to preserve leading order causality,   we must necessarily have  order unity corrections from extended scale vacuum fluctuations; i.e., from the vecro part of the vacuum wavefunctional.

\b

(iii) {\bf Are all horizons similar?}  The formal computations of black hole thermodynamics can be applied to different kinds of horizons: (a) the black hole horizon (b) the Rindler horizon  (c) cosmological horizons including the de Sitter horizon. Is the role of quantum gravity the same at these different horizons? In particular, observations of the sky show no large violations of semiclassical physics at the cosmological horizon; how should we reconcile this with the need for large corrections at the black hole horizon?

\b

{\it Resolution:} The Rindler horizon is obtained by choosing a coordinate patch in the vacuum; thus the associated energy is $M=0$. A large black hole horizon may look locally similar to the Rindler horizon, but this limit in obtained for $M$ going to {\it infinity}. The fuzzball paradigm says that these two limits are not the same; the large mass $M$ in the black hole case leads to the horizon region being filled with fuzzball structure, while the Rindler coordinates just parametrize empty space. 

The cosmological horizon and the black hole horizon both contain  nonzero energy in their interiors. But the physics at these two horizons is not the same because the vecro part of the vacuum wavefunctional is different. 
With the cosmology, the expansion limits the vecros radius to within the cosmological horizon radius $0<R_v\lesssim H^{-1}$, while asymptotically flat spacetime has vecros of all radii $R_v$. This difference in the vecro distribution function between the cosmology and the black hole  allows us to have large quantum gravity effects (leading to fuzzball formation) when a shell undergoes gravitational collapse in flat space but no large quantum gravity effects at the cosmological horizon: the vecros that become destabilized in the former case are absent in the latter. 

In section \ref{cosmopuzzle} we noted that the small corrections theorem allows us to make a precise map from the cosmological situation to the black hole puzzle. The black hole puzzle cannot be resolved (while maintaining leading order causality in gently curved spacetime) unless we have extended scale vecro type fluctuations in the wavefunctional. Thus we argue that we cannot resolve the conflict with cosmology if we do not use the order unity effects generated by differences in the vecro distribution functions $f$ between the cosmology and the black hole in asymptotically flat spacetime.

\b

We note that  the above resolution of (iii) has brought in the role of quantum gravity at the scale of the cosmological horizon;  we have speculated on some ways that such a role could be relevant in understanding other issues in cosmology including the cosmological constant problem.

 \section*{Acknowledgements}

I would like to thank for useful discussions Robert Brandenberger, Sumit Das, Patrick Dasgupta,  Bin Guo, G.~'t Hooft, Oleg Lunin, Emil Martinec, Ken Olum,  Mohammed Sami, Sudipta Sarkar, Anjan Sen, Joan Sola,  Terry Tomboulis, David Turton, Alex Vilenkin,  Amitabh Virmani and Robert Wald. This work is supported in part by DOE grant DE-SC0011726.

\pagebreak

\appendix

\section{The electric field of a collapsing shell}\label{appa}

Consider a particle with charge $q$ in $d+1$ spacetime dimensions. We single out one direction $x$ as the direction along which the charge will move; let the orthogonal space directions be labelled $y_1, \dots y_{d-1}$. We write $\rho=(y_1^2+\dots y_{d-1}^2)^\h$ for the distance of a point from the $x$-axis. 

\subsection{Electric field of a moving point charge}

First let the particle be at rest at the origin. The electric potential  at the point $(x, y_1, \dots y_{d-1})$ is
\be
\Phi=-A_t={kq\over (x^2+\rho^2)^{d-2\over 2}}
\ee
The electric field is given by $E_a=F_{at}=\p_a A_t$. Consider the field at the point $(x, y_1, y_2, \dots, y_{d-1})=(x, y, 0, \dots 0)$. We find
\be
E_x={(d-2)kqx\over (x^2+\rho^2)^{d\over 2}}, ~~~E_{y_1}={(d-2)kqy\over (x^2+\rho^2)^{d\over 2}}, ~~~E_{y_2}=\dots = E_{y_{d-1}}=0
\ee

Next, we wish to find the electric field of the charge $q$ when it is moving with speed $v$ along the positive $x$ direction, at the time when the charge is passing the origin. To get this field we go to a frame with coordinates $(t', x', y'_i)$ given through 
\be
t=\gamma(t'-vx'),~~~x=\gamma (x'-v t'), ~~~ y_i=y'_i
\ee
In the new frame we will compute the electric field at the point $P$ given by
\be
(t', x', y'_1, y'_2\dots y'_{d-1})~=~(0, X, Y, 0, \dots 0)
\ee
This corresponds to the following coordinates in the old frame
\be
t=-v\gamma X, ~~~~ x=\gamma X, ~~~y_1=Y, ~~~y_2=\dots = y_{d-1}=0
\ee
The electric field at $P$ will lie in the $x'-y'_1$ plane. It will have a component along the $x'$ direction which we call $E_{\parallel}$ since it is along the direction of motion of the charge, and a component along the $y'_1$ direction which we call $E_{\perp}$ since it is perpendicular to the direction of motion. We find
\be
E_{\parallel}=E'_{x'}=F'_{x't'}=F_{xt}={(d-2)kqx\over (x^2+\rho^2)^{d\over 2}}={(d-2)kq\gamma X\over (\gamma^2 X^2+\rho^2)^{d\over 2}}
\label{zthree}
\ee
\be
E_\perp=E'_{y'_1}=F'_{y'_1t'}=\gamma F_{y_1t}=\gamma {(d-2)kqy\over (x^2+\rho^2)^{d\over 2}}= {(d-2)kq\gamma Y\over (\gamma^2 X^2+\rho^2)^{d\over 2}}
\label{zfour}
\ee
where $\gamma=(1-v^2)^{-\h}$. 

\subsection{Electric field of a shell}

  \begin{figure}[htbp]
\begin{center}
\includegraphics[scale=.5]{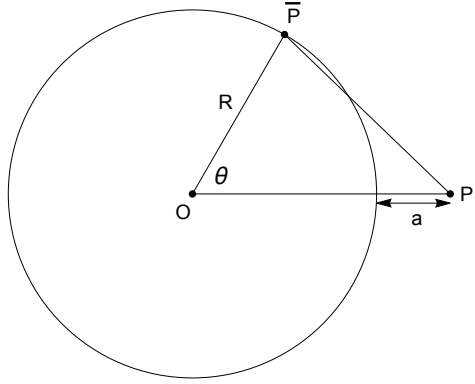}\hskip20pt
\end{center}
\caption{The circle denotes a spherical shell of charged particles falling radially inwards at speed $v$. We compute the contribution to the electric field from a point like $\bar P$ and then integrate over the position of $\bar P$.}
\label{fig4}
\end{figure}

Consider a shell with radius $R$, charge $Q$, collapsing with all charges moving radially inwards at speed $v$. The surface charge density is
\be
\sigma = {Q\over R^{d-1}\Omega_{d-1}}
\ee
We use radial coordinates $(r, \theta, \phi_k)$ where $0<\theta<\pi$ runs along a diameter of the sphere, and $\phi_k$ are coordinates on the $d-2$ dimensional sphere at each value of $\theta$. We wish to find the field at the point $P$ given by $(r, \theta, \phi_k)=(R+a,0, \vec 0)$; see fig.\ref{fig4}. Consider the contribution to this field by a charge $q$ on the shell at the point $\bar P$ with coordinates $(R, \theta, \vec 0)$. The origin, $P$ and $\bar P$ form a plane. All the vectors in the computation below lie in this plane, so we just write the polar coordinates $(r, \theta)$ of this plane in what follows, ignoring the $\phi_k$ until we have to integrate over the shell. It is also convenient to set up Cartesian corodinates on this plane, with $x_1$ along $OP$ and $x_2$ orthogonal to $x_1$.

\begin{figure}[htbp]
\begin{center}
\includegraphics[scale=.7]{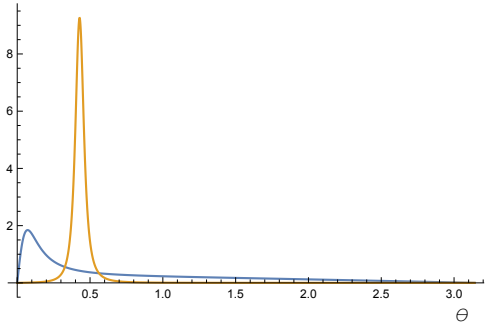}\hskip20pt
\end{center}
\caption{We plot the function (\ref{appfunction}) which gives the contribution from different angles $\theta$ to the field at $P$. The  curve with broad support describes infall with $v=0.2$, and the  curve with narrow support describes infall with $v=0.999$.}
\label{fig3}
\end{figure}

In these Cartesian coordinates, the separation vector from $\bar P$ to $ P$ is
\be
\vec W = (R(1-\cos\theta)+a, -R\sin\theta)
\ee
The unit vector along the direction of the velocity of the charge at $\bar P$ is 
\be
\hat n = (-\cos\theta, -\sin\theta)
\ee
Thus the component of the separation vector $\vec W$ along the velocity of the charge is
\be
X=\vec W \cdot \hat n =R-(R+a)\cos\theta
\ee
The component of $\vec W$ perpendicular to the velocity has magnitude
\be
Y=|\hat W \times \hat n|=(R+a)\sin\theta
\ee
Using (\ref{zthree}),(\ref{zfour}), we see that the parts of the field parallel and perpendicular to the motion of the charge are
\be
E_{||}={(d-2)kq\gamma [R-(R+a)\cos\theta]\over (\gamma^2 [R-(R+a)\cos\theta]^2+[(R+a)\sin\theta]^2)^{d\over 2}}
\ee
\be
E_\perp= {(d-2)kq\gamma[(R+a)\sin\theta]\over (\gamma^2 [R-(R+a)\cos\theta]^2+[(R+a)\sin\theta]^2)^{d\over 2}}
\ee
From the geometry of fig.\ref{fig4} we see that the component of the electric field at $P$ in the radially outward direction is
\be
E(\theta)=-E_{||}\cos\theta + E_\perp \sin\theta
\label{zfive}
\ee
When we integrate over all charges, the other components of $\vec E$ will cancel, so we just need to integrate  (\ref{zfive}) over different positions of $\bar P$. The charges in a strip around the angular position $\theta$ sum up to the charge
\be
dq= \sigma dA = \sigma R^{d-1}(\sin\theta)^{d-2}\Omega_{d-2}d\theta
\ee
Thus  the total field at $P$ is $\vec E= E_T \hat r$ with 
\be
E_T=\int_{\theta=0}^\pi d\theta [\sigma R^{d-1}(\sin\theta)^{d-2}\Omega_{d-2}] E(\theta)=
 {(d-2)kQ\over (R+a)^{d-1}}
\ee
This is just the electric field for a point charge $Q$ at the origin, as required by the Gauss law. Our interest however is in seeing which parts of the shell contribute significantly to this total field $E_T$. For this purpose we plot in fig.\ref{fig3} the integrand 
\be
f(\theta)=[\sigma R^{d-1}(\sin\theta)^{d-2}\Omega_{d-2}] E(\theta)
\label{appfunction}
\ee
 as a function of $\theta$, for different speeds $v$. We see that for small $v$ all values of $\theta$ contribute, while as $v$ tends towards $1$, the integrand gets peaked around a narrow ring at a specific value of  $\theta$. 
 
 The limit $v\r 1$ can be understood analytically as follows. We expand $f(\theta)$ as a series in ${1/\gamma}$. The leading term is at order $1/\gamma^2$, with a coefficient which diverges at 
\be
a=R \left ( {1\over \cos\theta}-1 \right )
\ee
Thus the contribution to the field gets peaked infinitely sharply at the value of $\theta$ given through the above relation. From the geometry of fig.\ref{fig4}, we see that this $\theta$ gives  precisely the  ring on the shell from which the Aichelburg-Sexl shock fronts contribute to  the field at $P$.

\section{Cosmological expansion}\label{appb}

We consider a flat dust cosmology in $d+1$ spacetime dimensions
\be
ds^2=-dT^2+a^2(T)(dr^2+r^2 d\Omega_{d-1}^2)
\label{qtenp}
\ee
The energy density of the dust, measured in a local orthonormal frame where the dust is at rest,   has the form
 \be
 T^0{}_0\equiv \rho_0={C\over a^d}
 \label{qtenq}
 \ee
 The cosmological metric has  the expansion
 \be
 a=a_0 T^{2\over d}
 \label{qten}
 \ee
where
\be
a_0=\left ( {4\pi dGC\over d-1}\right )^{1\over d} 
\ee
Consider the spacelike slice $T=T_0$.   Mark out a  ball around $\vec r=0$ of proper radius $R_b=a(T) r_b$. The mass of dust  in this ball is
\be
M_b=\rho_0 {\Omega_{d-1}\over d}(R_b)^{d}={C} {\Omega_{d-1} \over d}r_b^{d}
\label{qfourtb}
\ee
By the Birkoff theorem, the evolution of the dust in this ball is unaffected by the dust outside the ball. Let us replace the spacetime outside the ball with empty asymptotically flat space. The metric outside the dust ball is then the  Schwarzschild metric
 \be
 ds^2_S=-(1-\left ({R_0\over R}\right ) ^{d-2})dT^2+ {dR^2\over (1-\left ({R_0\over R}\right ) ^{d-1}} + R^2 d\Omega_{d-2}^2
 \label{qtenqq}
 \ee
 with
 \be
R_0=\left ({16\pi GM_b\over (d-1)\Omega_{d-1}}\right )^{1\over d-2}
\label{qthir}
\ee

  Our question is the following. Let $T_0$ be the time today. As we look towards earlier times $T<T_0$ we see a larger and larger sphere out of our dust universe. Suppose $T_1$ is the earliest time that we can get clear signals from. We  ask two questions: (a) What is the proper radius $R_p$ of the dust ball that we can see at time $T_1$?  (a) How big was the Schwarzschild radius $R_s$ of this dust ball ?  We will then compute the ratio
  \be
  {\mathcal C} ={R_p\over R_s}
  \ee
  Thus will tell us how compressed this observed dust ball is compared to its Schwarzschild radius. 
  
  \subsection{The particle horizon}
  
  To answer the first question, we  compute the particle horizon, which tells us how far we can see today. Consider an observer O at $T=T_0, \vec r=0$. Consider a light ray starting at times $T_1<T_0$ and travelling radially towards O. Along this ray
\be
dT=-a(T) dr
\ee
  Thus the comoving coordinate elapsed along the path of the light ray  is given by
\be
r_p=\int_{T'=T_1}^{T_0} {dT' \over a(T')}
\ee
For the expansion (\ref{qten}) we get
 \be
 r_p=\int_{T'=T_1}^{T_0} {dT' \over a(T')}={1\over a_0}{d\over d-2} [T_0^{{(d-2)\over d}}-T_1^{{(d-2)\over d}}]\approx {1\over a_0}{d\over d-2} T_0^{{(d-2)\over d}}
\label{qtwq}
\ee
where in the last step we have noted that we typically have  $T_1\ll T_0$.

\subsection{Horizon radius of the observed dust ball}

The dust ball that we can see, looking back to times $T_1$, has a radius in comoving coordinates given by   (\ref{qtwq}). The proper radius of this ball is
\be
R_p=a(T_1)r_p
\ee
 The mass inside this ball  is 
\be
M=\rho(T_1)  {\Omega_{d-1}\over d}(a(T_1) r_p)^{d}
\ee
The Schwarzschild radius for this mass is 
\be
R_s=\left ({16\pi GM\over (d-1)\Omega_{d-1}}\right )^{1\over d-2}
\ee

We then find
\be
{\cal C}\equiv {R_p\over R_s}=\left ({(d-2)\over 2}\right )^{2\over d-2}\left ( {T_1\over T_0}\right )^{2\over d}
\ee
 We set
 \be
T_0\approx 1.4\times 10^{10}\, yrs, ~~~T_1\approx 4\times 10^5 \, yrs
\ee
This gives
\be
{\cal C}\sim 2\times 10^{-4}
\ee
Thus we can observe a ball of matter in the sky that has been compressed to well within its Schwarzschild radius.

\subsection{The cosmological horizon}\label{cosmohorizon}

The cosmological horizon  radius is defined by
  \be
  H^{-1}={a\over \dot a} 
  \ee
  As we now recall, this is  the radius of the marginally anti-trapped surfaces in the cosmology. 
   Consider the sphere around $\vec r=0$ with comoving radius $\t r(T)$. The proper radius of this sphere is
\be
\t R(T)=a(T) \t r(T)
\ee
The area of the sphere is
\be
A=\Omega_{d-1} \t R^{d-1}=\Omega_{d-1} [a(T) \t r(T)]^{d-1}
\ee
Consider light rays travelling radially inwards from the sphere. Along such a ray
\be
a(T) dr=-dT, ~~~{dr\over dT}=-{1\over a(T)}
\ee
We compute the change in the area $A$ of the spherical surface as we follow these inward pointing radial geodesics:
\be
{dA\over dt}=\Omega_{d-1}(d-1)[a(T) \t r(T)]^{d-2}[\dot a (T) \t r(T)+a(T) \dot {\t r}(T)]
\ee
Thus the expansion of the outward directed null rays is zero when
\be
\dot a (T) \t r(T)+a(T) \dot {\t r}(T)=0
\ee
which gives
\be
{\dot a(T)\over a(T)}=-{\dot {\t r}(T)\over \t r(T)}={1\over a(T) \t r(T)}={1\over \t R(T)}
\ee
Thus the marginally anti-trapped surface appears at the radius
 \be
\t R=\left ( {\dot a(T)\over a(T)}\right )^{-1}\equiv H^{-1}
\label{eqhor}
\ee
In comoving coordinates, this radius  is
\be
\t r={1\over \dot a}
\ee
With the expansion (\ref{qten}),
\be
\t r={d\over 2 a_0} T^{d-2\over d}
\label{cosmohor}
\ee
Thus we see that the comoving horizon radius $\t r$ increases with time. As the universe expands, the cosmological horizons merge to create larger cosmological horizons. Note that the proper radius of the horizon is $a(T) \t r \sim T$. The Hubble relation says that the radius of the cosmological horizon is 
$R_{Hubble}\equiv H^{-1}\sim (G\rho)^\h$. The mass of the dust inside the cosmological horizon is $M_{Hubble}\sim \rho H^{-d}\sim G^{-1} H^{-(d-2)}$. Thus the matter inside a cosmological horizon is of the order needed to make a black hole with radius equal to the cosmological horizon
\be
{GM_{Hubble}\over R_{Hubble}^{d-2}}\sim 1
\label{eqcosmobh}
\ee

Let us now consider the observations of the cosmological horizon. We cannot see past the cosmological horizon today, since the radius of this horizon $H^{-1}$ is of the same order as the particle horizon $R_p$, which in turn is the furthest point from which light has been able to reach us since the big bang. But we can see regions in the sky which at an earlier time $T_1$  had a size equal to the cosmological horizon at that time.   The cosmological horizon radius at  time $T_1$ is, in comoving coordinates (eq. (\ref{cosmohor}))
\be
r_1(T_1)={d\over 2 a_0} T_1^{d-2\over d}
\ee
On the other hand (\ref{qtwq}) gives the comoving radius of the ball at time $T_1$ which we can we can observe today; i.e., light starting from the edge of this ball at time $T_1$ will reach us at time $T_0$.  
We find
 \be
\t  {\cal C}\equiv {r_1(T_1)\over r_p(T_1)}\sim \left ( {T_1\over T_0}\right ) ^{{d-2\over d}}
 \ee
 If we set
 \be
T_0\approx 1.4\times 10^{10}\, yrs, ~~~T_1\approx 4\times 10^5 \, yrs
\ee
then we find
\be
\t {\cal C}\sim {1\over 30}
\ee
Thus we can observe several patches in the sky which had the size of the cosmological horizon at the time that when they emitted the light by which we see them. Since we see no evidence of large violations of semiclassical physics in such observations, we can  say that semiclassical physics holds to leading order at cosmological horizons.



\newpage

\begin{thebibliography}{99}

 \bibitem{hawking}
  S.~W.~Hawking,
  Commun.\ Math.\ Phys.\  {\bf 43}, 199 (1975)
  [Erratum-ibid.\  {\bf 46}, 206 (1976)];
  S.~W.~Hawking,
  Phys.\ Rev.\  D {\bf 14}, 2460 (1976).
  
  \bibitem{bek}
J.~D.~Bekenstein,
Phys.\ Rev.\ D {\bf 7}, 2333 (1973).
%

\bibitem{bags} 
 J.A. Wheeler in  
  ``Relativité, Groupes et Topologie : Proceedings, Ecole d'été de Physique Théorique, Session XIII, Les Houches, France, Jul 1 - Aug 24, 1963,'' ed. C.~DeWitt and B.~DeWitt;
  D.~Marolf,
  Gen.\ Rel.\ Grav.\  {\bf 41}, 903 (2009)
  doi:10.1007/s10714-008-0749-7
  [arXiv:0810.4886 [gr-qc]];
  Z.~Fu and D.~Marolf,
  JHEP {\bf 1911}, 040 (2019)
  doi:10.1007/JHEP11(2019)040
  [arXiv:1909.02505 [hep-th]].

   
  \bibitem{monsters}
  S.~D.~H.~Hsu and D.~Reeb,
  Phys.\ Lett.\ B {\bf 658}, 244 (2008)
  doi:10.1016/j.physletb.2007.09.021
  [arXiv:0706.3239 [hep-th]];
  Y.~C.~Ong and P.~Chen,
  JHEP {\bf 1307}, 147 (2013)
  doi:10.1007/JHEP07(2013)147
  [arXiv:1304.3803 [hep-th]].

\bibitem{ho} 
  P.~M.~Ho and Y.~Matsuo,
  JHEP {\bf 1906}, 057 (2019)
  doi:10.1007/JHEP06(2019)057
  [arXiv:1905.00898 [gr-qc]].

\bibitem{cern}
  S.~D.~Mathur,
  Class.\ Quant.\ Grav.\  {\bf 26}, 224001 (2009)
  [arXiv:0909.1038 [hep-th]].
  
\bibitem{bubble} 
  E.~Witten,
  Nucl.\ Phys.\ B {\bf 195}, 481 (1982).
  
  
\bibitem{lm4} 
  O.~Lunin and S.~D.~Mathur,
  Nucl.\ Phys.\ B {\bf 623}, 342 (2002)
  doi:10.1016/S0550-3213(01)00620-4
  [hep-th/0109154];


  \bibitem{fuzzballs}
  O.~Lunin, J.~M.~Maldacena and L.~Maoz,
  hep-th/0212210;
  S.~D.~Mathur, A.~Saxena and Y.~K.~Srivastava,
  Nucl.\ Phys.\ B {\bf 680}, 415 (2004)
  doi:10.1016/j.nuclphysb.2003.12.022
  [hep-th/0311092].
 S.~D.~Mathur,
  Fortsch.\ Phys.\  {\bf 53}, 793 (2005)
  [arXiv:hep-th/0502050];\\
I.~Kanitscheider, K.~Skenderis and M.~Taylor,
  ``Fuzzballs with internal excitations,''
  arXiv:0704.0690 [hep-th];
 I.~Bena and N.~P.~Warner,
  ``Black holes, black rings and their microstates,''
  Lect.\ Notes Phys.\  {\bf 755}, 1 (2008)
  [arXiv:hep-th/0701216];
  B.~D.~Chowdhury and A.~Virmani,
  ``Modave Lectures on Fuzzballs and Emission from the D1-D5 System,''
  arXiv:1001.1444 [hep-th];
  I.~Bena, S.~Giusto, E.~J.~Martinec, R.~Russo, M.~Shigemori, D.~Turton and N.~P.~Warner,
  Phys.\ Rev.\ Lett.\  {\bf 117}, no. 20, 201601 (2016)
  doi:10.1103/PhysRevLett.117.201601
  [arXiv:1607.03908 [hep-th]].

  
\bibitem{gibbonswarner} 
  G.~W.~Gibbons and N.~P.~Warner,
  Class.\ Quant.\ Grav.\  {\bf 31}, 025016 (2014)
  doi:10.1088/0264-9381/31/2/025016
  [arXiv:1305.0957 [hep-th]].


\bibitem{mexample} 
  S.~D.~Mathur,
  Int.\ J.\ Mod.\ Phys.\ D {\bf 25}, no. 12, 1644018 (2016)
  doi:10.1142/S0218271816440181
  [arXiv:1609.05222 [hep-th]].

  \bibitem{tunnel}
  S.~D.~Mathur,
  arXiv:0805.3716 [hep-th];
  S.~D.~Mathur,
  Int.\ J.\ Mod.\ Phys.\  D {\bf 18}, 2215 (2009)
  [arXiv:0905.4483 [hep-th]];
  P.~Kraus and S.~D.~Mathur,
  Int.\ J.\ Mod.\ Phys.\ D {\bf 24}, no. 12, 1543003 (2015)
  doi:10.1142/S0218271815430038
  [arXiv:1505.05078 [hep-th]];
  I.~Bena, D.~R.~Mayerson, A.~Puhm and B.~Vercnocke,
  JHEP {\bf 1607}, 031 (2016)
  doi:10.1007/JHEP07(2016)031
  [arXiv:1512.05376 [hep-th]].
  
\bibitem{masoumi1} 
  A.~Masoumi and S.~D.~Mathur,
  Phys.\ Rev.\ D {\bf 90}, no. 8, 084052 (2014)
  doi:10.1103/PhysRevD.90.084052
  [arXiv:1406.5798 [hep-th]].

\bibitem{causality1} 
  S.~D.~Mathur,
  Gen.\ Rel.\ Grav.\  {\bf 51}, no. 2, 24 (2019)
  doi:10.1007/s10714-019-2505-6
  [arXiv:1703.03042 [hep-th]];
  
\bibitem{causality2} 
  S.~D.~Mathur,
  arXiv:1705.06407 [hep-th].

  
\bibitem{buchdahl} 
  H.~A.~Buchdahl,
  Phys.\ Rev.\  {\bf 116}, 1027 (1959).
  doi:10.1103/PhysRev.116.1027


\bibitem{observe} 
  B.~Guo, S.~Hampton and S.~D.~Mathur,
  JHEP {\bf 1807}, 162 (2018)
  doi:10.1007/JHEP07(2018)162
  [arXiv:1711.01617 [hep-th]].

\bibitem{afshordi} 
  J.~Abedi, H.~Dykaar and N.~Afshordi,
  arXiv:1612.00266 [gr-qc];
\bibitem{Bianchi:2019lmi} 
  M.~Bianchi, M.~Casolino and G.~Rizzo,
  arXiv:1904.01097 [hep-th];
  V.~Cardoso and P.~Pani,
  Living Rev.\ Rel.\  {\bf 22}, no. 1, 4 (2019)
  doi:10.1007/s41114-019-0020-4
  [arXiv:1904.05363 [gr-qc]].


 
 
 
 
 
 
 
 
 
  
   \bibitem{page}
  D.~N.~Page,
  ``Expected Entropy Of A Subsystem,''
  Phys.\ Rev.\ Lett.\  {\bf 71}, 1291 (1993)
  [arXiv:gr-qc/9305007].

   
\bibitem{hawkingreverse} 
  S.~W.~Hawking,
  Phys.\ Rev.\ D {\bf 72}, 084013 (2005)
  doi:10.1103/PhysRevD.72.084013
  [hep-th/0507171].
  
  
  
\bibitem{adscft} 
  J.~M.~Maldacena,
  ``The Large N limit of superconformal field theories and supergravity,''
  Adv.\ Theor.\ Math.\ Phys.\  {\bf 2}, 231 (1998)
  [hep-th/9711200];
  S.~S.~Gubser, I.~R.~Klebanov and A.~M.~Polyakov,
  ``Gauge theory correlators from noncritical string theory,''
  Phys.\ Lett.\ B {\bf 428}, 105 (1998)
  [hep-th/9802109];
  E.~Witten,
  ``Anti-de Sitter space and holography,''
  Adv.\ Theor.\ Math.\ Phys.\  {\bf 2}, 253 (1998)
  [hep-th/9802150].
  
  
\bibitem{pr} 
  K.~Papadodimas and S.~Raju,
  JHEP {\bf 1310}, 212 (2013)
  doi:10.1007/JHEP10(2013)212
  [arXiv:1211.6767 [hep-th]].
  
\bibitem{cool} 
  J.~Maldacena and L.~Susskind,
  Fortsch.\ Phys.\  {\bf 61}, 781 (2013)
  doi:10.1002/prop.201300020
  [arXiv:1306.0533 [hep-th]].

\bibitem{dual} 
  S.~D.~Mathur,
  arXiv:1402.6378 [hep-th].

\bibitem{dmcompare}
S.~R.~Das and S.~D.~Mathur,
Nucl.\ Phys.\ B {\bf 478}, 561 (1996)
[arXiv:hep-th/9606185];
J.~M.~Maldacena and A.~Strominger,
Phys.\ Rev.\ D {\bf 55}, 861 (1997)
[arXiv:hep-th/9609026].
%

 
\bibitem{giddings} 
  S.~B.~Giddings,
  ``Nonviolent information transfer from black holes: a field theory parameterization,''
  arXiv:1302.2613 [hep-th].

\bibitem{hps} 
  S.~W.~Hawking, M.~J.~Perry and A.~Strominger,
  Phys.\ Rev.\ Lett.\  {\bf 116}, no. 23, 231301 (2016)
  doi:10.1103/PhysRevLett.116.231301
  [arXiv:1601.00921 [hep-th]].

\bibitem{others} 
  G.~'t Hooft,
  Nucl.\ Phys.\ B {\bf 256}, 727 (1985).
  doi:10.1016/0550-3213(85)90418-3
  A.~Ashtekar and M.~Bojowald,
  Class.\ Quant.\ Grav.\  {\bf 22}, 3349 (2005)
  doi:10.1088/0264-9381/22/16/014
  [gr-qc/0504029];
T.~Vachaspati, D.~Stojkovic and L.~M.~Krauss,
  Phys.\ Rev.\ D {\bf 76}, 024005 (2007)
  doi:10.1103/PhysRevD.76.024005
  [gr-qc/0609024];
  L.~Mersini-Houghton,
  Phys.\ Lett.\ B {\bf 738}, 61 (2014)
  doi:10.1016/j.physletb.2014.09.018
  [arXiv:1406.1525 [hep-th]];
  L.~Buoninfante and A.~Mazumdar,
  Phys.\ Rev.\ D {\bf 100}, no. 2, 024031 (2019)
  doi:10.1103/PhysRevD.100.024031
  [arXiv:1903.01542 [gr-qc]].




 \bibitem{emission}
S.~D.~Mathur,
``Emission rates, the correspondence principle and the information  paradox,''
Nucl.\ Phys.\ B {\bf 529}, 295 (1998)
[arXiv:hep-th/9706151].
%


\bibitem{gas}
  T.~Banks and W.~Fischler,
  hep-th/0102077;
    T.~Banks and W.~Fischler,
  hep-th/0212113;
T.~Banks and W.~Fischler,
  Phys.\ Scripta T {\bf 117}, 56 (2005)
  [hep-th/0310288];
  R.~Brustein and G.~Veneziano,
  Phys.\ Rev.\ Lett.\  {\bf 84} (2000) 5695
  [hep-th/9912055];
   G.~Veneziano,
  JCAP {\bf 0403}, 004 (2004)
  [hep-th/0312182];
   D.~Bak and S.~-J.~Rey,
  Class.\ Quant.\ Grav.\  {\bf 17}, L83 (2000)
  [hep-th/9902173];
  A.~Masoumi and S.~D.~Mathur,
  Phys.\ Rev.\ D {\bf 91}, no. 8, 084058 (2015)
  doi:10.1103/PhysRevD.91.084058
  [arXiv:1412.2618 [hep-th]].

 
 
\bibitem{beyond} 
  S.~D.~Mathur,
  Annals Phys.\  {\bf 327}, 2760 (2012)
  doi:10.1016/j.aop.2012.05.001
  [arXiv:1205.0776 [hep-th]].



\bibitem{univexpand} 
  S.~D.~Mathur,
  Int.\ J.\ Mod.\ Phys.\ D {\bf 12}, 1681 (2003)
  doi:10.1142/S0218271803004031
  [hep-th/0305204].








\bibitem{cosmoessay1} 
  S.~D.~Mathur,
  Int.\ J.\ Mod.\ Phys.\ D {\bf 27}, no. 14, 1847004 (2018)
  doi:10.1142/S0218271818470041
  [arXiv:1805.09852 [hep-th]].

\bibitem{tunnelcompact}
  J.~J.~Blanco-Pillado, D.~Schwartz-Perlov and A.~Vilenkin,
  JCAP {\bf 0912}, 006 (2009)
  doi:10.1088/1475-7516/2009/12/006
  [arXiv:0904.3106 [hep-th]];
\bibitem{Brown:2011gt} 
  A.~R.~Brown and A.~Dahlen,
  Phys.\ Rev.\ D {\bf 85}, 104026 (2012)
  doi:10.1103/PhysRevD.85.104026
  [arXiv:1111.0301 [hep-th]].



\bibitem{nagpur} 
  S.~D.~Mathur,
  arXiv:1812.11641 [hep-th].

\bibitem{inflation}
  A.~H.~Guth,
  Phys.\ Rev.\ D {\bf 23}, 347 (1981)
  [Adv.\ Ser.\ Astrophys.\ Cosmol.\  {\bf 3}, 139 (1987)].
  doi:10.1103/PhysRevD.23.347;
  P.~J.~E.~Peebles and B.~Ratra,
  Rev.\ Mod.\ Phys.\  {\bf 75}, 559 (2003)
  doi:10.1103/RevModPhys.75.559
  [astro-ph/0207347];
  E.~J.~Copeland, M.~Sami and S.~Tsujikawa,
  Int.\ J.\ Mod.\ Phys.\ D {\bf 15}, 1753 (2006)
  doi:10.1142/S021827180600942X
  [hep-th/0603057];
  A.~Gomez-Valent, J.~Sola and S.~Basilakos,
  JCAP {\bf 1501}, 004 (2015)
  doi:10.1088/1475-7516/2015/01/004
  [arXiv:1409.7048 [astro-ph.CO]];
  M.~d.~C.~Bento, O.~Bertolami and A.~A.~Sen,
  Phys.\ Rev.\ D {\bf 67}, 063003 (2003)
  doi:10.1103/PhysRevD.67.063003
  [astro-ph/0210468];
  C.~Armendariz-Picon, V.~F.~Mukhanov and P.~J.~Steinhardt,
  Phys.\ Rev.\ Lett.\  {\bf 85}, 4438 (2000)
  doi:10.1103/PhysRevLett.85.4438
  [astro-ph/0004134];
  G.~R.~Dvali, G.~Gabadadze and M.~Porrati,
  Phys.\ Lett.\ B {\bf 485}, 208 (2000)
  doi:10.1016/S0370-2693(00)00669-9
  [hep-th/0005016];
  A.~De Simone, A.~H.~Guth, M.~P.~Salem and A.~Vilenkin,
  Phys.\ Rev.\ D {\bf 78}, 063520 (2008)
  doi:10.1103/PhysRevD.78.063520
  [arXiv:0805.2173 [hep-th]];
  S.~Kachru, R.~Kallosh, A.~D.~Linde and S.~P.~Trivedi,
  Phys.\ Rev.\ D {\bf 68}, 046005 (2003)
  doi:10.1103/PhysRevD.68.046005
  [hep-th/0301240];
  C.~Csaki, N.~Kaloper, J.~Serra and J.~Terning,
  Phys.\ Rev.\ Lett.\  {\bf 113}, 161302 (2014)
  doi:10.1103/PhysRevLett.113.161302
  [arXiv:1406.5192 [hep-th]];
  N.~Dadhich,
  Fundam.\ Theor.\ Phys.\  {\bf 187}, 73 (2017)
  doi:10.1007/978-3-319-51700-1-7
  [arXiv:1609.02138 [gr-qc]].


\bibitem{buchert} 
  T.~Buchert,
  Gen.\ Rel.\ Grav.\  {\bf 40}, 467 (2008)
  doi:10.1007/s10714-007-0554-8
  [arXiv:0707.2153 [gr-qc]].
  
\bibitem{branmajum} 
  R.~Brandenberger and A.~Mazumdar,
  JCAP {\bf 0408}, 015 (2004)
  doi:10.1088/1475-7516/2004/08/015
  [hep-th/0402205].

\bibitem{waldrebut} 
  A.~Ishibashi and R.~M.~Wald,
  Class.\ Quant.\ Grav.\  {\bf 23}, 235 (2006)
  doi:10.1088/0264-9381/23/1/012
  [gr-qc/0509108].

\end{thebibliography}
\end{document}